\documentclass[onecolumn,11pt]{article}
\usepackage{jheppub}

\usepackage[font={small}]{caption}
\usepackage{subfigure}
\usepackage{graphicx}
\usepackage{dcolumn}

\usepackage{array}
\usepackage{longtable}

\usepackage{float}
\usepackage[normalem]{ulem}

\usepackage{mathrsfs} 

\usepackage{multicol}
\usepackage{cancel}
\usepackage{mathtools}
\usepackage{hyperref}

\usepackage{xcolor}
\usepackage{simplewick}

\usepackage{amsmath}
\usepackage{amsfonts}
\usepackage{amssymb}
\usepackage{graphicx}
\usepackage{simplewick}
\usepackage{hyperref}
\usepackage{multirow}
\usepackage{enumitem}
\usepackage{physics}
\usepackage{booktabs}
\usepackage{xcolor}
\usepackage{dsfont}

\usepackage[compat=1.1.0]{tikz-feynman}
\usepackage{xcolor}
\usepackage[export]{adjustbox}
 \usepackage{slashed}
 \usetikzlibrary{shapes.geometric}

\usepackage{xcolor,pifont}
\newcommand*\colourcheck[1]{%
  \expandafter\newcommand\csname #1check\endcsname{\textcolor{#1}{\ding{52}}}%
}
\colourcheck{blue}
\colourcheck{green}
\colourcheck{red}

\newcommand*\colourmark[1]{%
  \expandafter\newcommand\csname #1mark\endcsname{\textcolor{#1}{\ding{56}}}%
}
\colourmark{blue}
\colourmark{green}
\colourmark{red}

\newcommand{\eg}{{\it e.g.,}\ }
\newcommand{\ie}{{\it i.e.,}\ }

\def\a{\alpha}

\def\Tr{{\rm Tr}}

\def\p{\partial}
\newcommand{\SU}{\text{SU}}

\makeatother

\newcommand{\beq}{\begin{equation}}
\newcommand{\eeq}{\end{equation}}
\newcommand{\bea}{\begin{eqnarray}}
\newcommand{\eea}{\end{eqnarray}}

\newtheorem{thm}{Theorem}[section]

\newtheorem{srule}[thm]{Selection rule}

\def\le{\left(}
\def\ri{\right)}

\def\tr           {\mathop{\rm tr}}

\def\Im           {{\rm Im\hskip0.1em}}

\renewcommand{\eqref}[1]{(\ref{#1})}

\begin{document}

\title{Aspects of Non-Relativistic Quantum Field Theories
}

\author{Stefano Baiguera$^{1}$}

\affiliation{$^1$ Department of Physics, Ben-Gurion University of the Negev, \\ David Ben Gurion Boulevard 1, Beer Sheva 84105, Israel \\}

\emailAdd{baiguera@post.bgu.ac.il}

\abstract{ \sloppy
Non-relativistic quantum field theory is a framework that describes systems where the velocities are much smaller than the speed of light.
A large class of those obey Schr\"{o}dinger invariance, which is the equivalent of the conformal symmetry in the relativistic world.
In this review, we pedagogically introduce the main theoretical tools used to study non-relativistic physics: null reduction and $c \rightarrow \infty$ limits, where $c$ is the speed of light. 
We present a historical overview of non-relativistic wave equations, Jackiw-Pi vortices, the Aharonov-Bohm scattering, and the trace anomaly for a Schr\"{o}dinger scalar.
We then review modern developments, including fermions at unitarity, the quantum Hall effect, off-shell actions, and a systematic classification of the trace anomaly.
The last part of this review is dedicated to current research topics.
We define non-relativistic supersymmetry and a corresponding superspace to covariantly deal with quantum corrections.
Finally, we define the Spin Matrix Theory limit of the AdS/CFT correspondence, which is a non-relativistic sector of the duality obtained via a decoupling limit, where a precise matching of the two sides can be achieved.  }

\maketitle

\setcounter{tocdepth}{1}

\section{Introduction and outline}

Non-relativistic theories received a lot of attention since the early days of physics due to their success in explaining most of the phenomena in Nature that can be experienced in our everyday life.
It was only discovered in the last century that the laws of physics were fundamentally Lorentz-invariant and that general relativity was needed to fully account for the observations in our Universe.
Despite this revolutionary change, non-relativistic theories continued to play an important role in the last century, mainly to describe phenomena happening at low energies (or speed) and to study applications to condensed matter systems.
The topic experienced a recent revival, starting with the seminal work by Son and Wingate on non-relativistic conformal invariance \cite{Son:2005rv}.
Their primary goal was the study of condensed matter systems, such as fermions at unitarity and the quantum Hall effect, that are invariant under a non-Lorentzian symmetry group and do not have a small parameter that can be used to define a perturbative expansion.\footnote{For previous applications of non-relativistic physics to condensed matter systems, \eg see the studies in nuclear physics \cite{Kaplan:1998tg}, settings with cold atoms \cite{Bedaque:1998kg}, and the Efimov effect \cite{Bedaque:1998km,Braaten:2004rn}. }
The major novelty of their approach lied in the application of modern geometric techniques adopted from high-energy physics to condensed matter systems.
These methods include the construction of actions using the general coordinate invariance under non-relativistic symmetries, and the investigation of Feynman diagrams in a QFT framework.
Following these ideas, several theoretical investigations of non-relativistic theories have been performed in the last decades.
A nice collection of reviews that cover these developments is \cite{Oling:2022fft,Hartong:2022lsy,Bergshoeff:2022eog,Grosvenor:2021hkn}.
While they provide an extensive analysis of gravity, string theory, non-Lorentzian symmetry algebras, particle actions and applications to fractons, a review specifically devoted to non-relativistic quantum field theory (QFT) is still missing in the literature.
Our goal is to fill the gap with the present paper.

The concept of non-relativistic theory is very broad.\footnote{The realm of possible non-relativistic field theories has been classified from an effective field theory perspective in \cite{Watanabe:2012hr,Mojahed:2021sxy,Mojahed:2022nrn}, based on either the study of non-Lorentzian Nambu-Goldstone bosons, or of soft theorems.}
While it usually refers to systems whose symmetry group is Galilei instead of Lorentz, sometimes this terminology is also adopted to describe theories invariant under \textit{Lifshitz} transformations
\beq
t' = e^{z \sigma} t \, , \qquad
x'_i = e^{\sigma} x_i \, ,
\label{eq:Lifhsitz_transformations}
\eeq
where $\sigma$ is a constant parameter, while $z$ is called \textit{dynamical exponent} and characterizes the anisotropy between the rescaling of space and time.
Lifshitz-invariant systems play an important role both from a theoretical and an experimental perspective.
Indeed, they provide an anisotropic analogue of scale invariance that describes quantum critical points for certain high-temperature superconductors \cite{Sachdev_2011}, the dynamics in the critical regime of strange metals \cite{coleman2005quantum,Sachdev:2011cs}, and transport properties \cite{Hoyos:2013eza,Hoyos:2013qna,Chapman:2014hja}.
Lifshitz scaling is also a property of certain covariant gravitational theories that have been suggested to admit a controlled UV behaviour (the so-called Horava-Lifshitz gravity) \cite{Horava:2009uw}.
In this review, we will mainly focus on the choice $z=2$, when the symmetries can be enhanced to the Schr\"{o}dinger group, which leaves the Schr\"{o}dinger equation invariant.
One convenient method to derive Schr\"{o}dinger-invariant theories is to perform a dimensional reduction along the null direction of a relativistic theory defined on a Lorentzian manifold with null isometry (this procedure is called \textit{null reduction}) \cite{Duval:1984cj}.
This technology will be extensively used in the present review to find an off-shell formulation for several non-relativistic models.
Other ways to generate non-relativistic theories from a relativistic parent consist of performing a limit $c \rightarrow \infty$ (where $c$ is the speed of light), or an expansion in powers of $c^{-2}$.
The reader can find a detailed treatment of the latter approach applied to gravity in the review \cite{Hartong:2022lsy}, while in this paper we will discuss the $c \rightarrow \infty$ limit and sometimes adopt it to build non-relativistic QFTs.

\vskip 5mm
The main reason to study non-relativistic systems using a QFT formulation is that is that a joint framework will allow to share methods and insights between the relativistic and non-relativistic communities.
Furthermore, QFT provides a theoretical setting where several more methods, compared to QM, are available.
In section \ref{sec:preliminaries}, we motivate the above-mentioned advantages and we introduce the basic ingredients necessary to study Schr\"{o}dinger-invariant systems, which are the non-relativistic counterpart of conformal field theories (CFTs).
We discuss the symmetry structure of the Schr\"{o}dinger group, a non-relativistic version of the state/operator correspondence, the form of correlation functions and the technology to deal with loop corrections.
The above-mentioned tools are sufficient to delve in section \ref{sec:historical_review} into a historical overview of the older applications: the study of non-relativistic fermions, gauge fields, and anyons. This includes exact solutions to the Jackiw-Pi model, the Aharonov-Bohm scattering, and the scale anomaly for a scalar theory with quartic interactions.

Section \ref{sec:modern_NR_QFT} reviews the most common methods used to study non-relativistic systems in the last decade: non-relativistic limits $c \rightarrow \infty$ (section \ref{ssec:nonrel_limit}) and null reduction (section \ref{ssec:null_red}).
We apply these methods in section \ref{sec:modern_applications}
to study several problems of interest in modern physics.
We begin with condensed matter applications, \ie fermions at unitarity and the quantum Hall effect.
Then we find an off-shell formulation for theories involving the non-relativistic fermions and gauge fields introduced above.
We consider a deformed version of null reduction to build QFTs invariant under the SU$(1,n)$ symmetry group, which play an important role in the non-relativistic limit of $M$--theory.
We discuss trace anomalies for Schr\"{o}dinger-invariant field theories.

Supersymmetry (SUSY) has been studied for several years as a candidate to uncover physics beyond the Standard Model, but has not been experimentally observed so far.
There are expectations that SUSY may arise as en emergent symmetry in the IR of certain systems, \eg the tricritical Ising model \cite{FRIEDAN198537}, topological superconductors \cite{Grover:2013rc}, optical lattices \cite{Yu:2010zv} and many others.
Therefore, one may be interested in combining SUSY with non-relativistic symmetries, which frequently arise in condensed matter physics.
From a theoretical perspective, SUSY puts strict constraints on the analytic structure of the effective action, and controls the running of physical couplings along the RG flow, leading to exact results and non-renormalization theorems \cite{Grisaru:1979wc,Seiberg:1993vc}.
Moreover, even if SUSY plays an indirect role in holography, most of the examples where the AdS/CFT correspondence is explicitly tested are supersymmetric.
It is then natural to ask whether the above-mentioned constraints on the quantum corrections are tied to the Lorentz group, or if they can be generalized to other spacetime symmetry groups, such as the Galilean case.
This topic is studied in section \ref{sec:SUSY},
where a superspace formulation is introduced. 
This tool allows for a simpler and systematic computation of the loop corrections to certain theories with super-Schr\"{o}dinger invariance.

The last part of the review is devoted to a recently growing research line, \ie the Spin Matrix Theory (SMT) limit of AdS/CFT duality, see section \ref{sec:SMT}.
The AdS/CFT correspondence between $\mathcal{N}=4$ super Yang-Mills (SYM) with gauge group $\mathrm{SU}(N)$ and type IIB string theory on $\mathrm{AdS}_5 \times S^5$ is the best-understood example which realizes the holographic principle, promising an understanding of how gravity emerges from a fundamental quantum theory \cite{Maldacena:1997re}. 
Evidence in favor of AdS/CFT duality has been accumulating over the last decades, including the impressive achievements of the integrability program when $N=\infty$ \cite{Minahan:2002ve,Beisert:2003tq,Beisert:2010jr}.
However, to get a full understanding of the non-perturbative regime, which includes essential objects such as black holes and D--branes, it is necessary to work at finite
$N$.
The SMT sectors of the AdS/CFT duality consist of decoupling limits that either approach zero-temperature critical points (in the grand-canonical ensemble), or unitarity bounds (in the microcanonical ensemble).
The outcome is that several degrees of freedom of $\mathcal{N}=4$ SYM decouple, and one is restricted to a subsector of the full theory where $N$ is fixed but arbitrary, and a non-perturbative matching between the two sides can be achieved.
An important feature of the SMT limit is that the resulting theories are non-relativistic.

We summarize the main points of the review and propose several future developments in section \ref{sec:discussion}.
Appendix \ref{app:conventions} collects the conventions and a list of acronyms.

Let us conclude with some caveats.
While we will incidentally discuss applications to holography in sections \ref{ssec:geometric_AdS_Schroed} and \ref{ssec:nonrel_holography}, the main focus of the review is on field-theoretical aspects.
The inexpert reader can find a pedagogical introduction of the main theoretical techniques used in non-relativistic physics in sections \ref{sec:preliminaries} and \ref{sec:modern_NR_QFT}.
These will allow the reader to delve into the applications discussed in the remainder of the paper.
Finally, we collect in table \ref{tab:intro} a brief primer for people interested in various topics: we mark with a green check $\,$ \greencheck $\,$ the sections which cover some material on condensed matter physics, supersymmetry, and contemporary research developments, respectively.

\begin{table}[ht]   
\begin{center}   
\begin{tabular}  {|p{35mm}|c|c|c|c|} \hline  
\textbf{Sections} & \textbf{Cond. matter} &\textbf{Supersymmetry} & \textbf{Current research}   \\ \hline
Preliminaries \newline
(section \ref{sec:preliminaries}) & \greencheck  & \redmark & \redmark \\
\rule{0pt}{4.9ex} Historical review \newline
(section \ref{sec:historical_review}) & \greencheck &  \greencheck & \redmark \\ 
\rule{0pt}{4.9ex} Modern techniques \newline (section \ref{sec:modern_NR_QFT}) & \greencheck & \redmark & \greencheck \\
\rule{0pt}{4.9ex}  Modern applications \newline
(section \ref{sec:modern_applications}) & \greencheck & \redmark &  \greencheck \\ 
\rule{0pt}{4.9ex} Supersymmetry \newline
(section \ref{sec:SUSY}) & \redmark & \greencheck & \greencheck \\
\rule{0pt}{4.9ex}  Spin Matrix Theory \newline
(section \ref{sec:SMT}) & \redmark & \greencheck & \greencheck
 \\[0.2cm]
\hline
\end{tabular}   
\caption{Topics discussed in various sections of the review. } 
\label{tab:intro}
\end{center}
\end{table}


\section{Preliminaries}
\label{sec:preliminaries}

In this section, we lay the foundations by introducing the key concepts that will be central throughout this review.
We begin in section \ref{ssec:why_NR_QFT} by motivating the need for non-relativistic quantum field theory to overcome the limitations of quantum mechanics.
We obtain via a limiting procedure the prototype of a Schr\"{o}dinger-invariant QFT in flat space in section \ref{ssec:NR_limit_QFT}, whose symmetry group is presented in section \ref{ssec:Schroed_group}.
This structure allows to build a non-relativistic version of state/operator correspondence and to find restrictions on the form of the correlation functions, as we present in sections \ref{ssec:nonrel_state_OPE} and \ref{ssec:NR_correlation_functions}, respectively. 
We discuss in section \ref{ssec:loop_corrections} the formalism to study the quantum corrections of a Schr\"{o}dinger QFT.
This section is based on the material collected in the lectures \cite{Shiralectures}.\footnote{These lectures were part of the \textit{1st school on Non-relativistic Quantum Field Theory, Gravity and Geometry} held online on the 23rd-27th August 2021. I thank Shira Chapman for sharing her notes. }

\subsection{Why non-relativistic quantum field theory?}
\label{ssec:why_NR_QFT}

QFT was born to reconcile quantum mechanics (QM) with relativity.
Historically, the Klein-Gordon equation was first considered by Schr\"{o}dinger in late 1925 to study the fine structure of the hydrogen atom, but this attempt failed because the spin of the electron was not properly taken into account.
This is what led Schr\"{o}dinger to publish \textit{his} equation in 1926 as a non-relativistic approximation able to determine the Bohr energy levels of the hydrogen atom without any fine structure.
Later on, Dirac formulated in 1928 a wave equation for spin one-half particles, which was consistent with special relativity and represented an extension of the Schr\"{o}dinger equation.
While the Dirac equation successfully combines the invariance under the Poincaré group with the axioms of quantum mechanics, the corresponding theory admits a spectrum of energies with negative modes, thus allowing particles to decay and making the system unstable.
The famous resolution to this problem proposed by Dirac was the existence of a \textit{sea} of electrons that already occupy the negative energy states, and that give rise to a hole whenever they are excited into a positive energy state.
This observation was later supported by the discovery of antiparticles in 1931.

The previous proposal led us from a quantum description of a single relativistic particle to a theory that contains a Dirac sea with an infinite number of particles.
Wigner proved in 1931 a theorem stating that \textit{all the unitary representations of the Poincaré group are infinite-dimensional}.
This formal result implies that an infinite number of particles is required to achieve consistency between quantum mechanics and special relativity.
The consequence is that QFT assigns an operator $\phi(\vec{x})$ (called \textit{field}) in the Schr\"{o}dinger picture to any point $\vec{x}$ in space.
In the Heisenberg picture, operators become time-dependent as\footnote{In this section we will keep the factors of $\hbar$ explicit. }
\beq
\phi (t,\vec{x}) = e^{\frac{i}{\hbar} Ht} \phi(\vec{x}) e^{-\frac{i}{\hbar} Ht} \, .
\eeq
In this way, the positions $(t,\vec{x})$ in the spacetime become labels on operators and they are treated on equal footing.
\textit{Second quantization} is the process that imposes canonical (anti)commutation relations between fields, leading to a quantum field theory.
This formalism deals with observables assigned to regions of space rather than individual particles. 

In this context, an important result for the consistency of QFT with QM is that \textit{quantum field theory reduces to ordinary non-relativistic quantum mechanics in a sector with fixed particle number}.
This can be shown as follows \cite{Srednicki:2007qs}.\footnote{Similar discussions can be found in many other textbooks. For instance, a complete treatment of the topic is discussed in chapter 17 of \cite{Dick2012AdvancedQM}. }
Let us consider the Schr\"{o}dinger equation in configuration space for $n$ particles with the same mass $m$ moving in an external potential $U(\vec{x})$ and interacting with each other through the potential $V(\vec{x}_i - \vec{x}_j)$
\beq
i \hbar \frac{\p}{\p t} \psi = \left[ \sum_{k=1}^n \le - \frac{\hbar^2}{2m} \bigtriangleup_k + U(\vec{x}_k)  \ri 
+ \sum_{k=1}^n \sum_{l=1}^{k-1} V(\vec{x}_k - \vec{x}_l)  \right] \psi \, ,
\label{eq:Schroed_eq_position}
\eeq
where $\psi = \psi(t, \vec{x}_1, \dots, \vec{x}_n)$ is the wavefunction in position space.
This is a partial differential equation that determines the complex function $\psi$. 

An alternative perspective is to consider $\psi$ as a quantum field and use the abstract form of the Schr\"{o}dinger equation
\beq
i \hbar \frac{\p}{\p t} \ket{\psi, t} = H \ket{\psi, t} \, ,
\label{eq:abstract_Schroed}
\eeq
which describes the evolution of a state $\ket{\psi,t}$ defined in a Hilbert space $\mathcal{H}$ under the action of the Hamiltonian operator $H$.  
To this aim, we introduce ladder operators satisfying the equal-time (anti)commutation relations
\beq
[a(\vec{x}), a (\vec{y})]_{\pm} = 0 \, , \qquad
[a^{\dagger}(\vec{x}), a^{\dagger} (\vec{y})]_{\pm} = 0 \, , \qquad
[a(\vec{x}), a^{\dagger} (\vec{y})]_{\pm} = \delta^d (\vec{x}-\vec{y}) \, ,
\label{eq:ladder_op_sec2}
\eeq
where $\delta^d (\vec{x})$ is the $d$--dimensional Dirac distribution and $\pm$ refers to either fermionic or bosonic statistics, respectively.
The vacuum $\ket{0}$ of the theory is the state annihilated by all the lowering operators, \ie $a(\vec{x}) \ket{0} = 0 $.
We define the Hamiltonian of the system
\beq
H \equiv \int d^d x \, a^{\dagger}(\vec{x}) \le - \frac{\hbar^2}{2m} \bigtriangleup + U(\vec{x}) \ri a(\vec{x}) 
+ \frac{1}{2} \int d^dx d^dy \, a^{\dagger}(\vec{x}) a^{\dagger}(\vec{y}) V(\vec{x}-\vec{y}) a(\vec{x}) a(\vec{y}) \, ,
\label{eq:Ham_QM}
\eeq
and a generic time-dependent multi-particle state as
\beq
\ket{\psi,t} \equiv \int d^d x_1 \dots d^d x_n \,
\psi (t,\vec{x}_1, \dots, \vec{x}_n) a^{\dagger}(\vec{x}_1) \dots a^{\dagger}(\vec{x}_n) \ket{0} \, ,
\label{eq:ket_QM}
\eeq
where $\psi$ is a profile function that parametrizes a superposition between the $n$ particles.
It can be proven that \textit{the Hamiltonian \eqref{eq:Ham_QM} and the state \eqref{eq:ket_QM} satisfy the abstract Schr\"{o}dinger equation \eqref{eq:abstract_Schroed}  if and only if the wavefunction $\psi(t,\vec{x}_1, \dots, \vec{x}_n)$ solves the differential equation \eqref{eq:Schroed_eq_position} in configuration space. }

We can now interpret the vacuum $\ket{0}$ as a state without particles, and $\ket{\psi,t}$ defined in eq.~\eqref{eq:ket_QM} as a state with $n$ particles created at positions $\vec{x}_1, \dots, \vec{x}_n$.
This counting can be performed with the number operator 
\beq
N \equiv \int d^d x \, a^{\dagger}(\vec{x}) a(\vec{x}) \, . 
\label{eq:number_op}
\eeq 
The Hamiltonian in eq.~\eqref{eq:Ham_QM} satisfies $[H,N]=0$, implying that the number of particles in a state remains constant during the time evolution.

At this point, one may wonder what are the advantages of the QFT formalism, compared to QM, to describe non-relativistic phenomena. 
An important reason is that working in the common framework given by QFT may help us to exchange the knowledge between relativistic and non-relativistic communities for all the cases when this is relevant, \eg to describe phenomena at low energies and when particles move at low speed.
Limits usually make a system more tractable.
Since non-relativistic theories can be achieved performing $c \rightarrow \infty$ limits (where $c$ is the speed of light), one may hope to get useful insights on non-relativistic systems, and then apply them to the relativistic case, too.
The methods of QFT provide several technical advantages to studying the non-relativistic realm, including anomalies, computations of cross-sections and off-shell action formulations in condensed-matter settings.\footnote{QFT methods include perturbation theory with Feynman diagrams, the path integral, the formulation of a model in terms of fields associated to points in the spacetime \textit{etc.} We will provide several examples of these techniques in section \ref{sec:historical_review}.}
QFT also conveniently accounts for when particles are indistinguishable.
Finally, let us mention that while particle number is usually conserved in non-relativistic systems, this may not always be the case.
The QFT formalism goes beyond this limitation, thus describing the phenomenology of the emission or absorption of particles.

\subsection{Non-relativistic limit of quantum field theory}
\label{ssec:NR_limit_QFT}

A simple way to generate non-relativistic QFTs is via a limiting procedure implemented on a Lorentz-invariant QFT.
We will analyze in detail the modern approaches to this problem in section \ref{sec:modern_NR_QFT}, but for the moment we review the standard approach that can be found in books (\eg see \cite{Zee:2003mt}) and older papers \cite{Bergman:1991hf}.
Consider the Lagrangian for a relativistic real scalar field with quartic interaction
\beq
S= \int dt d^d x \, \mathcal{L} \, , \qquad
\mathcal{L} = - \frac{1}{2} (\p_{\mu} \Phi)(\p^{\mu} \Phi) 
- \frac{m^2}{2} \Phi^2 - \frac{\lambda}{4!} \Phi^4 \, ,
\label{eq:Lagrangian_sec2}
\eeq
where $m$ is the physical mass and $\lambda$ the coupling constant.
We parametrize the relativistic scalar field as 
\beq
\Phi(t,\vec{x}) = \frac{1}{\sqrt{2m}} \le e^{-i m t} \varphi(t,\vec{x}) + e^{i m t} \varphi^{\dagger}(t,\vec{x})  \ri \, . 
\label{eq:ansatz_Phi_sec2}
\eeq
In this parametrization, we have in mind to interpret $\varphi, \varphi^{\dagger}$ as the non-relativistic counterparts of the original scalar field.
We assume that the kinetic energy is much smaller than the rest mass, which implies $(\p^2/\p t^2) \varphi \ll -2 i m  (\p/\p t) \varphi .$
Plugging the ansatz \eqref{eq:ansatz_Phi_sec2} inside eq.~\eqref{eq:Lagrangian_sec2} and discarding the terms which oscillate in the mass $m$ gives
\beq
\mathcal{L} = \varphi^{\dagger} \le i \p_t + \frac{\bigtriangleup}{2m} \ri \varphi - \frac{g}{4} |\varphi^{\dagger} \varphi|^2 \, ,
\label{eq:Lagrangian_NR_phi4_sec2}
\eeq
where $g \equiv \lambda/(4m^2)$.
This is the Lagrangian for a Schr\"{o}dinger field with quartic interactions.

Let us study the classical properties of this theory.
The dynamics of the scalar field $\varphi$ (and its hermitian conjugate) is given by the Schr\"{o}dinger equation with a potential term.
It is relevant to notice that the field $\varphi$ is complex and the Lagrangian \eqref{eq:Lagrangian_NR_phi4_sec2} enjoys a global $\mathrm{U}(1)$ invariance. This is interpreted as the conservation of particle number (or equivalently, of the mass) and it is a distinctive feature of physical realizations of non-relativistic systems.
We will see in section \ref{ssec:Schroed_group} that the corresponding conserved charge $N$, analogous to the number operator in eq.~\eqref{eq:number_op}, provides a central extension of the Galilei symmetry, called the Bargmann group.

The theory \eqref{eq:Lagrangian_NR_phi4_sec2} enjoys a larger symmetry, which includes scale invariance.
Despite the explicit appearance of the mass parameter $m$, this is possible because the speed of light disappears in the non-relativistic regime.\footnote{We can formally consider the non-relativistic regime to arise in the limit $c \rightarrow \infty$, or we can perform an expansion of a relativistic theory around $c=\infty$ and work at fixed order in $1/c$. These perspectives will be taken in section \ref{sec:modern_NR_QFT}. }
For this reason, the mass is not dimensionally equivalent to an inverse length, but we have the freedom to consider $m$ as an independent parameter.
Indeed, the length dimensions $L$ of the coordinates and of the fields in $d+1$ dimensions read
\beq
[t] = L^2 \, , \qquad
[\vec{x}] = L \, , \qquad
[\varphi]=[\varphi^{\dagger}] = L^{- \frac{d}{2}} \, , \qquad
[m]= L^0 \, , \qquad
[g] = L^{d-2} \, .
\label{eq:length_dimensions_sec2}
\eeq
The crucial novelty, compared to the relativistic case, is that the time coordinate scales \textit{twice} as much as the spatial one.
This is a special case of the Lifshitz transformations \eqref{eq:Lifhsitz_transformations}.

It turns out that the symmetry structure of the Lagrangian \eqref{eq:Lagrangian_NR_phi4_sec2} is even larger, comprising the invariance under the following transformations:
\beq
\begin{cases}
t \rightarrow t + \xi^t &  \mathrm{time \,\, translations} \\
x^i \rightarrow x^i + \xi^i & \mathrm{spatial \,\, translations} \\
x^i \rightarrow x^i + R^i_{\,\,j} x^j & \mathrm{spatial \,\, rotations} \\
x^i \rightarrow x^i + v^i t & \mathrm{Galilean \,\, boosts} \\
t \rightarrow e^{2 \sigma} t \, , \quad
x_i \rightarrow e^{\sigma} x_i  & \mathrm{dilatations} \\
t \rightarrow \frac{t}{1+ \eta t} \, , \quad
x^i \rightarrow \frac{x^i}{1+ \eta t} & \mathrm{special \,\, conformal \,\, transf. \,\,(\mathrm{SCT})} 
\end{cases}
\label{eq:transf_Schroed_coord}
\eeq
Correspondingly, the Schr\"{o}dinger field transforms as
\beq
\begin{cases}
\varphi \rightarrow  e^{i m \vec{v} \cdot R \vec{x} + \frac{i}{2} m \vec{v}^2 t } \, \varphi &  \mathrm{Galilean \,\, transformations} \\
\varphi \rightarrow  e^{-\frac{d}{2} \sigma} \, \varphi &  \mathrm{dilatations} \\
\varphi \rightarrow  \le 1-\eta t \ri^d e^{\frac{i}{2} \frac{m \eta \vec{x}^2}{1- \eta t}}
\, \varphi &  \mathrm{SCT} \\
\varphi \rightarrow e^{i \alpha} \, \varphi \, , \quad
\varphi^{\dagger} \rightarrow e^{-i \alpha} \, \varphi^{\dagger}  &  \mathrm{U(1) \,\, particle \,\, number}
\end{cases}
\label{eq:transf_Schroed_fields}
\eeq
where $R$ is a rotation matrix and the Galilean transformations in eq.~\eqref{eq:transf_Schroed_fields} include translations, rotations and boosts.
The full set of transformations listed above generates the so-called \textit{Schr\"{o}dinger group}, which can be thought of as a non-relativistic analog of the conformal group.
We will sometimes refer to a QFT invariant under the Schr\"{o}dinger group as a non-relativistic conformal field theory (NRCFT).
We analyze the conserved charges, their Lie brackets, and the embedding inside the conformal group in section \ref{ssec:Schroed_group}.

\subsection{The Schr\"{o}dinger group}
\label{ssec:Schroed_group}

The prototype of a non-relativistic theory described by a second-quantized Schr\"{o}dinger field $\psi_{\alpha} (\vec{x})$, where $\alpha$ labels a representation of the spin group, is given by the action
\beq
\begin{aligned}
S = & \int dt\,  d^d x \, \psi^{\dagger}_{\alpha}(t, \vec{x}) \le i \p_t + \frac{\bigtriangleup}{2m} \ri \psi_{\alpha}(t, \vec{x}) \\
& - \frac{1}{2} \int dt \, d^d x d^d y \,   \psi^{\dagger}_{\alpha}(t, \vec{x}) \psi^{\dagger}_{\beta}(t, \vec{y}) V(\vec{x}-\vec{y}) \psi_{\beta}(t, \vec{y}) \psi_{\alpha}(t, \vec{x}) \, .
\label{eq:Schroed_action_sec2}
\end{aligned}
\eeq
This expression generalizes eq.~\eqref{eq:Lagrangian_NR_phi4_sec2}, which is recovered in the case of a $\delta$--like potential $V(\vec{x}-\vec{y}) \propto \delta^d(\vec{x}-\vec{y})$.
We assume that the fields satisfy the (anti)commutation relations
\beq
[\psi_{\alpha}(\vec{x}), \psi^{\dagger}_{\beta}(\vec{y})]_{\pm} = \delta^d (\vec{x}-\vec{y}) \, \delta_{\alpha\beta} \, .
\eeq
The Noether charges generating the transformations \eqref{eq:transf_Schroed_coord} and \eqref{eq:transf_Schroed_fields} read \cite{Hagen:1972pd}
\beq
\begin{aligned}
& H = \frac{1}{2m} \int  d^d x \, \p_i \psi^{\dagger}_{\alpha}(\vec{x}) \p_i \psi_{\alpha}(\vec{x}) 
+ \frac{1}{2} \int  d^d x d^d y \,   \psi^{\dagger}_{\alpha}(\vec{x}) \psi^{\dagger}_{\beta}(\vec{y}) V(\vec{x}-\vec{y}) \psi_{\beta}(\vec{y}) \psi_{\alpha}(\vec{x}) \, , \\
& P_i = \int d^d x \,  j_i(\vec{x}) \, , \qquad
J_{ij} =  \int d^d x \,  \le x_i j_j(\vec{x}) - x_j j_i (\vec{x}) \ri \, , \qquad
  G_i = \int d^d x \,  x_i n(\vec{x}) \, , \\
& M= m N = m \int d^d x \,  n(\vec{x}) \, ,  \qquad
D = - \int d^d x \,  x_i j_i(\vec{x}) \, , \qquad
C = \int d^d x \,  \frac{\vec{x}^2}{2} n(\vec{x}) \, , &
\end{aligned}
\label{eq:repr_Schr}
\eeq
where the number and momentum density are defined as
\beq
n(\vec{x}) = \psi^{\dagger}_{\alpha}(\vec{x}) \psi_{\alpha}(\vec{x}) \, , \qquad
j_i(\vec{x}) = - \frac{i}{2} \le \psi^{\dagger}_{\alpha}(\vec{x}) \p_i \psi_{\alpha}(\vec{x}) - \p_i \psi^{\dagger}_{\alpha}(\vec{x}) \psi_{\alpha}(\vec{x}) \ri \, .
\label{eq:number_mom_densities}
\eeq
The currents \eqref{eq:number_mom_densities} satisfy the continuity equation $ \p_t n + \p_i j_i =0.$
We associate the symmetry transformations with the corresponding conserved charges in table~\ref{tab:generators}.
The generators satisfy the following Lie brackets (all the other commutators vanish)
\beq
\begin{aligned}
& [J^{ij} , J^{kl} ] =  i \le \delta^{ik} J^{jl} + \delta^{jl} J^{ik}  - \delta^{il} J^{jk}  - \delta^{jk} J^{il}  \ri \, ,  & \\
& [J^{ij} , P^{k} ] =  i \le \delta^{ik} P^{j} - \delta^{jk} P^{i} \ri \, , \qquad
[J^{ij} , G^{k} ] =  i \le \delta^{ik} G^{j} - \delta^{jk} G^{i} \ri \, ,   & \\
& [D, P^i] =  i P^i \, , \qquad
[D, G^{i}] = - i G^i \, ,  \qquad
[D,H] = 2i H \, , \qquad
[D,C] = -2i C \, , 
 & \\
& [H, G^i] = -i P^i \, , \qquad
[H,C] =  -i D \, , \qquad
 [P^i, G^j] =  -i \delta^{ij} M \, , \qquad  
 [P_i, C] = -i G_i  &
\end{aligned}
\label{eq:Schroedinger_algebra}
\eeq
composing the Schr\"{o}dinger algebra.
The mass generator is a central charge of the algebra, since it commutes with all the other generators.
Since $[H,M]=[H,N]=0,$ the mass (and the particle number) is conserved in a physical theory invariant under the Schr\"{o}dinger symmetry.
The subset $\lbrace H, P_i, J_{ij}, G_i, M \rbrace$ composes the Bargmann subalgebra, which is a central extension of the Galilean algebra.
The Schr\"{o}dinger group is the largest group of (bosonic) symmetries of the Schr\"{o}dinger equation.

\begin{table}[ht]   
\begin{center}   
\begin{tabular}  {|c|c|} \hline  
\textbf{Transformations} & \textbf{Generators}  \\ \hline
\rule{0pt}{4.9ex}  Time translations & Hamiltonian $H$  \\
\rule{0pt}{4.9ex}  Spatial translations & Momenta $P_i$  \\ 
 \rule{0pt}{4.9ex}  Spatial rotations & Angular momenta $J_{ij}$ \\
\rule{0pt}{4.9ex}  Galilean boosts & Generators $G_i$  \\
\rule{0pt}{4.9ex}  $\mathrm{U(1)}$ internal symmetry & Mass $M$  \\ 
\rule{0pt}{4.9ex}  Scale transformations & Dilatation operator $D$  \\
\rule{0pt}{4.9ex}  SCT & Generator $C$ 
 \\[0.2cm]
\hline
\end{tabular}   
\caption{Conserved charges (right column) in eq.~\eqref{eq:repr_Schr} generating the transformations of the Schr\"{o}dinger group listed in the left column.  } 
\label{tab:generators}
\end{center}
\end{table}

In the relativistic case, an open problem is the relation between scale and conformal symmetry, \eg see \cite{Nakayama:2013is} for a review.
There are several hints to believe that unitary plus scale invariance, together with other technical assumptions, imply conformal invariance. 
The validity of this statement was also proven in two and four dimensions  \cite{Polchinski:1987dy,Dymarsky:2013pqa}.
It is natural to ask whether the same relation exists between scale and Schr\"{o}dinger invariance, since the latter can be thought as the non-relativistic counterpart of conformal symmetry. 
First of all, let us comment that the Lifshitz algebra, which is associated to the dilatations \eqref{eq:Lifhsitz_transformations}, does not contain any special conformal transformation. Therefore, this example provides a class of non-Lorentzian theories where the enhancement is not possible, simply because it cannot exist for $z \ne 2$.
Instead, the non-trivial issue is the possibility to enhance Galilean plus scale invariance into the Schr\"{o}dinger symmetry.
To begin with, one can show that the Ward identity associated with Schr\"{o}dinger invariance implies 
\beq
2 T^{00} - \delta_{ij} T^{ij} = 0 \, ,
\label{eq:flat_Ward_identity}
\eeq
where the components of the stress tensor are specified below.\footnote{The Ward identity associated to Weyl invariance in curved space will be derived from the variation of the effective action in eq.~\eqref{eq:variationW_sec5}, which reduces to eq.~\eqref{eq:flat_Ward_identity} in flat space. }
More precisely, we define
\beq
H = \int d^d x \, T^{00} \, , \qquad
P^i = \int d^d x \, T^{0i} \, ,
\eeq
while the spatial components $T_{ij}$ of the energy-momentum tensor satisfy the conservation equation $\p_0 j_i + \p_j T_{ij} = 0$, where the previous currents and charges were defined in eqs.~\eqref{eq:repr_Schr} and \eqref{eq:number_mom_densities}, and we recognize $j_i=T^{0i}$. 
We do not report here $T_{ij}$ because its explicit expression is cumbersome and not required for the following discussion.

At this point, it was argued in \cite{Nakayama:2009ww} that any stress tensor whose trace satisfies
\beq
2 T^{00} - \delta_{ij} T^{ij} = \p_0 S + \p_i A^i \, , 
\label{eq:stress_tensor_notimproved}
\eeq
corresponds to a scale-invariant theory with dilatation operator 
\beq
D = \int d^d x \,  \le  x^0 T^{00} - \frac{1}{2} x_i T^{0i} - \frac{S}{2} \ri \, . 
\eeq
One can further show that a stress tensor of the form \eqref{eq:stress_tensor_notimproved} can be improved to be traceless if $S = \p_i \sigma^i$, in which case one can indeed build a conserved current associated to SCTs.
When this improvement is possible, scale and Galilean invariance leads to Schr\"{o}dinger symmetry.
This topic is still subject of research studies.

In the next subsection, we analyze the consequences of the Schr\"{o}dinger symmetry on the operator product expansion (OPE) and the correlation functions of a theory.

\subsection{Non-relativistic state/operator correspondence}
\label{ssec:nonrel_state_OPE}

\begin{figure}[ht]
\centering
\includegraphics[scale=0.5]{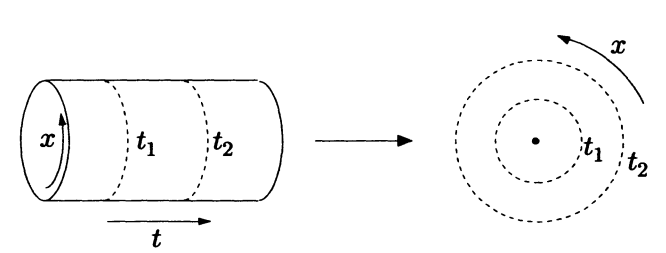}
\caption{Conformal mapping between $\mathbb{R} \times S^{D-1}$ (left) and $\mathbb{R}^D$ (right). The foliation induced by the time coordinate on the cylinder corresponds to a radial foliation on the Euclidean plane. Picture taken from \cite{DiFrancesco:1997nk}. }
\label{fig:state_OPE}
\end{figure}

A powerful tool in CFT is the state/operator correspondence, which relates\footnote{For a discussion of state/operator correspondence in CFT, some references are \cite{DiFrancesco:1997nk,Qualls:2015qjb,Rychkov:2016iqz,Simmons-Duffin:2016gjk}. A recent discussion on the state/operator correspondence defined on other compact manifolds than the sphere, such as the torus, can be found in \cite{Belin:2018jtf}.  }
\beq
\mathrm{local \,\, operators \,\, on} \,\, \mathbb{R}^D \quad
\Leftrightarrow \quad
\mathrm{states \,\, on\,\,} S^{D-1} \, .
\eeq
This is possible due to the conformal mapping between the Euclidean plane $\mathbb{R}^D$ and the cylinder $\mathbb{R} \times S^{D-1}$, up to a conformal factor:
\beq
ds^2 = d \rho^2 + \rho^2 d\Omega_{D-1}^2 
\quad
\underset{\rho = e^{\tau}}{\longrightarrow} \quad
ds^2 = e^{2 \tau} \le d\tau^2 + d\Omega_{D-1}^2 \ri \, .
\label{eq:relativistic_state_OPE}
\eeq
This transformation maps the dilatation operator $D= \rho \p_{\rho}$ on the plane to the Hamiltonian $H = \p_{\tau}$ on the cylinder.
On $\mathbb{R} \times S^{D-1},$ boundary conditions are chosen at the Euclidean time $\tau=-\infty$ to prepare a state at $\tau=0$. The space is foliated by time slices and this setting is called \textit{state picture}.
After the conformal transformation to the plane, the boundary conditions are mapped to the insertion of a local operator at $\rho=0$, while the state is now prepared at the sphere located at $\rho=1$. The dilatation operator naturally foliates the space into radial slices, and this setting is called \textit{operator picture}. 
The relation between the states created on the cylinder and the insertion of an operator at the origin is the realization of the \textit{state/operator correspondence}.
The essential features defining this relation are
\begin{itemize}
\item The bijective map between the cylinder and the Euclidean plane (fig.~\ref{fig:state_OPE}).
\item The identification between the energy eigenvalues of the Hamiltonian in the state picture and the scaling dimensions of the corresponding operators.
\end{itemize}

In particular, the first bullet can be regarded as a (Riemannian) geometric input that allows to identify a map between the eigenvalues of the Hamiltonian and of the dilatation operator in the two pictures.
It is natural to ask whether the state/operator correspondence also applies to the non-relativistic case.
The bijective map between two geometries with non-relativistic isometries is a non-trivial task and is the subject of ongoing research, as we will mention in section \ref{ssec:Lambert_QFT}.
Here we review the identification between the eigenvalues of a certain Hamiltonian and the dilatation operator \cite{Nishida:2007pj}.
Local operators $\mathcal{O}(t,\vec{x})$ in the Heisenberg picture can be defined at any point in spacetime  in terms of operators inserted at the origin $\mathcal{O}(0)$ as
\beq
\mathcal{O}(t,\vec{x}) = e^{i (Ht - P_i x^i)} \mathcal{O}(0)
e^{-i (Ht - P_i x^i)} \, .
\label{eq:local_operator_space}
\eeq
The subgroup of the Schr\"{o}dinger group which leaves invariant the origin corresponds to spatial rotations, the $U(1)$ global symmetry, dilatations, and SCTs.
The local operators at the origin form a representation of the reduced algebra. We declare that the scaling dimension $\Delta_{\mathcal{O}}$ and the mass eigenvalue $M_{\mathcal{O}}$ are given by
\beq
[D, \mathcal{O}(0)] = i \Delta_{\mathcal{O}} \mathcal{O}(0) \, , \qquad
[M, \mathcal{O}(0)] = M_{\mathcal{O}} \mathcal{O}(0) \, .
\label{eq:eigenvalues_dil_mass_sec2}
\eeq
The scaling dimension of local operators can be raised or lowered with the following generators
\beq
\begin{aligned}
& [D, [P_i, \mathcal{O}(0)]] = i \le \Delta_{\mathcal{O}} +1 \ri [P_i, \mathcal{O}(0)] \, , \qquad
[D, [G_i, \mathcal{O}(0)]] = i \le \Delta_{\mathcal{O}} -1 \ri [G_i, \mathcal{O}(0)] \, , & \\
& [D, [H, \mathcal{O}(0)]] = i \le \Delta_{\mathcal{O}} +2 \ri [H, \mathcal{O}(0)] \, , \qquad
[D, [C, \mathcal{O}(0)]] = i \le \Delta_{\mathcal{O}} -2 \ri [C, \mathcal{O}(0)] \, . &
\end{aligned}
\eeq
The lowest weights in the representation are called \textit{primary operators}, satisfying 
\beq
[G_i, \mathcal{O}(0)] = [C, \mathcal{O}(0)] = 0 \, .
\label{eq:primary_Schroed}
\eeq
They comprise an irreducible representation of the Schr\"{o}dinger algebra.
The tower of descendants is obtained by acting with the Hamiltonian and momentum generators
\beq
[H, \mathcal{O}(0)] = i \p_t \mathcal{O}(0) \, , \qquad
[P_i, \mathcal{O}(0)] = -i \p_i \mathcal{O}(0) \, .
\label{eq:actions_HP_opsec2}
\eeq
The above-mentioned representation structure only makes sense when the mass eigenvalue is non-vanishing, \ie $M_{\mathcal{O}} \ne 0$.
When $M_{\mathcal{O}}=0$, the Galilean algebra \eqref{eq:Schroedinger_algebra} implies $[P^i, G^j]=0,$ therefore the distinction between primaries and descendants breaks down \cite{Bekaert:2011qd}.
In particular, if $\mathcal{O}(0)$ is primary, so is also the descendant $[P_i, \mathcal{O}(0)]$.
From now on, we will assume to work in a sector with a non-vanishing mass eigenvalue.

Now we show that each primary operator of the Schr\"{o}dinger group corresponds to an energy eigenstate of the oscillator Hamiltonian \cite{Nishida:2007pj}
\beq
H_{\rm osc} = H+ C \, ,
\label{eq:oscillator_Ham}
\eeq
where we set the mass and the frequency of the oscillator to $m=\omega=1$ for convenience.
This system is a harmonic oscillator because the generator $C$ of the special conformal transformation can be represented in terms of a quadratic potential, see eq.~\eqref{eq:repr_Schr}.
Assuming that the operator $\mathcal{O}(0)$ is composed of annihilation operators, then its hermitian conjugate will act non-trivially on the vacuum $\ket{0}$, and it is possible to define the state
\beq
\ket{\Psi_{\mathcal{O}}} \equiv e^{-H} \mathcal{O}^{\dagger} \ket{0} \, ,
\label{eq:state_Nishida}
\eeq
which has mass eigenvalue $M_{\mathcal{O}^{\dagger}}.$
By direct computation, one can show that this is an eigenstate of the oscillator Hamiltonian such that
\beq
H_{\rm osc} \ket{\Psi_{\mathcal{O}}}= \Delta_{\mathcal{O}} \ket{ \Psi_{\mathcal{O}}} \, .
\eeq
This identifies the energy levels of an oscillator Hamiltonian as the scaling dimensions of the dilatation operator of the Schr\"{o}dinger algebra.
In this sense, the oscillator Hamiltonian \eqref{eq:oscillator_Ham} plays the same role as the radial quantization Hamiltonian in a relativistic CFT.
Since the eigenstates of a harmonic oscillator are organized into ladders with raising and lowering operators $L_{\pm} = H - C \pm i D $,
one can further show that the state $\ket{\Psi_{\mathcal{O}}}$ is the lowest weight of the representation, \ie 
\beq
L_- \ket{\Psi_{\mathcal{O}}} = 0 \, .
\eeq
Thus we have built a tower of states by inserting an operator at the origin and then acting with the ladder operators of the Hamiltonian for a harmonic oscillator.

To conclude, let us clarify that the previous arguments have shown that it is always possible to find a state associated to the action of a primary operator at the origin.
In order to properly define a state/operator correspondence, one need to show that the map is bijective, and possibly supply a geometric realization in terms of non-relativistic geometries.
These developments are an open research problem.

\subsection{Correlation functions}
\label{ssec:NR_correlation_functions}

In a relativistic CFT, conformal symmetry uniquely fixes the form of two and three-point functions involving primary operators, while the kinematics is not sufficient to fully restrict four-point functions.
In particular, there exist two conformal-invariant ratios of spacetime coordinates (called cross ratios) that can be used to construct the general form of four and higher-point functions.
The conformal bootstrap program is based on these facts and the idea that the OPE imposes crossing relations between correlation functions \cite{FERRARA1973161}.

Schr\"{o}dinger invariance is less constraining than the conformal case, in particular only two-point functions are fixed by symmetry \cite{Goldberger:2014hca}.
For any set of $n$ operators, we define a correlation function as
\beq
G_n (t_1, \vec{x}_1; \dots ; t_n, \vec{x}_n) \equiv 
\langle 0 | T \mathcal{O}_1 (t_1, \vec{x}_1) \dots \mathcal{O}_n (t_n, \vec{x}_n)   | 0 \rangle \, ,
\label{eq:general_correlation_function_sec2}
\eeq
where $T$ is the time-ordering and $\ket{0}$ is the vacuum state, annihilated by all the generators, including $M \ket{0} =0$.
Since local operators are required to have definite mass eigenvalue, this implies that the $n$--point functions are non-vanishing if and only if $\sum_{i=1}^n M_{\mathcal{O}_i} =0$, where $M_{\mathcal{O}_i}$ are the mass eigenvalues associated to the operators $\mathcal{O}_i$.

Let us now focus on the correlation functions of scalar primary operators, which we denote as $\phi_i(t_i, \vec{x}_i)$. 
We begin with the two-point functions.
Due to the symmetries of the theory, it is not restrictive to choose the operator basis such that $\phi_2(t,\vec{x})= \phi_1^{\dagger}(t,\vec{x})$. 
The structure of two-point functions is restricted by symmetries to the form
\beq
G_2(t_1, \vec{x}_1; t_2, \vec{x}_2) = \langle 0 | T \phi_1(t_1,\vec{x}_1) \phi_1^{\dagger}(t_2,\vec{x}_2) | 0 \rangle =  c |t_{12}|^{- \Delta_1 } \, \exp \le -i \frac{M_1}{2} \frac{\vec{x}_{12}^2}{t_{12}} \ri \, ,
\eeq
with the constraint $\Delta_1 = \Delta_2$ on the scaling dimensions of the fields.
Here $M_1$ denotes the mass eigenvalue of the scalar field, while $t_{12} \equiv t_1 - t_2$, $\vec{x}_{12} \equiv \vec{x}_1-\vec{x}_2$.

Similar techniques can be applied to restrict the form of higher-point functions.
The major difference with the relativistic case is that one can build either cross ratios involving only the time direction, or mixed combinations of time and space coordinates:
\beq
\begin{aligned}
& t_{ijkl} \equiv  \frac{t_{ij} t_{kl}}{t_{il} t_{jk}} \, , \qquad (i,j,k,l = 1, \dots , n) & \\
& v_{ijk} \equiv \frac{1}{2} \le \frac{\vec{x}_{jk}^2}{t_{jk}} + \frac{\vec{x}^2_{ij}}{t_{ij}} - \frac{\vec{x}^2_{ik}}{t_{ik}} \ri \, ,  \qquad  (i<j<k) & 
\end{aligned}
\eeq
Since these cross ratios are both invariant under the Schr\"{o}dinger symmetries, they can be used to build non-trivial correlation functions.
The most general form of the three and four-point functions involving scalar primaries is given by \cite{Henkel:1993sg,Volovich:2009yh,Fuertes:2009ex}\footnote{Recent studies of correlation functions for non-relativistic conformal field theory, even in the presence of a boundary, have been studied in \cite{Gupta:2022azd,Gupta:2022mer}.}
\beq
\begin{aligned}
& G_3 (1,2,3) = F(v_{123}) \, \exp \le -i \frac{M_1}{2} \frac{\vec{x}_{13}^2}{t_{13}} - i \frac{M_2}{2} \frac{\vec{x}_{23}^2}{t_{23}}  \ri  \prod_{i<j}  t_{ij}^{\frac{\Delta}{2}-\Delta_i-\Delta_j}  \, , & \\
& G_4(1,2,3,4) =   H(v_{124},v_{134},v_{234},t_{1234}) \, \exp \le -i \frac{M_1}{2} \frac{\vec{x}_{14}^2}{t_{14}} - i \frac{M_2}{2} \frac{\vec{x}_{24}^2}{t_{24}} - i \frac{M_3}{2} \frac{\vec{x}_{34}^2}{t_{34}}  \ri  \prod_{i<j}  t_{ij}^{\frac{\Delta}{6}-\frac{\Delta_i+\Delta_j}{2}}  \, , &
\end{aligned}
\eeq
where $\Delta \equiv \sum_i \Delta_i,$ while $F,H$ are conformal blocks, \ie analytic functions of the cross ratios.

Schr\"{o}dinger symmetry is less constraining than its relativistic counterpart, in fact, the structure of three-point functions is not entirely fixed.
For this reason, the conformal bootstrap approach is less constraining, and in general more freedom is allowed in the non-relativistic case \cite{Chen:2020vvn,Chen:2022jhx}. 
Relatedly, in any QFT the operator product expansion (OPE) states that products of nearby operators have an expansion in terms of local operators as
\beq
\lim_{t,x \rightarrow 0} \phi_1(t,\vec{x}) \phi_2(0) =
\sum_{\alpha} c_{\alpha}(t,\vec{x}, \p_t, \p_i) \mathcal{O}(0) \, ,
\eeq
where the summation involves a sum over primaries and descendants.\footnote{Descendants are obtained by acting with derivatives on primary operators, see eq.~\eqref{eq:actions_HP_opsec2}. }
This is an operator equation, which holds inside correlation functions and can be used to obtain consistency conditions between different channels.
In principle, the functions $c_{\alpha}$ can be determined by acting on the OPE with the generators of the non-relativistic conformal group and by exploiting the transformation properties of the operators, \eg see \cite{Goldberger:2014hca} for a concrete approach.

\subsection{Quantum corrections}
\label{ssec:loop_corrections}

Next, we move to the analysis of the quantum properties of a Schr\"{o}dinger-invariant field theory.
While the following discussion will be general, for a concrete example we will refer to the scalar model \eqref{eq:Lagrangian_NR_phi4_sec2}.
Free fields satisfy the Schr\"{o}dinger equation
\beq
\le i \p_t + \frac{\bigtriangleup}{2m} \ri \varphi = 0 \, ,
\eeq
therefore they can be written in terms of a Fourier expansion
\beq
\varphi(t,\vec{x}) = \int \frac{d^d k}{(2\pi)^d} \, a(\vec{k}) e^{-i (\omega_k t - \vec{k} \cdot \vec{x})} \, , \qquad
\varphi^{\dagger}(t,\vec{x}) = \int \frac{d^d k}{(2\pi)^d} \, a^{\dagger}(\vec{k}) e^{i (\omega_k t - \vec{k} \cdot \vec{x})} \, ,
\label{eq:Fourier_expansion_sec2}
\eeq
where $\omega_k \equiv \vec{k}^2/(2m).$
We canonically quantize the scalar fields by imposing the commutation relations
\beq
[\varphi(t,\vec{x}), \varphi(t,\vec{y})] = 
[\varphi^{\dagger} (t,\vec{x}), \varphi^{\dagger} (t,\vec{y})] =  0 \, , \qquad
[\varphi(t,\vec{x}), \varphi^{\dagger} (t,\vec{y})] = \delta^d (\vec{x}-\vec{y}) \, .
\eeq
The vacuum state is annihilated by the lowering operator $a(\vec{k})\ket{0}$.
The operator $a(\vec{k})$ destroys a particle with momentum $\vec{k}$, while the hermitian conjugate $a^{\dagger}(\vec{k})$ creates a particle with momentum $\vec{k}$.
It is important to notice that, contrarily to the relativistic case, the Fourier expansions \eqref{eq:Fourier_expansion_sec2} only contain ladder operators of one kind.
This is a consequence of the global $\mathrm{U}(1)$ symmetry, which implies the number of particles is conserved.
After the non-relativistic limit, the anti-particles of the original QFT have decoupled and they form a separate sector of the theory.

One can study quantum corrections using standard perturbative methods as in any standard QFT.
The free non-relativistic propagator reads
\beq
G_2 (t,\vec{x}) = \int \frac{d\omega d^{d}p}{(2\pi)^{d+1}} 
\frac{e^{-i (\omega t - \vec{p} \cdot \vec{x})}}{\omega - \frac{\vec{p}^2}{2m} + i \varepsilon} = 
- \frac{\Theta(t)}{(2\pi)^{d/2}} \, \exp \le \frac{i \vec{x}^2}{2 m t} \ri \, ,
\label{eq:nonrel_prop_sec2}
\eeq
where $\Theta$ is the Heaviside distribution.
As discussed in section \ref{ssec:NR_limit_QFT} below eq.~\eqref{eq:length_dimensions_sec2}, the dimensional counting in a non-relativistic theory should be performed by referring to the scaling dimensions of the data under Lifshitz dilatations.
In terms of the energy dimensions, for a Schr\"{o}dinger-invariant theory we have
\beq
[\omega] = E^2 \, , \qquad
[k_i] = E \, , \qquad
[m]= [g] = E^0 \, .
\eeq
The non-relativistic propagator has a retarded $i \varepsilon$ prescription which follows the order of the fields shown in eq.~\eqref{eq:fig_prop_scalar}.
This property will be crucial to simplify the study of loop corrections, as we will show below.

\paragraph{Feynman rules.}
We define the generating functional
\beq
\mathcal{Z}[J,J^{\dagger}] = \int [\mathcal{D}\varphi \, \mathcal{D}\varphi^{\dagger}] \, \exp \left[ i S + i \int dt d^d x \, \le J \varphi + J^{\dagger} \varphi^{\dagger} \ri \right] \, ,
\label{eq:generating_functional_sec2}
\eeq
where $S$ is the classical action for a (Schr\"{o}dinger) field, while $J,J^{\dagger}$ are sources for the scalar field and its hermitian conjugate.
By defining functional derivatives as
\beq
\frac{\delta J(t,\vec{x})}{\delta J(t', \vec{x}')} = \delta(t-t') \delta^d(\vec{x}-\vec{x}') \, , \qquad
\frac{\delta J^{\dagger}(t,\vec{x})}{\delta J^{\dagger} (t', \vec{x}')} = \delta(t-t') \delta^d(\vec{x}-\vec{x}') \, ,
\label{eq:functional_derivatives_sec2}
\eeq
we obtain the correlation functions via repeated action of the derivatives on the generating functional:\footnote{In the following expression, we explicitly use the observation that the total particle number of the fields inside the correlator must vanish, see discussion below eq.~\eqref{eq:general_correlation_function_sec2}. In this case, it simply means that $n$ is even. }
\beq
\begin{aligned}
& \langle \varphi (t_1,\vec{x}_1) \dots \varphi (t_{n/2}, \vec{x}_{n/2}) \varphi^{\dagger} (t_{n/2+1}, \vec{x}_{n/2+1}) \dots \varphi^{\dagger}(t_n, \vec{x}_n) \rangle = \\
& = \frac{1}{\mathcal{Z}[J,J^{\dagger}]} \frac{\delta}{i\, \delta J(t_1, \vec{x}_1)} \dots  \frac{\delta}{i\, \delta J(t_{n/2}, \vec{x}_{n/2})} \frac{\delta}{i\, \delta J^{\dagger}(t_{n/2+1}, \vec{x}_{n/2+1})} \dots  \frac{\delta}{i\, \delta J^{\dagger}(t_n, \vec{x}_n)} \mathcal{Z}[J,J^{\dagger}]\Big|_{J=J^{\dagger}=0} \, .
\end{aligned}
\eeq
We list the Feynman rules:
\begin{itemize}
\item \textbf{Propagator.}
It is obtained by inverting the kinetic operator. For a Schr\"{o}dinger scalar:
\beq
\includegraphics[valign=c, scale=1]{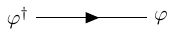}  =
\langle \varphi (\omega, \vec{p}) \varphi^{\dagger} (-\omega, -\vec{p}) \rangle  = 
\frac{i}{ \omega -\frac{\vec{p}^2}{2m} + i \varepsilon}
\label{eq:fig_prop_scalar}
\eeq
The arrow denotes the eigenvalue associated with the global $\mathrm{U(1)}$ particle number.
\item \textbf{Vertices.} They are directly read from the action. 
Energy and spatial momentum are conserved at every vertex.
Furthermore, in Schr\"{o}dinger-invariant QFTs the $\mathrm{U}(1)$ particle number is conserved, which implies at a visual level that the number of arrows entering and exiting a vertex has to match. In the case of the theory \eqref{eq:Lagrangian_NR_phi4_sec2}, there is only the following four-point vertex, with associated coupling $g$:
\beq
\includegraphics[valign=c, scale=1]{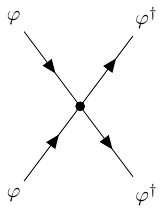}
\eeq
\end{itemize}
At this point, Feynman diagrams are built in the standard way:
\begin{enumerate}
\item Draw external lines for each ingoing and outgoing particle.
\item Connect the external lines using propagators and assign to them the corresponding energy, momentum, and $\mathrm{U}(1)$ particle number.
\item Impose conservation of energy, momentum, and particle number at each vertex.
\item Integrate over all loop momenta. Include a factor of $-1$ for each fermionic loop.
\item Draw all the topologically inequivalent graphs at the desired order.
\item Include combinatorial factors arising from the expansion of the interacting Lagrangian.
\end{enumerate}
Another familiar notion from relativistic QFT is the
effective action, which is the Legendre transform of the free energy.
The effective action plays an important role because it is the generating functional of \textit{one-particle irreducible} (1PI) diagrams, which are graphs remaining connected after cutting any of their lines.
The precise relation reads
\beq
\begin{aligned}
& \langle \varphi (t_1,\vec{x}_1) \dots \varphi (t_{n/2}, \vec{x}_{n/2}) \varphi^{\dagger} (t_{n/2+1}, \vec{x}_{n/2+1}) \dots \varphi^{\dagger}(t_n, \vec{x}_n) \rangle_{\rm 1PI} = \\
& = i \, \frac{\delta}{ \delta \varphi_0(t_1, \vec{x}_1)} \dots  \frac{\delta}{ \delta \varphi_0(t_{n/2}, \vec{x}_{n/2})} \frac{\delta}{ \delta \varphi_0^{\dagger}(t_{n/2+1}, \vec{x}_{n/2+1})} \dots  \frac{\delta}{ \delta \varphi_0^{\dagger}(t_n, \vec{x}_n)} \Gamma^{(n)}[\varphi_0, \varphi_0^{\dagger}]\Big|_{J=J^{\dagger}=0} \, .
\end{aligned}
\label{eq:relation_effectiveaction_1PI_sec2}
\eeq
Here we denoted with $(\varphi_0, \varphi^{\dagger}_0)$ the classical fields, and with $\Gamma^{(n)}[\varphi_0, \varphi_0^{\dagger}]$ the $k$--th order in the Taylor expansion of the effective action around the classical field configurations.
With these ingredients, loop corrections to the effective action can be systematically studied.

\begin{figure}[ht]
\centering
\includegraphics[scale=1.1]{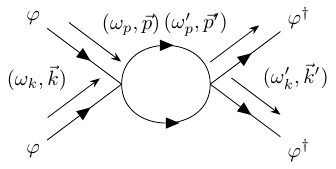}
\caption{One-loop contribution to the self-energy of the scalar field.}
\label{fig:selection_rule}
\end{figure}

\paragraph{Causal properties of the non-relativistic propagator.}
Let us focus on the causal properties of the non-relativistic propagator \eqref{eq:nonrel_prop_sec2}.
The retarded nature is responsible for the following selection rule:
\begin{srule}
\label{sel_rule1}
Any 1P-irreducible Feynman diagram with a negative superficial degree of divergence in the $\omega$ variable and whose arrows form a closed loop, vanishes.
\end{srule}
The superficial degree of divergence $\Delta_{\omega}$ is defined as the power of the $\omega$ variable at the numerator minus the power of $\omega$ at the denominator of a loop integral, including its measure.
It is used to determine the UV divergence of an integrand.
The arrows represent the flux of $\mathrm{U}(1)$ particle number across the diagram, and they form a closed loop when they are oriented in the same way (clockwise or counter-clockwise). 
To explain concretely the application of the selection rule, we consider the prototypical example represented by the one-loop correction to the four-point vertex for the action \eqref{eq:Lagrangian_NR_phi4_sec2} depicted in fig.~\ref{fig:selection_rule}.

We start for convenience from the case $d=2$.
The relevant integral reads
\beq
\int \frac{d\omega d^2 l}{(2\pi)^3} \, 
\frac{1}{\omega - \frac{\vec{p}^2}{2m} + i \varepsilon}
\frac{1}{(\omega_p + \omega_k -\omega) - \frac{(\vec{p} + \vec{k}-\vec{l})^2}{2m} + i \varepsilon} \, ,
\eeq
whose superficial degree of divergence is $\Delta_{\omega}=-1.$
There are two ways to prove that this integral vanishes:
\begin{itemize}
\item \textbf{Momentum space.}  
We perform the $\omega$ integration first.
Since the poles in $\omega$ are located on the same half-plane in complex space, the application of Jordan's lemma to close the integration contour in the region without poles gives a vanishing integral by means of the residue theorem.
\item \textbf{Configuration space.}
Direct use of the expression \eqref{eq:nonrel_prop_sec2} in position space leads to a product of two Heaviside step functions with opposite arguments. Since this result has support in a single point, we can impose that it vanishes by normal ordering.
\end{itemize}
The generalization of the previous argument to higher dimensions is trivial, since the identity \eqref{eq:nonrel_prop_sec2} is general.
The extension to generic loop integrals relies on the observation that Jordan's lemma can be applied to any meromorphic function whose superficial degree of divergence respects $\Delta_{\omega} <0,$ which is one of the hypotheses of the selection rule.
Since arrows forming a closed loop imply that the poles in $\omega$ always sit in the same half-plane, the proof follows by applying the residue theorem after choosing an integration contour located in that region of complex space.
The selection rule \ref{sel_rule1} was initially observed in \cite{Bergman:1991hf} and later applied in \cite{Klose:2006dd} and in more recent works (\eg see \cite{Auzzi:2019kdd,Arav:2019tqm,Chapman:2020vtn,Baiguera:2022cbp}) to prove the existence of non-renormalization theorems, as we will discuss in sections \ref{ssec:GED}, \ref{ssec:Galilean_WZ} and \ref{ssec:SGED}.


\section{Historical review}
\label{sec:historical_review}

We present a historical overview of the old developments in the realm of non-relativistic QFT, from the '60s until the last decades.
Recent methods (after the revival of the topic by Son and collaborators in 2005) and their applications will be discussed in section \ref{sec:modern_NR_QFT} and \ref{sec:modern_applications}, respectively.
Our journey begins in sections \ref{ssec:nonrel_Dirac} and \ref{ssec:Galilean_em} with the non-relativistic versions of Dirac equations and electromagnetism.
Particles with fractional statistics, the anyons, give rise to solvable systems when coupled with non-relativistic matter.
This leads to the Jackiw-Pi model (section \ref{ssec:Jackiw_Pi}), the Aharonov-Bohm scattering problem (section \ref{ssec:AB_scattering}), and the superconformal Chern-Simons theory (section \ref{ssec:SUSY_CS}).
Finally, we will show in section \ref{ssec:nonrel_scale_anomaly} that non-relativistic QFT methods allow us to compute the trace anomaly for a Schr\"{o}dinger scalar with quartic interaction.

\subsection{Non-relativistic Dirac equation}
\label{ssec:nonrel_Dirac}

It is often believed that spin is a property of relativistic theories and that the Dirac equation is the only framework able to reproduce the gyromagnetic ratio for a spin one-half particle.
Below we review, following
\cite{LvyLeblond1967NonrelativisticPA}, how some of these features are valid for non-relativistic systems, too.

\paragraph{Non-relativistic wave equation.}
By construction, the Schr\"{o}dinger equation is preserved by Galilean transformations, as it was shown for scalar fields in section \ref{ssec:NR_limit_QFT}.
To accommodate for spin, fields must transform in the following way under the Galilean transformations in eq.~\eqref{eq:transf_Schroed_coord} \cite{LvyLeblond1967NonrelativisticPA,Hagen:1972pd}:
\beq
\psi_{\alpha}'(t',\vec{x}') = e^{i f(t,\vec{x})} \sum_{\beta=-s}^s D^s_{\alpha\beta}(R) \, \psi_{\beta} (t,\vec{x}) \, , \qquad
f(t,\vec{x})= \frac{1}{2} m \vec{v}^2 t + m \vec{v} \cdot R \vec{x}  \, ,
\label{eq:transf_psi_sec3}
\eeq
where $\alpha,\beta$ are spinorial indices and $D^s$ is the $(2s+1)$--dimensional representation of the rotation group in Euclidean space.
Eq.~\eqref{eq:transf_psi_sec3} corresponds to a unitary ray representation of the Galilei group which preserves the local probability density $|\psi_{\alpha}'(t',\vec{x}')|^2 = |\psi_{\alpha} (t,\vec{x})|^2$.

Next, we show that the index $s$ can be interpreted as the spin of a non-relativistic particle, starting from the case $s=1/2$.
Let us search for a first-order wave equation linear in both the temporal and spatial derivatives
\beq
\mathcal{O} \psi \equiv \le i a \p_t + i \vec{b} \cdot \nabla + c \ri \psi = 0 \, ,
\label{eq:ansatz_Dirac_sec3}
\eeq
where $\mathcal{O}$ denotes the differential operator, $(a, \vec{b}, c)$ are generic linear operators, and $\psi$ is the spin one-half field.
To find the explicit form of the linear operator $\mathcal{O}$ in eq.~\eqref{eq:ansatz_Dirac_sec3}, we impose that its square gives the Schr\"{o}dinger operator, \ie $ \mathcal{O}^2 = 2 i m \p_t + \bigtriangleup $.\footnote{Of course, the analogous requirement in the relativistic case is $\mathcal{O}^2 = \square$.}
In particular, one can find a representation of $(a, \vec{b}, c)$ in terms of $4 \times 4$ complex matrices such that $\vec{b}$ satisfy the Clifford algebra (see \cite{LvyLeblond1967NonrelativisticPA} for more details). 
An explicit computation gives the following wave equation in Fourier space
\beq
\begin{cases}
E \chi + (\vec{\sigma} \cdot \vec{p}) \xi = 0 \\
(\vec{\sigma} \cdot \vec{p}) \chi + 2m \xi = 0 \, ,
\end{cases}
\label{eq:nonrel_Dirac_sec3}
\eeq
where $(E,\vec{p})$ are the energy and momentum of the fields, while $\vec{\sigma}$ are the Pauli matrices.
In order to achieve the form in eq.~\eqref{eq:nonrel_Dirac_sec3}, we chose a particular irreducible representation of the matrices composing the set $\vec{b}$.
In four spacetime dimensions, the field $\psi$ has four components and decomposes into two-component complex spinors $ \psi = (\xi, \chi)$. 
One can show that $\chi$ is the only dynamical component, while $\xi$ plays the role of an auxiliary field that can be integrated out by solving the second equation in \eqref{eq:nonrel_Dirac_sec3}.
In particular, $\chi$ satisfies the Schr\"{o}dinger equation and the wave equations \eqref{eq:nonrel_Dirac_sec3} are invariant under Galilean transformations, provided that the fields transform as
\beq
\begin{pmatrix}
\xi'(t',\vec{x}') \\
\chi'(t',\vec{x}')
\end{pmatrix} =
e^{i f(t,\vec{x})} 
\begin{pmatrix}
- \frac{\vec{\sigma} \cdot \vec{v}}{2} D^{1/2}(R)  &  D^{1/2}(R) \\
0 & D^{1/2}(R)
\end{pmatrix} 
\begin{pmatrix}
\xi(t,\vec{x}) \\
\chi(t,\vec{x})
\end{pmatrix} \, ,
\label{eq:transf_spinors_sec3}
\eeq
with the same function $f$ entering eq.~\eqref{eq:transf_psi_sec3}, and $D^{1/2}(R)$ is the two-dimensional representation of the rotation group.
In particular, the transformation law of the spinor $\chi$ does not mix with the auxiliary field $\xi$. 
Comparing it with eq.~\eqref{eq:transf_psi_sec3}, we recognize that the spinor $\chi$ describes a non-relativistic particle with spin $1/2$, and the $4 \times 4$ matrix in the right-hand side of eq.~\eqref{eq:transf_spinors_sec3} provides a faithful representation of the homogeneous Galilei group.

\paragraph{Gyromagnetic ratio.}
One of the striking applications of the relativistic Dirac equation is the prediction of the Landé gyromagnetic factor $g_s=2$ for the electron.
Remarkably, this result can also be inferred from the non-relativistic wave equation \eqref{eq:nonrel_Dirac_sec3}.
Let us minimally couple the system to an external electromagnetic field, \ie we make the derivatives covariant as $\p_{\mu} \rightarrow D_{\mu} = \p_{\mu} -i q A_{\mu} $, where $q$ is the electric charge.
After integrating out the auxiliary field $\xi$, we obtain
\beq
\left[ E - q A_0 - \frac{1}{2m} \le \vec{p} - q \vec{A} \ri^2 - \frac{q}{2m} \vec{\sigma} \cdot \vec{B} \right] \chi = 0 \, ,
\label{eq:nonrel_wave_coupling_sec3}
\eeq
where $\vec{B}$ is the magnetic field.
The standard form of the interaction between a charged particle and a magnetic field is given by the Pauli term
\beq
H =  \gamma \, \vec{S} \cdot \vec{B} = - \frac{e g_s}{2m} \, \le \frac{\hbar}{2} \vec{\sigma} \ri \cdot \vec{B} \, ,
\label{eq:ham_gyromagnetic_ratio}
\eeq
where $\vec{S}$ is the spin, $\gamma$ the gyromagnetic ratio, $g_s$ the Landé factor and we momentarily restored the factor of $\hbar$.
After some manipulations and comparing with 
eq.~\eqref{eq:ham_gyromagnetic_ratio} using $q=e$, one finds that $g_s=2$, as anticipated.
Therefore, \textit{a non-relativistic theory for a spin one-half particle predicts the correct value for the intrinsic magnetic moment of the electron}.

The Landé factor for the electron can also be derived from a non-relativistic limit of the Dirac equation in momentum space\footnote{In fact, this is the common procedure used in textbooks to derive the gyromagnetic ratio for the electron. }
\beq
\begin{cases}
( \mathcal{E} - m) \chi + (\vec{\sigma} \cdot \vec{p}) \xi = 0 \\
(\vec{\sigma} \cdot \vec{p}) \chi + (\mathcal{E}+m) \xi = 0 \, ,
\end{cases}
\label{eq:rel_Dirac_sec3}
\eeq
where now $\mathcal{E}= E+m$ represents the kinetic energy $E$ plus the rest mass contribution $mc^2$ (in the previous equation, we set $c=1$).
After minimally coupling the system to an external electromagnetic field and working in the non-relativistic regime $E \ll m,$ eq.~\eqref{eq:rel_Dirac_sec3} precisely reduces to eq.~\eqref{eq:nonrel_Dirac_sec3}, thus leading to the same identification $g_s=2$.

\paragraph{Conclusive remarks.}
We have just shown that \textbf{(1)} one can build from first principles a non-relativistic first-order wave equation for a spin one-half particle; \textbf{(2)} the spin representation naturally arises when requiring Galilean invariance, and \textbf{(3)} in this setting one obtains the correct prediction for the gyromagnetic ratio.

However, non-relativistic physics is not sufficient to describe all the phenomenology.
To reproduce the spin-orbit interaction and the Darwin term (essential for the fine structure of the atoms), the fully relativistic Dirac equation is needed.
Another difference with the relativistic case is that the \textit{spin-statistics} theorem does not necessarily hold. While we found spinful representations of Galilean-invariant particles, the kinematical group also allows for fermionic fields that behave as scalars under spatial rotations.
This will be manifest in supersymmetric examples, discussed in section \ref{ssec:nonrel_SUSY_algebra}.
Finally, while we built here a non-relativistic wave equation for a spin one-half field, it would be desirable to derive this result from a variational principle.
We will discuss in section \ref{ssec:nonrel_fermions} the Lagrangian formulation of this theory.

\subsection{Galilean electromagnetism}
\label{ssec:Galilean_em}

The main motivation to study a non-relativistic version of electromagnetism lies in the desire to distinguish the truly relativistic aspects from the features which are in common with its Galilean-invariant version.
The non-relativistic limit is useful when the theory simplifies after $c \rightarrow \infty$ is implemented, while at the same time the model is sufficient to reproduce experimental results.
Moreover, these systems provide a non-trivial example of dynamical theories to which charged non-relativistic matter should couple.

For convenience, we momentarily restore the factors in the speed of light and we work with the MKSA system of units, where the vacuum permittivity $\varepsilon_0$ and permeability $\mu_0$ are independent constants satisfying $\varepsilon_0 \mu_0 c^2 =1$.
We focus on the Lorentz transformation of any vector $U^{\mu} = (U^0, \vec{U})$, with particular interest in the current vector denoted as $j^{\mu} = (c \rho, \vec{j})$.
There are two different ways to perform a non-relativistic limit of Maxwell's equations \cite{osti_4476686}:
\begin{itemize}
\item \textbf{Timelike (electric) limit.}
When $|\vec{U}| \ll |U^0|,$ the vector is largely timelike and its Lorentz transformations reduce to 
\beq
(U^0)' = U^0 \, , \qquad
\vec{U}' = \vec{U} - \frac{U^0}{c} \, \vec{v} \, .
\eeq
This is also called electric limit because when the current vector is largely timelike $c |\rho| \gg |\vec{j}|$, then the electric field dominates the magnetic field $|\vec{E}| \gg c |\vec{B}|.$
Applying these limits to Maxwell's equations $\p_{\mu} F^{\mu\nu} = j^{\nu},$ we get in components
\beq
\begin{aligned}
& \nabla \cdot \vec{B} = 0 \, , \qquad
\nabla \cdot \vec{E} = \frac{\rho}{\varepsilon_0} \, ,  &  \\
& \nabla \times \vec{B} = \mu_0 \vec{j} + \mu_0 \varepsilon_0 \p_t \vec{E} \, , \qquad
\nabla \times \vec{E} = 0 \, .  &
\end{aligned}
\label{eq:electric_limit_EOM_Levy}
\eeq
The main difference with the relativistic case is that the Faraday term is missing in the last equation. 
Finally, the Lorentz force in this setting is given by
\beq
\vec{F} = \int d^d x \, \rho(\vec{x}) \, \vec{E}(\vec{x}) \, .
\eeq
A magnetic field is present, but it does not have enter the expression of any physical observables.
\item \textbf{Spacelike (magnetic) limit.}
When $|\vec{U}| \gg |U^0|,$ the vector is largely spacelike and the Lorentz transformations become 
\beq
(U^0)' = U^0 - \frac{1}{c} \, \vec{v} \cdot \vec{U} \, , \qquad
\vec{U}' = \vec{U} \, .
\eeq
This is called magnetic limit because $|\vec{j}| \gg c |\rho|$, thus the magnetic field dominates compared to the electric one, \ie $c |\vec{B}| \gg |\vec{E}|.$
Maxwell's equations reduce to
\beq
\begin{aligned}
& \nabla \cdot \vec{B} = 0 \, , \qquad
\nabla \cdot \vec{E} = \frac{\rho}{\varepsilon_0} \, ,  &  \\
& \nabla \times \vec{B} = \mu_0 \vec{j}  \, , \qquad
\nabla \times \vec{E} = - \p_t \vec{B} \, .  &
\end{aligned}
\label{eq:magnetic_limit_EOM_Levy}
\eeq
The displacement current term vanishes and the Lorentz force reads
\beq
\vec{F} = \int d^d x \, \vec{j}(\vec{x}) \times \vec{B}(\vec{x}) \, .
\eeq
The electric field is non-vanishing, but it does not contribute to any physical effect, since the observables can only be expressed in terms of the magnetic field.
\end{itemize}
One can understand the existence of two Galilean limits of electromagnetism by inspecting the identity $\varepsilon_0 \mu_0 c^2 =1$.
In the non-relativistic limit $c \rightarrow \infty,$ it is not possible to keep both the permittivity and the permeability finite.
In the electric limit, we express $\mu_0= (\varepsilon_0 c^2)^{-1}$ and we take the limit $c \rightarrow \infty$ with $\varepsilon_0$ fixed, while at the same time rescaling the magnetic field as $\vec{B}' = c^2 \vec{B}$.
In the magnetic limit, we get rid of the permittivity using $\varepsilon_0 = (\mu_0 c^2)^{-1}$ and we send $c \rightarrow \infty$ with $\mu_0$ fixed.
\textit{En passant}, we mention that the existence of two distinct ways to perform an expansion of a relativistic theory around $c=\infty$ is a general feature that also applies in curved backgrounds \cite{Hansen:2020pqs}.

In this section, we worked at the level of the equations of motion (EOM), but it would be desirable to find an action formulation that encodes the two limits.
One can build two Lagrangians (containing different auxiliary fields) whose Euler-Lagrange equations reproduce the electric and magnetic limits \cite{Santos:2004pq,de2006electrodynamics}.
Interestingly, one can perform a third limiting procedure, which leads to an off-shell formulation without auxiliary fields \cite{Bagchi:2014ysa,Bergshoeff:2015sic,Festuccia:2016caf}.
This QFT model is referred to as \textit{Galilean electrodynamics} (GED) and will be treated in more detail in section \ref{ssec:GED}.

\subsection{Jackiw-Pi model}
\label{ssec:Jackiw_Pi}

Anyons are particles without definite statistics that are ubiquitous in condensed matter systems, such as the quantum Hall effect or topological superconductors \cite{Read:1999fn}.
Being described by Chern-Simons (CS) theories, they do not depend explicitly on the background in which they are located, but only on its topology.
For this reason, CS theory is intrinsically neither Lorentzian nor Galilean, but its dynamics depends on the matter to which the gauge field couples.
In this section, we review an example where the coupling of anyons to non-relativistic matter provides a simpler problem compared to the relativistic counterpart: the Jackiw-Pi model \cite{Jackiw:1990tz}.

\paragraph{Non-linear Schr\"{o}dinger equation.}
A relevant problem in mathematical physics is the so-called 1+1 dimensional \textit{non-linear Schr\"{o}dinger equation}:
\beq
i \frac{\p}{\p t} \varphi(t,x) = - \frac{1}{2m} \frac{\p^2}{\p x^2} \varphi(t,x) + g \varphi^*(t,x) \varphi(t,x) \varphi(t,x) \, . 
\label{eq:nonlinear_Schrodinger} 
\eeq
One can approach this problem in two ways.
\textbf{(1)} The first method is to consider $\psi$ as a classical complex function, solve the differential equation \eqref{eq:nonlinear_Schrodinger} with the techniques of integral calculus, and then quantize the solitonic solution.
Indeed, it turns out that this wave equation is integrable and the solitonic solutions are classified \cite{Zakharov1974OnTC}.\footnote{Solitons are wave solutions to non-linear (partial) differential equations that preserve their form and speed along the propagation, and they emerge unchanged (up to a change in phasis) after a collision with other solitons \cite{1985RRS,Manton:2004tk}.
In particular, they are excitations with finite energy.  }
\textbf{(2)} The second approach uses the methods of non-relativistic QFT in a sector with fixed particle number $N$, as presented in section \ref{ssec:why_NR_QFT}.
By defining the Hamiltonian \eqref{eq:Ham_QM} and the quantum states \eqref{eq:ket_QM} in the abstract formulation of QM, one obtains the form \eqref{eq:Schroed_eq_position} for the Schr\"{o}dinger equation in configuration space, with potentials $U(x)=0$ and $V(x-y)= g \, \delta(x-y)$.
In this form, the many-body problem is solvable using QFT methods and its S-matrix and energy spectrum are known.
The results obtained in this way are completely equivalent to the quantization of the solitonic solutions obtained by direct analysis of eq.~\eqref{eq:nonlinear_Schrodinger}.

Exact solitonic solutions to this problem are not known in higher dimensions, unless we couple the matter field to a gauge potential.
Intuitively, simplifications in higher dimensions occur because of the existence of certain BPS-like conditions that reduce the problem to the study of a first-order differential equation, instead of a second-order one (see eq.~\eqref{eq:Jackiw_Pi_self_dual_condition} below).
This is only possible when there is an intertwining between the mass and the charge of the particles. 
In higher dimensions, the wave equations become
\beq
i D_t \varphi = -\frac{1}{2m} \vec{D}^2 \varphi + g \varphi^* \varphi \varphi \, ,
\label{eq:2p1_waveeq_sec3}
\eeq
where we defined the covariant derivatives as $D_{\mu} \equiv \p_{\mu}-i e A_{\mu}$, with $e$ the electric charge.
Focusing on 2+1 dimensions, there are ways to make the gauge field dynamical: including Maxwell or CS kinetic terms. 
The former contribution can be neglected at low energies,
thus we only impose the EOM of the Abelian CS gauge theory 
\beq
\frac{\kappa}{2} \epsilon^{\mu\nu\rho} F_{\nu\rho} = j^{\mu} \, , \qquad \mathrm{where} \,\,\,
j^0 \equiv \rho = \varphi^* \varphi \, , \quad
j^i = \Im \le \varphi^* D_i \varphi \ri \, .
\label{eq:CS_eqs_motion_sec3}
\eeq
Here $\kappa$ is a coupling constant and $(j^0,j^i)$ is the matter current.
The classical wave equations \eqref{eq:2p1_waveeq_sec3} and \eqref{eq:CS_eqs_motion_sec3} can be obtained as the Euler-Lagrange equations of the action
\beq
S= \int dt d^2x \, \mathcal{L} \, , \qquad
\mathcal{L} = \frac{\kappa}{4} \epsilon^{\mu\nu\rho} A_{\mu} F_{\nu\rho} + i \varphi^* D_t \varphi  - \frac{1}{2m} |\vec{D} \varphi|^2 - \frac{g}{4} |\varphi^* \varphi|^2 \, .
\label{eq:nonrel_CS_action_sec3}
\eeq
The first contribution in the Lagrangian is the CS term, which is local and gauge-invariant (up to boundary terms) \cite{Dunne:1998qy}.
Eq.~\eqref{eq:nonrel_CS_action_sec3} provides an example of a Schr\"{o}dinger-invariant gauge theory, first proposed in \cite{PhysRevD.31.848}.

\paragraph{Static solutions.}
Static solutions to the 2+1 dimensional Abelian CS model are fully classified.
By exploiting the Galilean symmetries of the theory, one can show that static solutions carry zero energy and momentum.
This reduces the second-order differential equation \eqref{eq:2p1_waveeq_sec3} to a first-order one, thanks to the explicit expression of the Hamiltonian
\beq
H = \frac{1}{2} \int d^2 x \, \left[ |(D_x -i \, \mathrm{sign}(\kappa) D_y ) \varphi|^2 + \le \frac{g}{2} + \frac{e^2}{m  c|\kappa|} \ri \rho^2 \right]  \, ,
\eeq
where we momentarily restored the units of speed of light and a boundary term $\int d^2 x \, \nabla \times \vec{j}$ has been neglected by assuming regularity conditions at the boundary.
Since the first term and $\rho^2$ are semidefinite-positive, a non-trivial configuration with vanishing energy only exists  when $m c g |\kappa| \geq - 2 e^2.$
The problem simplifies when the bound is saturated, leading to the BPS-like condition 
\beq
g = - \frac{2 e^2}{m c \kappa}  \, . 
\label{eq:Jackiw_Pi_self_dual_condition}
\eeq
In this case, the potential $V(\vec{x}-\vec{y})= g \, \delta^2 (\vec{x}-\vec{y})$ is attractive, and one can focus on the case $\kappa >0$ without loss of generality.
Imposing eq.~\eqref{eq:Jackiw_Pi_self_dual_condition}, combined with the requirement that the Hamiltonian vanishes, one gets a self-duality condition plus the EOM \eqref{eq:CS_eqs_motion_sec3} for the CS theory
\begin{subequations}
\beq
(D_x - i D_y) \psi = 0 \, , 
\eeq
\beq
B = - \frac{1}{\kappa} (\varphi^* \varphi) \, , \qquad
E^i = \frac{1}{\kappa} \epsilon^{ij} \, \Im \le \varphi^* D_j \varphi \ri \, , 
\eeq
\end{subequations}
where $E_i \equiv F_{0i}$ and $B = \epsilon^{ij} F_{ij} $ are the electric and magnetic fields, respectively.
This system of equations is analytically solved by
\begin{subequations}
\beq
\varphi (\vec{x}) = \frac{2 \sqrt{\kappa} \mathcal{N}}{r} \left[ \le \frac{r_0}{r} \ri^{\mathcal{N}} + \le \frac{r}{r_0} \ri^{\mathcal{N}}  \right]^{-1} \, , 
\eeq
\beq
\vec{A}(t,\vec{x}) = - \frac{1}{2 \pi \kappa} \int d^2 x' \, \nabla \theta(\vec{x}-\vec{x}') \rho(t,\vec{x}') \, , \qquad
\theta(\vec{x}) = \mathrm{arctan} \, \le \frac{y}{x} \ri \, ,
\label{eq:solutionA_Jackiw_Pi} 
\eeq
\end{subequations}
where $\mathcal{N} \in \mathbb{N}$ and $r_0$ is a positive real constant.
This solution describes $\mathcal{N}$ solitons superimposed at the origin, all of them characterized by the same scale $r_0$.
They represent vortices carrying $2 \mathcal{N}$ units of magnetic flux.
The most general solution consists of separating the previous coincident $\mathcal{N}$ solitons into distinct objects located at different points.
Notice that $\vec{A}$ is not a pure gauge, since the integrand in eq.~\eqref{eq:solutionA_Jackiw_Pi} is multi-valued and thus the solution presents a non-trivial topology.

Solitonic solutions to the wave equations in CS theories coupled to non-relativistic fermions were considered in \cite{Duval:1995fa}.
Reference \cite{Horvathy:2008hd} provides a broader review on the topic of Jackiw-Pi vortices.
The Abelian CS model can also be studied in a relativistic setting, by coupling the gauge field to a Klein-Gordon scalar.
One can show that the critical value $m c g |\kappa|=2 e^2$ corresponds to a non-relativistic limit, where the theory also enjoys an emergent supersymmetry, as we will see in section \ref{ssec:SUSY_CS}.
For all these reasons, the non-relativistic CS model is integrable and easier to study compared to its relativistic counterpart.

\subsection{Aharanov-Bohm scattering}
\label{ssec:AB_scattering}

The Aharonov-Bohm (AB) effect consists of the scattering of an electron beam by a magnetic field in the limit where the size of the magnetic field region approximates zero radius, while keeping at the same time the flux fixed \cite{PhysRev.115.485}.
Flux-charge composite objects in this setting acquire fractional statistics and are then described by anyons \cite{PhysRevLett.48.1144}.
For this reason, the above-mentioned scattering problem can be equivalently formulated as the scattering of two non-relativistic particles coupled to a CS gauge field, described by the action \eqref{eq:nonrel_CS_action_sec3}.
We will focus here on the Abelian case, but we refer the reader to \cite{PhysRevD.33.407} for the analysis of the non-Abelian AB effect.

\paragraph{QM approch.}
The original computation of the AB scattering was performed in 2+1 dimensions by directly studying the following wave equation in cylindrical coordinates
\beq
\left[ \frac{\p^2}{\p r^2} + \frac{1}{r} \frac{\p}{\p r} + \frac{1}{r^2} \le \frac{\p}{\p \theta} + i \alpha \ri^2 + k^2  \right] \varphi(r) = 0 \, ,
\label{eq:AB_wave_equation}
\eeq
where $\vec{k}$ is the wave vector of the incident particle,  $\alpha= e^2/(2\pi \kappa)$ and $\kappa$ is the coupling constant of the CS term.
Eq.~\eqref{eq:AB_wave_equation} is obtained by coupling the non-linear Schr\"{o}dinger equation \eqref{eq:nonlinear_Schrodinger} to the gauge potential, and then working in the Coulomb gauge and in the center of mass frame \cite{PhysRev.115.485,RUIJSENAARS19831,PhysRevD.41.2015,JACKIW199083}.
The scattering amplitude of identical particles in this setting reads
\beq
\mathcal{A}_{\rm AB} (k,\theta) = -i (\pi k)^{-1/2} \, \sin \le \pi \alpha \ri \left[ \cot \theta -i \, \mathrm{sign} \, (\alpha)  \right] \, .
\label{eq:AB_amplitude_QM}
\eeq
Since the momentum $\vec{k}$ only enters through the kinematical factor, this dependence is a symptom of the scale invariance of the theory, as we revisit below from a QFT perspective.

\paragraph{QFT perturbative computation.}
It is possible to recover the result \eqref{eq:AB_amplitude_QM} in the QFT framework.
Previous perturbative attempts were tried in \cite{Feinberg,10.1119/1.11155,10.1119/1.11390},
but they failed because the $s$--wave contribution in the Born approximation is singular, as later observed in \cite{PhysRevD.29.2396}.
While other \textit{ad-hoc} approaches can be taken to reproduce the scattering process of two anyons at lowest order \cite{PhysRevD.44.2533,Chou1992PerturbativeAS,SEN1991397}, the resolution to the problem was finally found from a one-loop analysis in \cite{Bergman:1993kq}. The result was later generalized to all orders in perturbation theory in \cite{Kim:1996rz}.

In the remainder of this subsection, we will mainly follow the procedure developed in reference \cite{Bergman:1993kq}.
The starting point is the action \eqref{eq:nonrel_CS_action_sec3} describing the coupling between a Schr\"{o}dinger scalar and anyons through a gauge potential in 2+1 dimensions.
The contact interaction $g |\varphi^{\dagger} \varphi|^2$ was omitted in the other references, but is crucial to compute the AB scattering amplitude.
One can explicitly check that the theory is invariant under the Lifshitz scaling transformations \eqref{eq:Lifhsitz_transformations} with dynamical exponent $z=2$, but loop corrections can break this invariance by introducing a dimensionful renormalization scale.
To properly define the path integral and study perturbation theory around a saddle point, we need to add a gauge-fixing term to the Lagrangian
$ \mathcal{L}_{\rm g.f.} = \frac{1}{\xi} (\nabla \cdot \vec{A})^2 $, where $\xi$ is a free parameter.
Since all the physical observables are gauge-invariant, from now on we will set $\xi=0$ for convenience (Landau gauge).
To deal with divergent momentum integrals, one defines renormalized quantities in terms of bare ones as
\begin{subequations}
\beq
\varphi_B = Z_{\varphi}^{1/2} \varphi \equiv \le 1+ \frac{1}{2} \delta_{\varphi} \ri \varphi \, ,  \qquad
m_B = Z_m m \equiv \le 1+ \delta_m \ri m \, , \qquad
g_B =  Z_g g \equiv \le 1+ \delta_g \ri g \, ,
\label{eq:renormalized_data_sec3}
\eeq
\beq
A^B_{\mu} = Z_{A} A_{\mu} \equiv \le 1 + \delta_A \ri A_{\mu} \, , \quad
e_B = Z_{\kappa} \kappa \equiv \le 1+ \delta_{e} \ri e \, , \quad
\kappa_B = Z_{\kappa} \kappa \equiv \le 1+ \delta_{\kappa} \ri \kappa \, ,
\eeq
\end{subequations}
which determine the counterterms $\delta_{\dots}$ by requiring that UV divergences are cancelled.
Integrals in momentum space are evaluated as follows: one performs the $\omega$ integration (which is always convergent) first, and then the remaining spatial integrals are regularized introducing a spatial UV cutoff $\Lambda.$
The Feynman rules of the theory read
\begin{subequations}
\beq
\includegraphics[valign=c, scale=1]{Figures/prop_scalar.pdf}  =
\frac{i}{ \omega -\frac{\vec{p}^2}{2m} + i \varepsilon} \, ,
\qquad
\includegraphics[valign=c, scale=1]{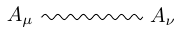}  =
\begin{pmatrix}
0 &  \frac{\epsilon^{ij} p_j}{ \kappa \, \vec{p}^2}\\
- \frac{\epsilon^{ij} p_j}{ \kappa \, \vec{p}^2} & 0 
\end{pmatrix} \, ,
\eeq
\beq
\includegraphics[scale=0.7, valign=c]{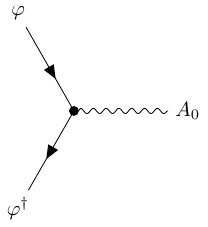} = i e \, , \qquad
\includegraphics[scale=0.7, valign=c]{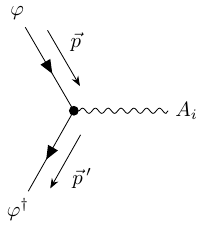} = - \frac{i e}{2m} (p_i + p'_i) \, ,
\eeq
\beq 
\includegraphics[scale=0.7, valign=c]{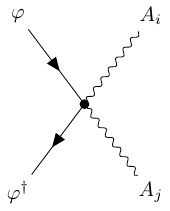} =  \frac{i e^2}{m} \, \delta_{ij} \, , \qquad
\includegraphics[scale=0.7, valign=c]{Figures/4pt_vertex} = i g \, .
\eeq
\end{subequations}
Let us focus on 1PI diagrams, which contribute to the effective action according to eq.~\eqref{eq:relation_effectiveaction_1PI_sec2}.
A remarkable feature of the theory is the non-renormalization of the propagators. 
For instance, a class of diagrams contributing to the self-energy of the scalar field is given by
\beq
i \Gamma^{(2)}[\varphi_0, \varphi_0^{\dagger}]  =
\includegraphics[scale=0.8, valign=c]{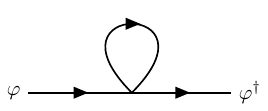}  + 
\includegraphics[scale=0.8, valign=c]{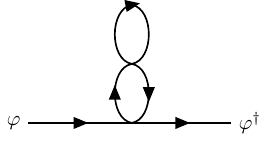}  
+ \dots
\eeq
Arrows in each diagram always form a closed loop, and the superficial degree of divergence in $\omega$ is always negative.
A direct application of selection rule \ref{sel_rule1} implies that at all orders, the previous contributions vanish.
A similar analysis can be performed for the diagrams involving gauge fields in the loops. Therefore, one concludes that $ \Gamma^{(2)} [\varphi_0, \varphi_0^{\dagger}] = 0$.
Similar steps can be followed for the self-energy of the gauge field.\footnote{We will see another application of this argument in section \ref{ssec:GED}, for the study of Galilean electrodynamics.}
The non-renormalization of the propagators implies that 
\beq
Z_{\varphi} = Z_A = Z_m = Z_e = Z_{\kappa} = 1 \, .
\eeq 
This is the first example of a \textit{non-renormalization theorem}, which is a consequence of the causal structure of the non-relativistic propagator.

To study the AB effect, we need to consider the $2 \rightarrow 2$ scattering of scalar particles.
Up to one-loop, the contributions to the corresponding vertex are given by
\begin{subequations}
\beq
i \Gamma^{(4)}_{\rm tree} = \includegraphics[scale=0.6, valign=c]{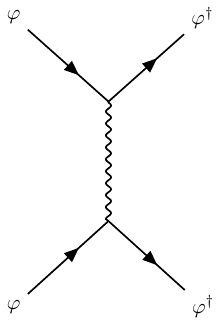} + 
\includegraphics[scale=0.8, valign=c]{Figures/4pt_vertex} 
\eeq
\beq
i \Gamma^{(4)}_{\rm 1loop} =
\includegraphics[scale=0.6, valign=c]{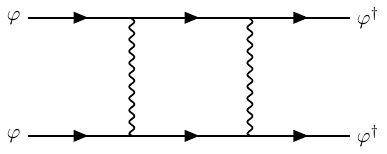} + 
\includegraphics[scale=0.6, valign=c]{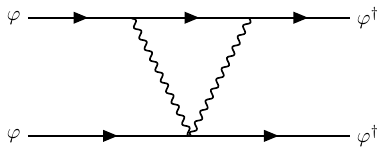} + 
 \includegraphics[scale=0.6, valign=c]{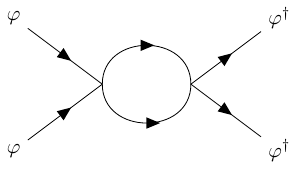} 
\eeq
\end{subequations}
One can show that the divergences coming from the previous diagrams can be renormalized by choosing the 
counterterm (here $\mu$ is an energy scale)
\beq
\delta_g =  \frac{1}{4 \pi g} \le g^2 - \frac{4 e^4}{m^2 \kappa^2} \ri \log \le \frac{\Lambda}{\mu} \ri + \mathcal{O}(g^2, e^6) \, .
\label{eq:counterterm_deltag_sec3}
\eeq
This leads to the following total amplitude (tree level plus one-loop) 
\beq
\mathcal{M}(p,\theta,\mu) = - \frac{2i e^2}{m \kappa} \cot \theta - g + \frac{m}{8 \pi} \le g^2 - \frac{4e^4}{m^2 \kappa^2} \ri \left[ \log \le \frac{\mu^2}{p^2} \ri + i \pi \right] \, ,
\label{eq:intermediate_amplitude_sec35}
\eeq
where $\theta$ is the scattering angle and $p$ the magnitude of the relative momentum.

Since this expression involves contributions from both the scalar and gauge fields, it is possible to fine-tune the coupling constants such that the $\mu$--dependent term vanishes.
The critical point corresponds to the choice
\beq
g = \pm \frac{2 e^2}{m c |\kappa|} \, ,
\label{eq:critical_point_sec34}
\eeq
where we momentarily restored the factors in the speed of light.
We take $\kappa>0$ without loss of generality and we select the upper sign, corresponding to a repulsive contact interaction.
Notice that this critical value coincides with  eq.~\eqref{eq:Jackiw_Pi_self_dual_condition}, which was the condition to find self-dual solutions for the Jackiw-Pi model in section \ref{ssec:Jackiw_Pi}.\footnote{The conditions \eqref{eq:Jackiw_Pi_self_dual_condition} and \eqref{eq:critical_point_sec34} differ by a sign because in the former case one considers an attractive potential, while in the latter a repulsive one.}
Using eq.~\eqref{eq:critical_point_sec34} inside \eqref{eq:intermediate_amplitude_sec35}, the amplitude  becomes
\beq
\mathcal{M}_{\rm AB} (p,\theta) = - i \frac{4 \pi \alpha}{m} \left[ \cot \theta - i \, \mathrm{sign} \, (\alpha) \right] \, ,
\label{eq:AB_effect_QFT_sec3}
\eeq
where $\alpha$ was defined below eq.~\eqref{eq:AB_wave_equation}.
After including the appropriate kinematical factor, this amplitude coincides with $\mathcal{A}_{\rm AB}$ in eq.~\eqref{eq:AB_amplitude_QM}.

\paragraph{Concluding remarks.}
We showed that a perturbative QFT computation reproduces the amplitude of the AB scattering computed from the non-linear Schr\"{o}dinger equation.
To achieve this matching, we had to impose a choice of the coupling constants to preserve the scale invariance of the model, which was broken by renormalization.
The critical value \eqref{eq:critical_point_sec34} coincides with the self-dual condition for the non-linear Schr\"{o}dinger equation coupled to a CS term.
Had we considered a fermionic field instead of the Schr\"{o}dinger scalar, the Lagrangian would have contained a Pauli term instead of the quartic scalar interaction.
It turns out that the Pauli interaction fixes the quartic vertex describing the AB scattering and that the condition \eqref{eq:critical_point_sec34} is automatically satisfied.
This fact is a manifestation of the $\mathcal{N}=2$ supersymmetry enjoyed by the model, as we will discuss in section \ref{ssec:SUSY_CS}.

\subsection{Supersymmetric Chern-Simons theory}
\label{ssec:SUSY_CS}

The Jackiw-Pi model reviewed in section \ref{ssec:Jackiw_Pi} admits self-dual solitonic solutions in correspondence with the critical point \eqref{eq:Jackiw_Pi_self_dual_condition}.
The same fine-tuning of the coupling constants preserves the scale invariance of the theory at one loop in the perturbative computation and correctly reproduces the AB effect \eqref{eq:AB_amplitude_QM}.
As anticipated at the end of section \ref{ssec:AB_scattering}, the critical value \eqref{eq:Jackiw_Pi_self_dual_condition} is automatically imposed when fermionic particles are scattered in the presence of an electromagnetic field.
In this section, we will show that the previous phenomena arise because Schr\"{o}dinger matter coupled to a CS term enjoys a $\mathcal{N}=2$ supersymmetric extension \cite{Leblanc:1992wu}.

\paragraph{Non-relativistic limit of $\mathcal{N}=2$ Chern-Simons model.}
There are three main approaches to building the action for the supersymmetric version of the Jackiw-Pi model:
\begin{enumerate}
\item One can consider the symmetry structure of the theory as fundamental: in this case, one first derives the graded Galilean algebra, and then builds the action by imposing invariance under the transformations of the fields.
To build the algebra, one can either directly construct the graded version of the Galilean algebra in 2+1 dimensions \cite{Puzalowski:1978rv}, or take an Inon\"{u}-Wigner contraction $c \rightarrow \infty$ of the relativistic SUSY algebra in the same number of dimensions \cite{deAzcarraga:1991fa}.
Another way to derive the algebra, based on the null reduction of the 3+1 dimensional super-Poincaré symmetries, will be developed in section \ref{ssec:nonrel_SUSY_algebra}.
\item One builds a superspace formalism, finds the corresponding representations of the algebra, and then the invariant action in terms of superfields. This method will be discussed in section \ref{ssec:nonrel_superspace}. Its implementation for the $\mathcal{N}=2$ Jackiw-Pi model was done in \cite{Nakayama:2009ku}.
\item In this section, we review the method in \cite{Leblanc:1992wu} where the action is taken as the fundamental object, and it is constructed via a $c \rightarrow \infty$ limit of the relativistic $\mathcal{N}=2$ supersymmetric CS system in 2+1 dimensions \cite{Lee:1990it}.
The SUSY algebra is derived afterward from the symmetry transformations of the fields.
\end{enumerate}
Let us begin with the Lagrangian of the relativistic $\mathcal{N}=2$ model in 2+1 dimensions \cite{Lee:1990it}
\beq
\begin{aligned}
\mathcal{L} & = \frac{\kappa}{4 c} \epsilon^{\mu\nu\rho} A_{\mu} F_{\nu\rho} + |D_{\mu} \Phi|^2 + i \bar{\Psi} \gamma^{\mu} D_{\mu} \Psi \\
& - \le \frac{e^2}{\kappa c^2} \ri^2 |\Phi|^2 \le |\Phi|^2 - v^2 \ri^2 + \frac{e^2}{\kappa c^2} \le 3 |\Phi|^2 - v^2 \ri \bar{\Psi} \Psi \, ,
\end{aligned}
\label{eq:lagrangian_relCS_sec36}
\eeq
where $\Phi$ is a complex scalar, $\Psi$ a two-component Dirac spinor, $v$ a constant and we defined the covariant derivatives as $D_{\mu} \equiv \p_{\mu} - \frac{ie}{c} A_{\mu}$. The factors in the speed of light have been restored to take the non-relativistic limit.
To implement the latter, we parametrize the matter fields as
\beq
 \Phi = \frac{1}{\sqrt{2m}} \le e^{-i m c^2 t} \varphi + e^{i m c^2 t} \hat{\varphi}^{\dagger} \ri \, , \qquad
   \Psi = \frac{1}{\sqrt{2m}} \le e^{-i m c^2 t} \psi + e^{i m c^2 t} \sigma_2 \hat{\psi}^{\dagger} \ri \, ,  
\label{eq:ansatz_fields_sec36}
\eeq
where $\sigma_2$ is the second Pauli matrix, $(\varphi, \psi)$ are the non-relativistic fields associated with the particle sector, and $(\hat{\varphi}^{\dagger}, \hat{\psi}^{\dagger})$ refer to the anti-particles.\footnote{The ansatz for the scalar field is similar to the procedure proposed in eq.~\eqref{eq:ansatz_Phi_sec2}, but the difference is that $\hat{\varphi}^{\dagger}, \hat{\psi}^{\dagger}$ are not the hermitian conjugates of $\varphi, \psi$.}
Since the dynamics of the gauge field is only governed by the topological CS term, there are no degrees of freedom associated with it.
For this reason, there is no need to modify the parametrization of the gauge field.

At leading order in $1/c$ one of the two components of the fermionic fields becomes non-dynamical, and can be integrated out. 
Moreover, particles and anti-particles are separately conserved in the Lagrangian \eqref{eq:lagrangian_relCS_sec36} after plugging in the ansatz \eqref{eq:ansatz_fields_sec36}, therefore we can simply focus on the particle sector by setting $\varphi^{\dagger}=\psi^{\dagger}=0$.
Taking into account these considerations, we perform the limit $c \rightarrow \infty$ and we drop terms which oscillate. The result reads
\beq
\begin{aligned}
\mathcal{L} & = \frac{\kappa}{4 c} \epsilon^{\mu\nu\rho} A_{\mu} F_{\nu\rho} 
+ i \varphi^{\dagger} D_t \varphi
+ i \chi^{\dagger} D_t \chi
- \frac{1}{2m} |\vec{D} \Phi|^2
- \frac{1}{2m} |\vec{D} \chi|^2 \\
& + \frac{e}{2 m c} B |\chi|^2 
+ \frac{\lambda_1}{4} |\varphi|^4 + \frac{\lambda_2}{4} |\varphi|^2 |\chi|^2 \, ,
\end{aligned}
\label{eq:lagrangian_nonrelCS_sec36}
\eeq
where $\chi$ is the dynamical (complex) component of the non-relativistic fermionic field $\psi=(\xi,\chi)$, and $B$ is the only independent component of the magnetic field $B_{ij}= B \epsilon_{ij}$ in 2+1 dimensions.
The first term in the second line is a Pauli interaction, analog to eq.~\eqref{eq:ham_gyromagnetic_ratio}, that arises after integrating out the non-dynamical component $\xi$ of the fermion.
The coupling constants are given by 
\beq
\lambda_1 = - \frac{2 e^2}{m c \kappa} \, , \qquad
\lambda_2 = 6 \lambda_1 \, .
\eeq
In particular, $\lambda_1$ coincides with the self-dual condition \eqref{eq:Jackiw_Pi_self_dual_condition} in the Jackiw-Pi model, as promised.
The model \eqref{eq:lagrangian_nonrelCS_sec36} describes scalar matter minimally coupled to a CS term through an attractive $\delta$--function potential of strength $\lambda_1$, complemented with the interactions for the fermionic superpartner.

\paragraph{Classical symmetries.}
The classical symmetry group associated with the Lagrangian \eqref{eq:lagrangian_nonrelCS_sec36} can be determined by computing the Noether charges and their Lie brackets.
The bosonic symmetries of the theory form the Schr\"{o}dinger group with commutation rules \eqref{eq:Schroedinger_algebra}.
The model further enjoys a supersymmetry invariance generated by two complex supercharges $Q_{\alpha}$ (with $\alpha \in \lbrace 1,2 \rbrace$) satisfying the anti-commutation rules
\beq
\begin{aligned}
&  \lbrace Q_1, Q_1^{\dagger}  \rbrace = \sqrt{2} H \, , \quad
\lbrace Q_2, Q_2^{\dagger}  \rbrace = \sqrt{2} M \, , & \\
& \lbrace Q_1, Q_2^{\dagger} \rbrace = -  (P_1 -i P_2) \, , \quad
\lbrace Q_2, Q_1^{\dagger} \rbrace = -  (P_1 + i P_2) \, .
\end{aligned}
\label{commu_superGalileo2_sec36} 
\eeq
This structure strictly resembles the relativistic case, except that some components close into the mass generator.
We will see that there is a natural way to derive this structure using null reduction in section \ref{ssec:nonrel_SUSY_algebra}.
Here we only comment that the supercharges transform as spin one-half objects and have fermionic statistics.

The Lagrangian \eqref{eq:lagrangian_nonrelCS_sec36} enjoys even more symmetries, complemented by a bosonic generator $R$ for $\mathrm{U}(1)$ R-symmetry and by the superpartner $S$ of the SCTs generated by $C$.
The additional graded brackets that close the algebra are given by
\beq
\begin{aligned}
& \lbrace S, S^{\dagger} \rbrace = C \, , \qquad
\lbrace S, Q_2^{\dagger} \rbrace = - (G_1 + i G_2) \, , \qquad
&  \\
& [J,Q_1] =  \frac12 Q_1 \, , \quad
[J,Q_2] = - \frac12 Q_2 \, , \quad
[Q_1, G_1 - i G_2] = -i Q_2 \, , & \\
& 
[H, S^{\dagger}] = i Q_1^{\dagger} \, , \qquad
[J,S] = - \frac{1}{2} S \, , \qquad
[D, S] = - i S \, , \qquad
[C, Q_2] = - i S \, , & \\
& \lbrace S, Q_1^{\dagger}  \rbrace = \frac{i}{2} \le i D - J  +\frac{3}{2} R \ri \, , \qquad
[R, Q_{\alpha}] = - Q_{\alpha}  \, , \qquad
[R,S] = - S \, , &
\end{aligned}
\label{eq:superconformal_algebra_sec36}
\eeq
where $J_{ij}= J \epsilon_{ij}$ denotes the only independent component of the rotation generator in two spatial dimensions.
The (anti)commutators \eqref{eq:Schroedinger_algebra}, \eqref{commu_superGalileo2_sec36} and \eqref{eq:superconformal_algebra_sec36} form the $\mathcal{N}=2$ super-Schr\"{o}dinger algebra in 2+1 dimensions \cite{Duval:1984cj,Duval:1993hs,Julia:1994bs,Duval:1990hj}.
We will analyze the quantum properties of two models with this symmetry group in sections \ref{ssec:Galilean_WZ} and \ref{ssec:SGED}.
Finally, let us comment that other non-relativistic and supersymmetric realizations of CS theory have been realized, including different gauge groups and a different number of supercharges \cite{Nakayama:2008qz,Nakayama:2008td,Lopez-Arcos:2015cqa,Nakayama:2009cz,Tong:2015xaa,Tong:2016kpv}.

\subsection{Non-relativistic scale anomaly}
\label{ssec:nonrel_scale_anomaly}

We apply the second-quantization framework to the  Schr\"{o}dinger scalar field \eqref{eq:Schroed_action_sec2} with contact interaction in 2+1 dimensions \cite{Bergman:1991hf}, described by the Lagrangian \eqref{eq:Lagrangian_NR_phi4_sec2}.
This setting corresponds to the non-linear Schr\"{o}dinger system in eq.~\eqref{eq:nonlinear_Schrodinger}, but in one higher dimension.

\paragraph{QM approach.}
The scattering amplitude between two non-relativistic particles interacting via the contact potential $V(\vec{x}-\vec{y}) = g \, \delta(\vec{x}-\vec{y})$ can be computed in the Born approximation
\beq
\mathcal{M} (p,g,\mu) = g \left[ 1 - \frac{g m}{8 \pi} \le \log \le \frac{\mu^2}{p^2} \ri + i \pi \ri + \mathcal{O}(g^2) \right] \, ,
\label{eq:exact_amplitude_Born}
\eeq
where $\mu$ is an arbitrary energy scale. 
This result can be obtained by setting $e=0$ in the  $2 \rightarrow 2$ scattering amplitude \eqref{eq:intermediate_amplitude_sec35} for the AB process studied in section \ref{ssec:AB_scattering}.
In this case, one can actually proceed further and sum the perturbative expansion to all orders, getting
\beq
\mathcal{M} (p,g,\mu) = g \left[ 1 + \frac{g m}{4 \pi} \le \log \le \frac{\mu}{p} \ri + \frac{i \pi}{2} \ri \right]^{-1} \, .
\label{eq:exact_amplitude_Born}
\eeq
Non-trivial physics in this setting corresponds to the existence of a bound state with energy $ E_B = - \frac{\mu^2}{2m} e^{\frac{8\pi}{m g}} $, determined from the pole of the scattering amplitude. The bound state only arises in the case of an attractive $\delta$--like potential.

\paragraph{QFT approach.}
In a second-quantization framework, the starting point  is the Lagrangian \eqref{eq:Lagrangian_NR_phi4_sec2}.
Feynman rules and loop corrections can be obtained by setting to zero the gauge field and the electric charge $e$ in all the computations performed for the AB scattering problem in section \ref{ssec:AB_scattering}.
An immediate consequence of this identification is the non-renormalization of the wavefunction and of the mass parameter at any order, \ie
\beq
Z_{\varphi} =1 \, , \qquad
Z_m = 1 \, .
\eeq
Next, we proceed with the study of loop corrections to the four-point vertex. 
Selection rule \ref{sel_rule1} comes to rescue, since all the diagrams involving exchanges in the $t$ and $u$ channels vanish due to the arrows of particle number symmetry forming a closed loop.
There is only one non-trivial contribution at each loop order, therefore the full set of diagrams is the following:
\beq
i \Gamma^{(4)} =
\includegraphics[scale=0.7, valign=c]{Figures/4pt_vertex.pdf}  + \includegraphics[scale=0.7, valign=c]{Figures/4pt_vertex_1loop.pdf} 
+ \includegraphics[scale=0.7, valign=c]{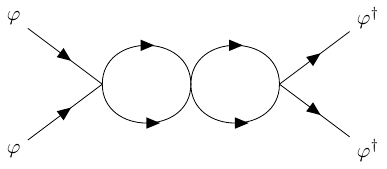} 
+ \dots
\eeq
Using the recursive structure of the Feynman diagrams, one can perform the exact summation to get the effective action
\beq
i \Gamma^{(4)} = i g \left[ 1 + \frac{g \, m}{8 \pi} \log \le \frac{4 \mu^2}{(\vec{p}_1 + \vec{p}_2)^2 - 4 m^2 (\omega_1 + \omega_2)} \ri  \right]^{-1} \, ,
\label{eq:effective_action4_sec3}
\eeq
where $(\omega_i, \vec{p}_i)$ with $i=1,2$ refer to the momenta of the incoming particles, and $\mu$ is an arbitrary energy scale.
The renormalized coupling coincides with eq.~\eqref{eq:counterterm_deltag_sec3} with $e=0$.
Taking the momenta in eq.~\eqref{eq:effective_action4_sec3} to be on-shell and going to the center of mass frame with relative momentum $\vec{p}$, one precisely obtains the scattering amplitude \eqref{eq:exact_amplitude_Born}.
This result shows the agreement between the first and the second-quantization methods, but the latter makes evident the important role played by the causal structure of the non-relativistic propagator.

\paragraph{Trace anomaly.}
In the QFT framework, it is clear that the running of the coupling constant is interpreted as the signal of a trace anomaly.\footnote{Anomalies can also arise in purely QM systems, as shown in \cite{Gaiotto:2017yup}. However, the QFT framework is very convenient to study them.}
We recall that the Lagrangian \eqref{eq:Lagrangian_NR_phi4_sec2} is classically scale invariant under the Lifshitz transformations \eqref{eq:Lifhsitz_transformations} with $z=2$.
The invariance under dilatations in a relativistic QFT implies that the energy-momentum tensor is traceless, \ie $T^{\mu}_{\,\, \mu} =0$.
By computing the Noether current associated with Lifshitz scale transformations and imposing its conservation, we find
$ z T^0_{\,\, 0} - T^i_{\,\, i} = 0$.
When $z=2$, this identity corresponds to the Schr\"{o}dinger version of the conservation equation. 

Loop corrections break the scale invariance by introducing 
a renormalization energy scale, leading to a traceful stress-tensor (\textit{trace anomaly}).
In flat space, the quantum violation of the classical scale symmetry for a Schr\"{o}dinger-invariant theory is quantified by
\beq
2 T^0_{\,\, 0} - T^i_{\,\, i} = \beta^i \mathcal{O}_i \, ,
\label{eq:quantum_breaking_betafunction}
\eeq
where $\beta^i$ are the beta functions of the coupling constants associated with the operators $\mathcal{O}_i(t,\vec{x})$.
The identity \eqref{eq:quantum_breaking_betafunction} can be 
derived from the renormalization group (RG) equations and the scale dependence of the effective action, \eg see \cite{Bergman:1991hf}.

In the present case, the only running coupling is $g$, whose associated beta function reads
$ \beta(g) = \mu \frac{\p g}{\p \mu} = \frac{g^2}{4 \pi} $,
where eq.~\eqref{eq:counterterm_deltag_sec3} was used.
Since the associated operator in the Lagrangian \eqref{eq:Lagrangian_NR_phi4_sec2} is the scalar $|\varphi^{\dagger} \varphi|^2$, a direct application of the identity \eqref{eq:quantum_breaking_betafunction} gives
\beq
2 T^0_{\,\, 0} - T^i_{\,\, i} = \frac{g^2}{16 \pi} |\varphi^{\dagger} \varphi|^2 \, .
\eeq
This is a remarkable achievement: we found an exact expression for the non-relativistic trace anomaly of a theory.
In a relativistic setting, this kind of result is usually reachable in supersymmetric theories only, for instance using holomorphy \cite{Seiberg:1993vc}.
The causal structure of the retarded propagator introduced several simplifications that ultimately led to this result.


\section{Modern techniques}
\label{sec:modern_NR_QFT}

The historical review of section \ref{sec:historical_review} showed the power of non-relativistic QFT in many instances, \eg to compute solitonic solutions, scale anomalies, and measurable effects such as the Aharanov-Bohm scattering.
While the approaches adopted in the previous section were successful in describing several phenomenological problems, they have two main limitations.
\textbf{(1)} The older techniques developed in the 1960s gave the EOM for non-relativistic fermions and gauge fields, but it would be desirable to derive them from an action formulation admitting a non-relativistic general coordinate invariance. This also includes the coupling to a corresponding geometry.
\textbf{(2)} Interesting applications in condensed matter systems (such as the quantum Hall effect) and theoretical physics (such as the classification of the trace anomaly) necessarily require coupling to curved space.

To solve the above-mentioned issues, we need \textbf{(1)} a systematic way to derive non-relativistic theories and \textbf{(2)} a covariant description of Newtonian gravity, called Newton-Cartan (NC) geometry \cite{Cartan:1923zea,Cartan2}, that we review in section \ref{ssec:NC_geometry}.
While it is possible in principle to build non-relativistic actions by imposing the invariance under the Schr\"{o}dinger group, it is practically more convenient to start from a relativistic parent model.
In this section we present the two main mechanisms: the implementation of a well-defined $c \rightarrow \infty$ limit in the presence of a background electromagnetic field (section \ref{ssec:nonrel_limit}) and the dimensional reduction along a null direction of a Lorentzian theory (section \ref{ssec:null_red}).
We will show several applications of this modern technology in sections \ref{sec:modern_applications} and \ref{sec:SUSY}.

The techniques described in this section generate field theories coupled to type I NC gravity (see section \ref{ssec:NC_geometry} below for the precise definition), which is relevant for all the physical applications considered in this review.
Alternatively, one can perform a covariant expansion in powers of $c^{-2}$ of a field theory coupled to a (pseudo-)Riemannian background geometry.
The truncation at next-to-leading order of this expansion provides the fields composing type II NC geometry and their transformation rules.
The algebra satisfied by the symmetry generators is different from Bargmann \cite{Hansen:2019pkl,Hansen:2019vqf}.
Since in this review we focus on backgrounds associated with Schr\"{o}dinger-invariant field theories, type II NC geometries will not be considered here.\footnote{However, let us point out that there is no distinction between type I and II NC geometries when $d\tau=0$, with $\tau$ the one-form defined above eq.~\eqref{eq:constraints_NC}. }
We refer the interested reader to references \cite{Dautcourt:1996pm,VandenBleeken:2017rij,Hansen:2020pqs} and to the recent review \cite{Hartong:2022lsy} for a discussion of actions for point particles or field theories coupled to these backgrounds.

\subsection{Newton-Cartan geometry}
\label{ssec:NC_geometry}

NC geometry was born as a coordinate-independent way to describe Newtonian gravity, in the same fashion as general relativity deals with local observers that see the laws of special relativity in inertial frames.
NC geometry has a long history, which started from the seminal works by Cartan in 1923 \cite{Cartan:1923zea,Cartan2}, evolved into a modern axiomatic formulation by Trautman \cite{Trautman1,Trautman2,Trautman3}, and finally experienced a recent revival after the construction of NC geometry from a gauging of the Bargmann algebra \cite{Andringa:2010it}.
An exhaustive review of the historical development of the topic can be found in \cite{Hartong:2022lsy}.
In order to be self-contained, in this work we introduce the relevant notions to construct non-relativistic QFTs, either in flat or curved space.
We mainly follow the notation used in references \cite{Hartong:2022lsy,Hansen:2021xhm}.

\paragraph{Metric data and local Galilean boosts.}
The metric data of NC geometry in $d+1$ dimensions consist of a nowhere-vanishing clock one-form $\tau_{\mu}$ that designates the time direction and a degenerate symmetric contravariant two-tensor $h^{\mu\nu}$ with rank $d$.
We supplement these objects by an inverse velocity vector $v^{\mu}$ and a spatial metric $h_{\mu\nu}$ such that
\beq
\tau_{\mu} h^{\mu\nu} = 0 \, , \qquad
v^{\mu} h_{\mu\nu} = 0 \, , \qquad
\tau_{\mu} v^{\mu} = -1 \, , \qquad
h^{\mu\rho} h_{\rho\nu} = \delta^{\mu}_{\nu} + v^{\mu} \tau_{\nu} \, .
\label{eq:constraints_NC}
\eeq
As a byproduct of the introduction of the inverse data, we can define an invertible metric $\gamma_{\mu\nu} \equiv \tau_{\mu} \tau_{\nu} + h_{\mu\nu}$ with inverse $\gamma^{\mu\nu} = v^{\mu} v^{\nu} + h^{\mu\nu}$ and with vielbein determinant $e \equiv \sqrt{\mathrm{det} \, \gamma_{\mu\nu}}$.
In flat space, the metric data become
\beq
\tau_{\mu} dx^{\mu} = -dt \, , \quad
v^{\mu} \p_{\mu} = \p_{t} \, , \quad
h_{\mu\nu} dx^{\mu} dx^{\nu} = \delta_{ij} dx^i dx^j \, , \quad
h^{\mu\nu} \p_{\mu} \p_{\nu} = \delta^{ij} \p_i \p_j \, .
\label{eq:flat_NC_space}
\eeq
NC geometry gives a covariant formulation of Newtonian gravity where observers in inertial frames experience Galilean relativity.
As such, its data transform under \textit{local Galilean boosts} (sometimes also called \textit{Milne boosts} in the literature) as follows\footnote{The original notion of non-relativistic diffeomorphisms discussed in \cite{Son:2005rv} and developed in the later works \cite{Son:2013rqa,Geracie:2014nka} corresponds to fixing of the local Galilean symmetry, as pointed out in  \cite{Jensen:2014wha,Jensen:2014aia}.  }
\beq
\delta \tau_{\mu} = 0 \, , \qquad
\delta v^{\mu} = h^{\mu\nu} \lambda_{\nu} \, , \qquad
\delta h^{\mu\nu} = 0 \, , \qquad
\delta h_{\mu\nu} = \tau_{\mu} \lambda_{\nu} + \tau_{\nu} \lambda_{\mu} \, ,
\label{eq:local_Galilean_boosts}
\eeq
where $\lambda_{\mu}$ is a boost parameter satisfying $v^{\mu} \lambda_{\mu} = 0$.
Notice that $\tau_{\mu}$ and $h^{\mu\nu}$ do not transform under local Galilean boosts.
However, they are not sufficient alone to determine uniquely the inverse data $v^{\mu}$ and $h_{\mu\nu}$, and indeed the ambiguity in solving the conditions \eqref{eq:constraints_NC} precisely correspond to the transformations \eqref{eq:local_Galilean_boosts}.
In flat space, Galilean boosts act as
\beq
t \rightarrow t \, , \qquad
x^i \rightarrow x^i + v^i t \, ,
\eeq
where here $v^i$ are the relative velocities between the two frames.

Starting from the covariant tensors $(\tau_{\mu}, h_{\mu\nu})$, it is possible to uniquely find the other pair $(v^{\mu}, h^{\mu\nu})$  and viceversa, since they combine into the invertible metric $\gamma_{\mu\nu}$.
Finally, let us stress that the NC formalism enjoys full covariance under diffeomorphisms, in other words any scalar function built out of the geometric data listed above eq.~\eqref{eq:constraints_NC} is invariant under general coordinate transformations.

\paragraph{Connection.}
To define the transport of vectors across spacetime, we need to introduce a notion of connection.
We require that the boost-invariant objects are covariantly constant with respect to the covariant derivative $\nabla$ associated with the affine connection $\Gamma$, \ie
\beq
\nabla_{\mu} \tau_{\nu} = \p_{\mu} \tau_{\nu} - \Gamma^{\sigma}_{\mu\nu} \tau_{\sigma} = 0 \, , \qquad
\nabla_{\mu} h^{\nu\rho}  = \p_{\mu} h^{\nu\rho} 
+ \Gamma^{\nu}_{\mu\sigma} h^{\sigma\rho}  + \Gamma^{\rho}_{\mu\sigma} h^{\nu\sigma}  = 0 \, ,
\label{eq:compatibility_NC_metric}
\eeq
while in general $\nabla_{\mu} v^{\nu}$ and $\nabla_{\mu} h_{\nu\rho}$ are non-vanishing because of the degenerate metric structure.
The conditions \eqref{eq:compatibility_NC_metric} is a requirement that any NC metric-compatible connection has to satisfy.

The major differences compared to the relativistic case are that \textbf{(1)} there is not a unique compatible connection, and \textbf{(2)} the intrinsic torsion is a natural feature of NC geometry, since the first constraint in eq.~\eqref{eq:compatibility_NC_metric} implies $(d\tau)_{\mu\nu} = \tau_{\sigma} \Gamma^{\sigma}_{[\mu\nu]}$, which only vanishes when the one-form $\tau$ is closed.\footnote{The intrinsic torsion is defined as the non-zero value of the first Cartan's structure equation.    }
The simplest connection compatible with the metric data reads
\begin{equation}
    \check\Gamma^\rho_{\mu\nu}=-v^\rho\partial_\mu\tau_\nu+\frac{1}{2}h^{\rho\sigma}\left(\partial_\mu h_{\nu\sigma}+\partial_\nu h_{\mu\sigma}-\partial_\sigma h_{\mu\nu}\right)\,,
    \label{eq:checkGamma1}
\end{equation}
which resembles a Levi-Civita connection built from the spatial metric $h_{\mu\nu}$, except that there is an additional temporal part contributing to the torsion.
By explicit computation, we find
\beq
\check\nabla_{\mu} v^{\nu} = \frac{1}{2} h^{\nu\sigma} \mathcal{L}_v h_{\sigma\mu} \, , \qquad
\check\nabla_{\mu} h_{\nu\rho} = \tau_{(\nu} \mathcal{L}_v h_{\rho) \nu} \, ,
\eeq
where $\mathcal{L}_v$ denotes the Lie derivative with respect to the velocity field $v^{\mu}$.
Neither the affine connection \eqref{eq:checkGamma1} nor its induced covariant derivate are Galilean boost-invariant, but all the physical quantities built with them are required to only appear through boost-invariant combinations.

The torsion depends on the properties of the
clock one-form $\tau$.
The general classification distinguishes three classes of NC geometries \cite{Christensen:2013lma,Christensen:2013rfa,Hartong:2014pma,Figueroa-OFarrill:2020gpr}:
\begin{itemize}
\item $\boldsymbol{d\tau=0}$. This is the torsionless case, which was initially proposed by Cartan.
Inside this class of spacetimes, it is possible to choose an absolute notion of time $t$ (in the spirit of Newton's original idea) by picking $\tau_{\mu} = \p_{\mu} t$.
\item $\boldsymbol{\tau \wedge d\tau=0}$.
Backgrounds of this kind are called twistless-torsional Newton-Cartan (TTNC) geometries. They have non-vanishing torsion, but admit an integrability structure because they satisfy the Frobenius condition $\tau \wedge d\tau=0$. The clock one-form $\tau_{\mu}$ is used to define hypersurfaces whose tangent vectors are spacelike vectors $k^{\mu}$, \ie they satisfy $\tau_{\mu} k^{\mu}=0$. 
\item $\boldsymbol{\tau \wedge d\tau \ne 0}$.
They are called torsional Newton-Cartan (TNC) geometries. Due to the absence of an integrability structure, they are acausal because it is possible to find closed timelike curves; moreover, any two points can be connected with spacelike trajectories \cite{Geracie:2015dea}.
\end{itemize}
Other features arise if we equip the NC geometry with additional structure: this leads to type I and type II NC geometry, which we review below.

\paragraph{Type I Newton-Cartan geometry.}
Type I NC geometry corresponds to an enhancement of the local frame symmetry from the Galilean to the Bargmann group, \ie the metric data contain an additional $\mathrm{U}(1)$ gauge field $m_{\mu}$ which is associated to the particle number conservation.
This field transforms non-trivially both under local Galilean boosts $ \delta m_{\mu} = \lambda_{\mu} $ and under local $\mathrm{U}(1)$ transformations $\delta_{\sigma} m_{\mu} = \p_{\mu} \sigma ,$ where $\sigma$ is the gauge parameter. 
This allows to build the boost-invariant combinations
\beq
\hat{v}^{\mu} \equiv v^{\mu} - h^{\mu\nu} m_{\nu} \, , \qquad
\bar{h}_{\mu\nu} \equiv h_{\mu\nu} - 2 \tau_{(\mu} m_{\nu)} \, , \qquad
\hat{\Phi} = - v^{\mu} m_{\mu} + \frac{1}{2} h^{\mu\nu} m_{\mu} m_{\nu} \, .
\label{eq:boost_invariant_NCdata}
\eeq
While $\bar{h}_{\mu\nu}$ is now boost-invariant and not degenerate, it transforms non-trivially under $\mathrm{U}(1)$ gauge transformations.

The connection \eqref{eq:checkGamma1} is invariant under gauge transformations, but not under local Galilean boosts.
It is possible to construct a connection with the opposite properties (boost but not gauge-invariant) by using the data \eqref{eq:boost_invariant_NCdata} as follows
\begin{equation}
    \bar{\Gamma}^\rho_{\mu\nu}=- \hat{v}^\rho\partial_\mu\tau_\nu+\frac{1}{2}h^{\rho\sigma}\left(\partial_\mu \bar{h}_{\nu\sigma}+\partial_\nu \bar{h}_{\mu\sigma}-\partial_\sigma \bar{h}_{\mu\nu}\right)\,.
    \label{eq:checkGamma2}
\end{equation}
More generally, there does not exist any NC connection composed of type I fields which is metric-compatible and invariant under boosts and gauge transformations at the same time.
Furthermore, affine connections satisfying the constraints \eqref{eq:compatibility_NC_metric} and the condition $d\tau=0$ differ by an ambiguity proportional to the field strength $F=dm$.

Type I NC geometry can be obtained either from a large speed of light limit $c \rightarrow \infty$ of a Lorentzian geometry with an electromagnetic gauge field, or from the dimensional reduction of a higher-dimensional Lorentzian manifold with a null isometry, as we will discuss in sections \ref{ssec:nonrel_limit} and \ref{ssec:null_red}.
It can also be obtained by gauging the Bargmann algebra \cite{Andringa:2010it,Hartong:2015zia,Bergshoeff:2022eog}.

\paragraph{Type II Newton-Cartan geometry.}
Type II NC is the geometry that arises at next-to leading order from a covariant expansion in powers of $c^{-2}$ of general relativity.\footnote{The series expansion of general relativity, including both odd and even powers of $c^{-1}$, was studied in \cite{Dautcourt:1996pm,Ergen:2020yop}.  }
The field content is the same as type I NC geometry, except for the presence of an additional field denoted with $\pi^a_{\mu}$ in section 2.2 of the original reference \cite{Hansen:2020pqs}, and a new term in the transformation rule of the $\mathrm{U}(1)$ gauge connection.

\subsection{Non-relativistic limits}
\label{ssec:nonrel_limit}

In this section, we build a QFT coupled to type I NC geometry from a limit $c \rightarrow \infty$ of a relativistic parent theory in the same number of dimensions. We mainly follow reference \cite{Jensen:2014wha}, but we complement the discussion with the insights coming from the covariant expansion of general relativity in powers of $c^{-2}$ recently considered in  \cite{Hansen:2020pqs}, and we further refer the reader to the review \cite{Bergshoeff:2022eog} for other details.

\paragraph{Non-relativistic action.}
Let us start from a Lorentzian background in $d+1$ dimensions. We decompose the metric by designating a particular timelike one-form $T_{\mu}$ as follows:
\begin{subequations}
\beq
g_{\mu\nu} = -c^2 T_{\mu} T_{\nu} + \Pi_{\mu\nu} \, , \qquad
g^{\mu\nu} = - \frac{1}{c^2} V^{\mu} V^{\nu} + \Pi^{\mu\nu} \, ,
\label{eq:metric_PNR}
\eeq
\beq
T_{\mu} V^{\mu} = -1 \, , \qquad
T_{\mu} \Pi^{\mu\nu} = 0 \, , \qquad
V^{\mu} \Pi_{\mu\nu} = 0 \, , \qquad
\Pi^{\mu\sigma} \Pi_{\sigma\nu} = \delta^{\mu}_{\nu} + V^{\mu} T_{\nu} \, .
\label{eq:identities_PNR}
\eeq
\end{subequations}
This is sometimes referred to as the pre non-relativistic (PNR) parametrization \cite{Hansen:2020pqs}.
So far, this is only a rewriting of a Lorentzian metric in terms of other geometric data.
Next we assume that these objects admit an analytic expansion in powers of $c^{-2}$ given by
\beq
T_{\mu} = \tau_{\mu} + \mathcal{O}(c^{-2}) \, , \quad
V^{\mu} = v^{\mu} + \mathcal{O}(c^{-2}) \, , \quad
\Pi_{\mu\nu} = h_{\mu\nu} + \mathcal{O}(c^{-2}) \, , \quad
\Pi^{\mu\nu} = h^{\mu\nu} + \mathcal{O}(c^{-2}) \, . 
\label{eq:expansions_metric_data_Jensen}
\eeq
To get type I NC geometry, we need to introduce a background gauge field $C_{\mu}$.
While its origin may seem obscure, we can understand the need for a gauge connection by starting from the action of a relativistic free scalar coupled to a curved background \cite{Janiszewski:2012nb}
\beq
S= - \frac{1}{2} \int dt \, d^{d}x \, \sqrt{-g} \, \le  g^{\mu\nu} \mathcal{D}_{\mu} \Phi^{\dagger} \mathcal{D}_{\nu} \Phi  + m^2 c^2 |\Phi|^2 \ri \, ,
\label{eq:action_Jensen}
\eeq
where we introduced covariant derivatives $\mathcal{D}_{\mu} = \p_{\mu} - i C_{\mu} \Phi .$
Plugging the PNR parametrization \eqref{eq:metric_PNR} with the expansions \eqref{eq:expansions_metric_data_Jensen} inside the action \eqref{eq:action_Jensen}, we notice that the rest mass term is divergent.
This contribution can be canceled by the chemical potential $C_{\mu}$ if we require
\beq
C_{\mu} = m c^2 \tau_{\mu} + m \, m_{\mu} + \mathcal{O} (c^{-2}) \, .
\label{eq:ansatz_Cmu_Jensen}
\eeq
Now we plug this ansatz inside the relativistic action \eqref{eq:action_Jensen}, we rescale the scalar field as $\phi = \sqrt{mc} \, \Phi$ and we take the limit $c\rightarrow \infty$ to get
\beq
\lim_{c \rightarrow \infty} S = - \int dt \, d^d x \, e \, \left[ \frac{i v^{\mu}}{2} \le \phi^{\dagger} D_{\mu} \phi - \phi D_{\mu} \phi^{\dagger} \ri  + \frac{h^{\mu\nu}}{2m} D_{\mu} \phi^{\dagger} D_{\nu} \phi  \right] \, ,
\label{eq:nonrel_action_jensen}
\eeq
where $e$ is the vielbein determinant introduced below eq.~\eqref{eq:constraints_NC}, while $D_{\mu} \phi = \p_{\mu} \phi - i m \, m_{\mu} \phi.$
We recognize the appearance of the metric data $(v^{\mu}, h^{\mu\nu}, m_{\mu})$ introduced in section \ref{ssec:NC_geometry}.
This is the action of a Schr\"{o}dinger scalar field coupled to a type I NC background.
The classical symmetries of the model are given by $\mathrm{U}(1)$ gauge transformations, and local Galilean boosts.

The leading-order expansion of the local Lorentz transformations acting on local frames reproduces the local Galilean boosts \eqref{eq:local_Galilean_boosts}, including the transformation of the gauge field reported above eq.~\eqref{eq:boost_invariant_NCdata}.
Alternatively, one finds that there is an ambiguity in the decompositions \eqref{eq:metric_PNR}, \eqref{eq:expansions_metric_data_Jensen} and \eqref{eq:ansatz_Cmu_Jensen}, because one can re-define the NC metric data while at the same time preserving the identities \eqref{eq:identities_PNR}.
In the limit $c \rightarrow \infty$, this ambiguity precisely amounts to the local Galilean boosts collected in eq.~\eqref{eq:local_Galilean_boosts}.
From an algebraic perspective, these results correspond to the statement that the Inon\"{u}-Wigner contraction of the Poincaré algebra gives the Galilean algebra.

In flat space \eqref{eq:flat_NC_space}, the action~\eqref{eq:nonrel_action_jensen} becomes
\beq
\lim_{c \rightarrow \infty} S =  \int dt \, d^d x \, e \, \left[ \frac{i }{2} \le \phi^{\dagger} D_{t} \phi - \phi D_{t} \phi^{\dagger} \ri  - \frac{\delta^{ij}}{2m} D_{i} \phi^{\dagger} D_{j} \phi  \right] \, ,
\label{eq:nonrel_action_jensen_flat}
\eeq
which describes a Schr\"{o}dinger scalar minimally coupled to a gauge field $m_{\mu}$.

\paragraph{Connection.}
The connection plays an important role in NC geometry, despite its definition was not needed to compute the scalar action \eqref{eq:nonrel_action_jensen}.
The natural starting point is to consider the Levi-Civita connection $\Gamma$ compatible with the metric \eqref{eq:metric_PNR}, which reads
\begin{subequations}
\beq
\Gamma^{\rho}_{\mu\nu} = c^2 \overset{(-2)}{S^{\rho}_{\mu\nu}} +  \sum_{k=0}^{1} c^{-2k} \overset{(2k)}{S^{\rho}_{\mu\nu}} \, , 
\label{eq:splitting_connection_sec4}
\eeq
\beq
\overset{(-2)}{S^{\rho}_{\mu\nu}}  =  \frac{1}{2}T_\mu\Pi^{\rho\sigma}\left(\partial_\sigma T_\nu-\partial_\nu T_\sigma\right)+\frac{1}{2}T_\nu\Pi^{\rho\sigma}\left(\partial_\sigma T_\mu-\partial_\mu T_\sigma\right)\,.
\eeq
\end{subequations}
Since $\Gamma$ is torsionless, one necessarily gets a torsionless connection $\overset{(2k)}{S^{\rho}_{\mu\nu}}$ at each order in an expansion in powers of $c^{-2}$.
Since the leading term in eq.~\eqref{eq:splitting_connection_sec4} is divergent in the limit $c \rightarrow \infty$, to achieve a regular connection one needs to add a tensorial term involving the field strength $\mathcal{F}_{\mu\nu} = 2 \p_{[\mu} C_{\nu]}$ of the gauge potential \eqref{eq:ansatz_Cmu_Jensen}.
There are two prescriptions:
\beq
(\Gamma^{(1)})^{\rho}_{\mu\nu} \equiv \Gamma^{\rho}_{\mu\nu}  + \frac{1}{m c^2} \, g^{\rho\sigma} \tau_{(\nu} \mathcal{F}_{\mu) \sigma} \, , \qquad
(\Gamma^{(2)})^{\rho}_{\mu\nu} \equiv \Gamma^{\rho}_{\mu\nu}  + \frac{1}{m^2 c^2} \, g^{\rho\sigma} C_{(\nu} \mathcal{F}_{\mu) \sigma} \, .
\label{eq:two_connections_limit}
\eeq
Both these connections are engineered to give a regular limit when $c \rightarrow \infty$, where the former object $\Gamma^{(1)}$ is gauge-invariant, while $\Gamma^{(2)}$ is invariant under local Galilean boosts.
After performing the actual limit, they reduce to the torsionless parts of the connections \eqref{eq:checkGamma1} and \eqref{eq:checkGamma2}, respectively.
In this approach, the torsionful NC connections are obtained with further shifts by terms proportional to $g^{\rho\sigma} \mathcal{F}_{\mu\nu} \tau_{\sigma}$ or $g^{\rho\sigma} \mathcal{F}_{\mu\nu} C_{\sigma}$, respectively.

\paragraph{Conclusive remarks.}
We derived a non-relativistic action coupled to non-relativistic gravity with a limiting procedure $c \rightarrow \infty$.
One begins with the decomposition \eqref{eq:metric_PNR} of a pseudo-Riemannian metric in a form that designates a preferred time direction, and then introduces an additional background potential to get rid of the divergences in $c^2$.
A similar recipe, applied to the Levi-Civita connection compatible with the metric $g_{\mu\nu}$ in eq.~\eqref{eq:metric_PNR}, gives a regular NC connection in the limit $c \rightarrow \infty$.
Torsion needs to be added via a shift of the connection.

We mention that in the covariant expansion of general relativity in powers of $c^{-2}$, one does not introduce any background gauge field.
Instead the relativistic connection $\Gamma$ is allowed to start at order $c^2$, and matter fields are also expanded in powers of $c^{-2}$.
The torsionful NC connection is obtained from the expansion of a PNR connection, which is used instead of the Levi-Civita one.
Finally, there is a systematic procedure to determine invariant actions from an appropriate truncation at each order in the expansion \cite{Hansen:2020pqs,Hartong:2022lsy}.

\subsection{Null reduction}
\label{ssec:null_red}

Type I NC geometry in $(d+1)$ dimensions can be obtained 
from the dimensional reduction along a null direction of a Lorentzian geometry with a null isometry in $(d+2)$ dimensions \cite{PhysRevD.31.1841}.
This procedure is referred to as \textit{null reduction} or sometimes as \textit{discrete light-cone quantization} (DLCQ).\footnote{More precisely, the null reduction requires to have a Lorenztian background with null isometry along a certain coordinate $u$, but a priori $u$ need not to be compact. The DLCQ quantization consists in the further step where $u$ is required to be compact.    }
Given a background with a null isometry, it is possible to choose, without loss of generality, adapted coordinates $x^M=(u,x^{\mu})=(u,v,x^i)$ such that the metric reads
\beq
ds^2 = g_{MN} dx^M dx^N = 2 \tau_{\mu} dx^{\mu} \le du  - m_{\nu} dx^{\nu} \ri + h_{\mu\nu} dx^{\mu} dx^{\nu} \, ,
\label{eq:nullred_metric_sec4}
\eeq
where all the objects entering the line element are independent of the null coordinate $u$, and $x^{\mu}$ collects the coordinates of the $(d+1)$-dimensional Newton-Cartan (NC) geometry that arises from null reduction.
The other null coordinate $v$ in the parent spacetime is interpreted as the time direction in the lower-dimensional geometry.
In this coordinate system, the Killing vector $n^M$ parametrizing the null isometry reads
\beq
n^M \p_M = \p_u \, , \qquad
n_M dx^M = \tau_{\mu} dx^{\mu} \, . 
\label{eq:null_isometry_sec4}
\eeq
The tensors entering the metric \eqref{eq:nullred_metric_sec4} can be identified as the NC data introduced in section \ref{ssec:NC_geometry}, where $\tau$ is the no-where vanishing clock one-form, $m$ is a $\mathrm{U}(1)$ gauge connection and $h$ a spatial metric with rank $d$.
In particular, they satisfy the constraints \eqref{eq:constraints_NC}.
The metric determinant reads
\beq
\sqrt{-g} = \sqrt{\mathrm{det} (\tau_{\mu} \tau_{\nu} + h_{\mu\nu})} = \sqrt{\mathrm{det} (\gamma_{\mu\nu})} = e  \, ,
\eeq
and thus coincides with the determinant of the invertible metric $\gamma_{\mu\nu}$ introduced below eq.~\eqref{eq:constraints_NC}.
In components, the parent metric and its inverse can be written as
\beq
g_{MN} = \begin{pmatrix}
0 &  \tau_{\nu} \\
\tau_{\mu} & \bar{h}_{\mu\nu}
\end{pmatrix} \, , \qquad
g^{MN} = \begin{pmatrix}
2 \hat{\Phi} &  -\hat{v}^{\nu}\\
-\hat{v}^{\mu} & h^{\mu\nu} 
\end{pmatrix} \, ,
\label{eq:nullred_metric_TNC}
\eeq
where we used the boost-invariant combinations \eqref{eq:boost_invariant_NCdata}.
While the metric components are invariant by construction under local Galilean boosts, their decomposition into $(v^{\mu}, h_{\mu\nu}, m_{\mu})$ is ambiguous.
The freedom in re-defining these data precisely leads to the transformations \eqref{eq:local_Galilean_boosts} together with the variation of the gauge field reported above eq.~\eqref{eq:boost_invariant_NCdata}.

Diffeomorphisms along the $(d+1)$ dimensions of the non-relativistic geometry are obviously inherited from the diffeomorphisms of the parent geometry, while $\mathrm{U}(1)$ gauge transformations arise from the coordinate reparametrizations along the null direction $u$ in the parent spacetime.
For these reasons, null reduction provides a convenient mechanism that automatically encodes all the symmetries required for type I NC geometry \cite{Jensen:2014aia}.

One can also perform the null reduction in the first-order formalism, which is essential to build fermionic actions. The vielbeine read 
\beq
e_M^A = 
\begin{pmatrix}
	1 & -m_{\mu}\\
	0 & \tau_{\mu}\\
	0 & e_{\mu}^a
\end{pmatrix} \, , \qquad
e_A^M = 
\begin{pmatrix}
	1 & -v^\mu m_{\mu} & e^\mu_a m_{\mu} \\
	0 & -v^{\mu} & e^{\mu}_a 
	\end{pmatrix} \, ,
\label{eq:nullred_vielbein}
\eeq
where we introduced flat coordinates $x^A=(x^-, x^{\alpha})=(x^-,x^+,x^a)$ for local frames.
In terms of the metric of the tangent space in light-cone coordinates
\beq
g_{AB} = g^{AB} = \begin{pmatrix}
0 & \mathbf{1} \\
\mathbf{1} & \mathbf{0}
\end{pmatrix} \, ,
\eeq
one gets the usual consistency conditions
\beq
 g_{MN}= e^{A}_{\,\,\, M} g_{AB} e^{B}_{\,\,\, N}  \, ,  
\qquad  g_{AB}= e^{M}_{\,\,\, A} g_{MN} e^{N}_{\,\,\, B} \, ,  
\qquad
  e^{A}_{\,\,\, M} e^{M}_{\,\,\, B} = \delta^{A}_{\,\,\, B}  \, , \qquad
  e^{M}_{\,\,\, A} e^{A}_{\,\,\, N} = \delta^{M}_{\,\,\, N}  \, .
\eeq
Plugging the decomposition \eqref{eq:nullred_vielbein} inside these identities ultimately leads to the constraints
\begin{equation}
	 e^\mu_a e_\mu^b = \delta_a^b \, , \qquad
	 e^\mu_a \tau_\mu = 0 , \qquad 
	 v^\mu e_\mu^a = 0 \, , \qquad
	  h^{\mu \nu} = e^\mu_a e^\nu_a , \quad h_{\mu \nu} = e_\mu^a e_\nu^a \, . 
\end{equation}
The vielbeine \eqref{eq:nullred_vielbein} can be used to build the spin connection directly in the parent theory.

\paragraph{Connection.}
Null reduction naturally provides a connection.
Starting from the Levi-Civita connection $\Gamma_g$ compatible with the metric \eqref{eq:nullred_metric_sec4}, the restriction along the directions of the NC submanifold gives the torsionless part of the boost-invariant connection \eqref{eq:checkGamma2}, \ie
\beq
(\Gamma_g)^{\rho}_{\mu\nu} = \bar{\Gamma}^{\rho}_{(\mu\nu)}  =- \hat{v}^\rho\partial_{(\mu} \tau_{\nu)}+\frac{1}{2}h^{\rho\sigma}\left(\partial_\mu \bar{h}_{\nu\sigma}+\partial_\nu \bar{h}_{\mu\sigma}-\partial_\sigma \bar{h}_{\mu\nu}\right)  \,.
\eeq
The torsion part and the gauge-invariant connection \eqref{eq:checkGamma1} can be obtained with appropriate shifts of $\Gamma_g$ by terms defined in the higher-dimensional parent space. 
This is the counterpart of the similar procedure that can be implemented for the limit $c \rightarrow \infty$ of a relativistic theory explained below eq.~\eqref{eq:two_connections_limit}.

\paragraph{Null reduction of the algebra.}
From an algebraic perspective, it is possible to obtain a type I NC geometry by null reduction of a Lorentzian manifold because the Schr\"{o}dinger group $\mathrm{Sch}(d)$ in $d$ spatial dimensions is a subgroup of the conformal group $\mathrm{SO}(d+2,2)$ \cite{Son:2008ye}.
In fact, one can start with the conformal algebra in $(d+2)$ dimensions
\beq
\begin{aligned}
& [\mathcal{J}^{MN} , \mathcal{J}^{PQ} ] =  i \le \eta^{MP} \mathcal{J}^{NQ} + \eta^{NQ} \mathcal{J}^{MP}  - \eta^{MQ} \mathcal{J}^{NP}  - \eta^{NP} \mathcal{J}^{MQ}  \ri \, ,  & \\
& [\mathcal{J}^{MN} , \mathcal{P}^{Q} ] =  i \le \eta^{MQ} \mathcal{P}^{N} - \eta^{NQ} \mathcal{P}^{M} \ri \, , \qquad
 [\mathcal{J}^{MN} , \mathcal{K}^{Q} ] =  i \le \eta^{MQ} \mathcal{K}^{N} - \eta^{NQ} \mathcal{K}^{M} \ri \, ,   & \\
& [\mathcal{D}, \mathcal{P}^M] =  i \mathcal{P}^M \, , \qquad
[\mathcal{D}, \mathcal{K}^{M}] = -i \mathcal{K}^M \, ,  \qquad [\mathcal{P}^M, \mathcal{K}^N] =  2 i \le \eta^{MN} \mathcal{D} - \mathcal{J}^{MN} \ri \, , &
\end{aligned}
\label{eq:conformal_algebra}
\eeq
where we denoted with $\mathcal{J}$ the rotations, $\mathcal{P}$ translations, $\mathcal{D}$ dilatations and $\mathcal{K}$ special conformal transformations.

By identifying the null momentum $\mathcal{P}_u= \mathcal{P}^v $ as the $\mathrm{U}(1)$ mass operator $M$ of the non-relativistic theory, the Schr\"{o}dinger algebra is obtained by selecting all the generators commuting with $P_u$.
The identification is the following:
\beq
\begin{aligned}
& M = \mathcal{P}^v \, , \qquad
H = \mathcal{P}^u \, , \qquad
P^{i} = \mathcal{P}^{i} \, , \qquad
J^{ij} = \mathcal{J}^{ij} \, ,  & \\
& K^{i} = \mathcal{J}^{i v} \, , \qquad
 D = \mathcal{D} - \mathcal{J}^{vu} \, , \qquad
 C = \frac{1}{2} \mathcal{K}^v \, .   &
\end{aligned}
\label{eq:map_generators_Son}
\eeq
These generators compose the Schr\"{o}dinger algebra, and indeed they satisfy the commutation relations \eqref{eq:Schroedinger_algebra}.

\paragraph{Null reduction of field theories.}
Null reduction is realized by compactifying the null direction $u$ of the $(d+2)$--dimensional Lorentzian manifold on a small circle of radius $R$.
For convenience, one rescales $u \rightarrow u/R$ to make the null coordinate dimensionless, and accordingly $v \rightarrow R v$ to keep the metric tensor dimensionless.
A generic field $\Phi_l$ in the relativistic parent theory (where $l$ denotes a spin representation) can be Fourier-expanded along the null direction.
If the field has a definite Bargmann mass, then we can keep only a single mode with fixed momentum $P_u=m$ among the Kaluza-Klein (KK) modes, where $m$ is the eigenvalue of the mass generator.
One can then decompose the field as
\beq
\Phi_l (x^M) = e^{i m u} \varphi_l (x^{\mu}) \, ,
\label{eq:ansatz_fields_nullred_sec4}
\eeq
where $\varphi_l$ is interpreted as a non-relativistic field in $(d+1)$ dimensions.
This ansatz breaks the original Poincaré symmetries into 
the Bargamann subgroup.
For non-vanishing spin $l \ne 0$, there can be an additional decomposition according to the Bargmann representation that the non-relativistic field has to fit. This will be presented in detail for fermions and gauge fields in sections \ref{ssec:nonrel_fermions} and \ref{ssec:GED}, respectively.

We apply the null reduction to a relativistic Weyl-invariant scalar in curved space, \ie
\beq
S = -  \int d^{d+2}x  \, \sqrt{-g} \, 
\le  g^{MN} \partial_M \Phi^{\dagger} \partial_N \Phi - \xi R \Phi^{\dagger} \Phi \ri \, ,
\label{eq:free_scalar_sec4} 
\eeq
where $R$ is the Ricci scalar and $\xi = \frac{d}{4(d+1)}$ is the conformal coupling in $(d+2)$ dimensions.
By plugging the metric \eqref{eq:nullred_metric_sec4} and the field ansatz \eqref{eq:ansatz_fields_nullred_sec4} inside this action, one gets
\beq
S = - \int d^{d+1} x \, \sqrt{g} \, \left\lbrace i m v^{\mu} \le \varphi^{\dagger} D_t \varphi - D_{\mu} \varphi^{\dagger} \varphi  \ri + h^{\mu\nu} D_{\mu} \varphi^{\dagger} D_{\nu} \varphi - \xi R \varphi^{\dagger} \varphi  \right\rbrace \, ,
\label{eq:action_schr_scalar_sec4}
\eeq
where the covariant derivative is defined by $D_{\mu} \varphi = (\p_{\mu} - i m \, m_{\mu}) \varphi.$\footnote{We clarify that $R$ in eq.~\eqref{eq:action_schr_scalar_sec4} is the Ricci scalar of the Lorentzian metric with null isometry upon which the reduction is performed. By construction, it only depends on the $(d+1)$ coordinates of the lower-dimensional non-relativistic manifold. }
Up to an overall normalization, this is the action \eqref{eq:nonrel_action_jensen} for a Schr\"{o}dinger scalar complemented with a conformal coupling to type I NC geometry.

Null reduction provides a systematic method to build Schr\"{o}dinger-invariant actions starting from a relativistic parent theory.
The major advantage compared to other techniques is that non-relativistic symmetries are automatically preserved, and that no additional compensating background fields need to be added to make the action well-defined (see section \ref{ssec:nonrel_limit}).
Finally, let us clarify that null reduction does not necessarily require curved space to be implemented.
One can directly consider a field theory in Minkowski space with light-cone coordinates and apply the same technology described above, resulting in a non-relativistic theory coupled to type I NC geometry with data \eqref{eq:flat_NC_space}.


\section{Modern applications}
\label{sec:modern_applications}

The interest in non-relativistic physics from a high-energy perspective received a great boost after the seminal paper by Son and Wingate \cite{Son:2005rv}, which introduced some of the technology reviewed in sections \ref{sec:preliminaries} and \ref{sec:modern_NR_QFT}, in particular the coupling to NC geometry.\footnote{Let us emphasize that in the following developments, the requirement of general coordinate invariance on a curved background was a remarkable insight and novelty of the program developed in \cite{Son:2005rv}. }
These methods can be used to approach several recent problems.
We present applications in condensed-matter physics, focusing on fermions at unitarity and the quantum Hall effect, in sections \ref{ssec:fermions_unitarity} and \ref{ssec:QHE}, respectively.
This program suggests a dual holographic description of Schr\"{o}dinger symmetry, that we present in section \ref{ssec:geometric_AdS_Schroed}.
Next, we move to the off-shell formulation of Schr\"{o}dinger-invariant fermions (section \ref{ssec:nonrel_fermions}), gauge fields (section \ref{ssec:GED}) and general field theories with $\mathrm{SU}(1,n)$ spacetime symmetry obtained from a deformation of the standard null reduction (section \ref{ssec:Lambert_QFT}).
Finally, we summarize recent investigations of the non-relativistic trace anomaly in section \ref{ssec:nonrel_trace_anomalies}.

\subsection{Fermions at unitarity}
\label{ssec:fermions_unitarity}

The revival of non-relativistic physics by Son and Wingate originated from the need to study new kinds of fermionic superfluids that were realized in ultracold atomic gases in 2004 using the Feshbach resonance \cite{Regal:2004zza,Zwierlein:2004zz}.\footnote{Interactions between ultracold atoms can be controlled by a magnetic field, due to the coupling of free atoms to a molecular state in which they are tightly bounded. Feshbach resonances occur when the molecular level is close to the energy of the two free atoms, in which case the interactions between the atoms are strong. }
Particular interest lies in cold and dilute atomic gases, where features of \textit{universality} arise: the potential between different atoms is approximated by a contact interaction with vanishing range, whose precise details are not important.
The phase diagram of this system is governed by the dimensionless parameter $|a k_F|$, where $a$ is the scattering length and $k_F$ is the Fermi momentum.
At small $|a k_F| \ll 1$, two regimes appear:
\begin{itemize}
\item When $a k_F <0$, the system is a weak interacting Fermi gas.
\item When $a k_F >0$, couples of fermions form bound molecules and the system behaves like a weakly interacting Bose gas.
\end{itemize}
Since these two regimes are smoothly connected without phase transitions, one concludes that the ground state is a superfluid for any value of $|a k_F|$.
The \textit{unitarity} limit corresponds to the strongly-interacting case where $a k_F \rightarrow \infty$, \ie the scattering length becomes infinite.
The major difficulty for an analytic treatment of fermions at unitarity is the absence of any small dimensionless quantity.
A resolution to this limitation was proposed in \cite{Nishida:2006eu,Nishida:2007pj}, based on dimensional regularization around $d=4-\varepsilon$ or $d=2+\varepsilon$ spatial dimensions, which allows for a perturbative expansion in terms of a small parameter $\varepsilon$.
The physical case is achieved by extrapolating the series expansions to $\varepsilon=1$.
For a review on the topic, see \cite{Nishida:2010tm};
for recent developments, we refer to \cite{Chowdhury:2023ahp}.

The reason why $d=2,4$ play a special role is that at infinite scattering length, the two-body wavefunction has a leading $\psi(r) \sim r^{2-d}$ behaviour when the separation $r$ between two fermions becomes small.
Therefore, the integral for its normalization reads
\beq
\int d^{d} x \, |\psi(r)|^2 \sim \int_0^{\infty} \frac{dr}{r^{d-3}} \, ,
\eeq
which shows two different regimes \cite{Baumgardt:2004tp}:
\begin{itemize}
\item When $d=2$, the integrand is regular and the system reduces to a non-interacting Fermi gas. In fact, any attractive potential in $d \leq 2$ admits at least one bound state, which corresponds to a vanishing potential when the scattering length is tuned to infinity. 
\item When $d=4$, the two-body wavefunction is concentrated at the origin, and the fermion pair behaves like a composite boson. The system looks like a non-interacting Bose gas.
\end{itemize}
These two cases match the experimental settings mentioned above when $|a k_F| \ll 1$.
The physical case $d=3$ lies between them and can be approached from both sides.

\paragraph{Perturbative analysis.}
When $d=4-\varepsilon$, the Lagrangian reads
\beq
\mathcal{L}_{4 - \varepsilon} = i \psi^{\dagger}_{\sigma} \p_t \psi_{\sigma} - \frac{1}{2m} |\nabla \psi_{\sigma}|^2 + i \phi^{\dagger} \p_t \phi - \frac{1}{4m} |\nabla \phi|^2 
+ g \psi_{\uparrow}^{\dagger} \psi_{\downarrow}^{\dagger} \phi + g \psi_{\downarrow} \psi_{\uparrow} \phi^{\dagger} 
- \frac{g^2}{c_0} \phi^{\dagger} \phi \, ,
\eeq
where $\psi_{\sigma}= (\psi_{\uparrow}, \psi_{\downarrow})$ is a spin--$\frac{1}{2}$ fermion with mass $m$, $\phi$ is a scalar (it is the bound state of two fermions with opposite spin) and $g$ a coupling constant.
The parameter $(c_0)^{-1}$ is set to 0 in dimensional regularization, corresponding to the fine-tuning to infinite scattering length.
Perturbative corrections can now be studied using Feynman diagrams and the selection rule discussed in section \ref{ssec:loop_corrections} \cite{Nishida:2007pj}.

There is a fixed point at $g=8 \pi^2 \varepsilon$ in correspondence with which the theory enjoys Schr\"{o}dinger invariance and describes fermions at unitarity.
At the fixed point, we can use the non-relativistic state/operator correspondence outlined in section \ref{ssec:nonrel_state_OPE} to compare the scaling dimensions of the fields with the energy levels of a harmonic oscillator.
The scaling dimension of the fermionic field $\psi$ does not run along the RG flow and is $\Delta_{\psi} = d/2$.
This matches with the existence of a one-fermion state with energy $d \omega/2$ for a harmonic oscillator in $d$ spatial dimensions ($\omega$ is the frequency of the oscillator).
The scaling dimension of the bosonic operator $\phi$ at the fixed point, after the quantum corrections, is $\Delta_{\phi}=2.$
The ground state of two particles at unitarity in a harmonic potential is described by the wavefunction
\beq
\phi(\vec{x}, \vec{y}) \sim \frac{e^{-\frac{x^2+y^2}{2}}}{|\vec{x}- \vec{y}|^{d-2}} \, ,
\eeq
whose energy is $2 \omega,$ thus matching with the prediction from the state/operator correspondence.
One can use Feynman diagrams to compute the scaling dimension of more complicated operators and then map the problem to the study of energy levels in a harmonic oscillator using the non-relativistic state/operator correspondence.
One finally extrapolates the results to $\varepsilon=1$ and compares with Monte-Carlo simulations, \eg see \cite{PhysRevLett.95.060401}.

When working in $d=2+\bar{\varepsilon}$ spatial dimensions, the Lagrangian is
\beq
\mathcal{L}_{2 + \bar{\varepsilon}} = i \psi^{\dagger}_{\sigma} \p_t \psi_{\sigma} - \frac{1}{2m} |\nabla \psi_{\sigma}|^2 + 
\bar{g}^2 \psi^{\dagger}_{\uparrow} \psi^{\dagger}_{\downarrow} \psi_{\downarrow} \psi_{\uparrow} \, .
\eeq
In this case, the fixed point describing fermions at unitarity is located at $\bar{g}^2 = 2 \pi \bar{\varepsilon}.$
Again, one can study the scaling dimensions of the fields perturbatively and finally extrapolate to $d=3$.
The analytic approach provides a good approximation of the numerical data \cite{Nishida:2007pj}.

\subsection{Quantum Hall effect}
\label{ssec:QHE}

One of the hardest problems in condensed matter physics is the fractional quantum Hall (FQH) effect in 2+1 dimensions, which consists in the quantization of the Hall conductivity due to the topological properties of the ground state in the presence of a magnetic field, \eg see the lecture notes \cite{Tong:2016kpv}.
While the integer quantum Hall (IQH) effect is described by the dynamics of free electrons moving in a uniform magnetic field in the presence of impurities, instead the FQH effect relies crucially on the interactions of particles with a single Landau level.
Imposing the constraint with the lowest Landau level makes the problem difficult to treat theoretically because several proposed models inevitably break the constraint at some stage. 
In this context, Son and collaborators developed an effective field theory coupled to NC geometry, which allows to derive currents and Ward identities covariantly, thus preserving the constraint \cite{Son:2013rqa,Geracie:2014nka}.

At scales much larger than the magnetic length, universal information about the Hall conductivity is encoded by a pure CS theory (this is called the \textit{hydrodynamic theory}) \cite{PhysRevB.46.2290}.
The hydrodynamic regime is described by the action \eqref{eq:nonrel_CS_action_sec3}, which enjoys Schr\"{o}dinger invariance at the critical point \eqref{eq:Jackiw_Pi_self_dual_condition}.
In this case, one can study the scaling dimensions of the fields using Feynman diagrams and then map to the energy levels in a harmonic potential, thanks to the state/operator correspondence described in section \ref{ssec:nonrel_state_OPE}.
For instance, the composite scalar operators $\varphi^N$  have scaling dimensions \cite{Nishida:2007pj}
\beq
\Delta_{\varphi^N} = N \pm \frac{N(N-1)}{2} \frac{1}{\pi \kappa} \, ,
\eeq
which match with the energy levels of the $N$--anyon states in a harmonic potential. 
A similar analysis can be performed when fermionic matter is coupled to the CS term.

However, it was argued in \cite{Geracie:2014nka} that transport properties related to the Hall viscosity require additional structure and are not encoded by the hydrodynamic theory. The appropriate microscopic action reads 
\beq
S = \int d^3 x \, \sqrt{-g} \, \left[ \frac{i}{2} v^{\mu} \le \psi^{\dagger} \nabla_{\mu} \psi - \nabla_{\mu} \psi^{\dagger} \psi \ri - \frac{1}{2m} \le h^{\mu\nu} + \frac{i \mathfrak{g}}{2} \epsilon^{\mu\nu} \ri \nabla_{\mu} \psi^{\dagger} \nabla_{\nu} \psi - \lambda |\psi|^4 \right] \, ,
\label{eq:action_Geracie_Son}
\eeq
where $v^{\mu}, h^{\mu\nu}$ are the velocity and inverse spatial metric of the NC geometry, $\psi$ a field with spin $s$, $\mathfrak{g}$ is the Landé factor for the coupling to a magnetic field, and $\epsilon^{\mu\nu}$ the two-dimensional Levi-Civita symbol.
The covariant derivative is defined by $\nabla_{\mu} = \p_{\mu} - i A_{\mu} + i s \omega_{\mu}$, where $A_{\mu}$ is a gauge field and $\omega_{\mu}$ the spin connection.
Eq.~\eqref{eq:action_Geracie_Son} provides a framework that preserves the constraint with the lowest Landau level, includes the information about the Hall viscosity, and allows to compute covariant Ward identities that follow from the spacetime symmetries in an arbitrary TNC background.

\subsection{Holographic dual of Schr\"{o}dinger symmetry}
\label{ssec:geometric_AdS_Schroed}

A major role in the development of theoretical physics in the last decades was played by AdS/CFT duality, which relates a relativistic CFT in $d$ dimensions to a gravitational theory in $(d+1)$--dimensional AdS spacetime \cite{Maldacena:1997re}.
Given our focus on Schr\"{o}dinger-invariant QFTs, it is natural to ask whether there is a non-relativistic bulk realization of holography.
The idea is to perform a deformation of AdS spacetime which reduces its isometries from the conformal group down to Schr\"{o}dinger symmetry. This leads to the $(d+3)$--dimensional metric \cite{Son:2008ye}
\beq
ds^2 = - \frac{2 (dx^+)^2}{z^4} + \frac{-2 dx^+ dx^- + dx_i dx^i + dz^2}{z^2} \, ,
\label{eq:Son_AdS_metric}
\eeq
where $(x^-,x^+,x^i)$ are light-cone coordinates in $(d+2)$--dimensional Minkowski space, and $z$ is a radial direction.
The first term breaks the isometries of the Poincaré AdS geometry encoded by the second term down to the Schr\"{o}dinger group.
In order to identify $P^+=M$ as in eq.~\eqref{eq:map_generators_Son}, we need to assume that the null direction $x^-$ is compactified.
This allows us to get a discrete mass spectrum, as it happens in non-relativistic theories.
This background is a special case of a more general class proposed in \cite{Balasubramanian:2008dm}, which depends on the critical exponent defined in eq.~\eqref{eq:Lifhsitz_transformations} and admits Galilean and scale invariance.
When the dynamical exponent is 2, the metric \eqref{eq:Son_AdS_metric} is recovered and the symmetry group is enhanced to Schr\"{o}dinger.

Next, let us study the relation between mass and dimension of the dual fields in this putative non-relativistic AdS/CFT correspondence, in analogy to the relativistic case \cite{Klebanov:1999tb,Ammon:2015wua}.
Consider a massive scalar field minimally coupled to gravity
\beq
S= - \int d^{d+3} x \, \sqrt{-g}\, \le g^{\mu\nu} \p_{\mu} \Phi^{\dagger} \p_{\nu} \Phi + m_0^2 \Phi^{\dagger} \Phi \ri \, ,
\eeq
where $g_{\mu\nu}$ is the metric \eqref{eq:Son_AdS_metric}.
The scalar is supposedly dual to a scalar operator $\mathcal{O}$ defined on the boundary QFT.
We perform a null reduction along $(d+2)$--dimensional Minkowski space by assuming that the dimensionless coordinate $x^- \sim x^- + 2 \pi$ is periodic, identifying $t \equiv x^+$ and requiring that the scalar satisfies the ansatz \eqref{eq:ansatz_fields_nullred_sec4}. The action becomes
\beq
S= 2\pi \int d^{d+1} x \, dz \, \frac{1}{z^{d+3}} \le 2 i M z^2 \varphi^{\dagger} \p_t \varphi - z^2 \p_i \varphi^{\dagger} \p_i \varphi - m^2 \varphi^{\dagger} \varphi \ri \, , 
\eeq
where $m^2 = m_0^2 = 2 M^2$.
As it happens in standard AdS/CFT, the EOM admit two independent solutions.
They correspond to different boundary conditions: either the scalar is a source for $\mathcal{O}$ in the boundary theory, or a vacuum expectation value of such operator.
The two possibilities correspond to the following scaling dimension for the dual operator
\beq
\Delta_{\mathcal{O}} = \frac{d}{2} + 1 \pm \nu \, , \qquad
\nu \equiv \sqrt{\le \frac{d}{2} +1\ri^2 + m^2} \, .
\label{eq:scaling_dimensions_nonrel_AdSCFT}
\eeq
One can check that the smallest possible case is $\Delta_{\mathcal{O}} \geq d/2$, which matches with the expectations from the state/operator correspondence discussed in section \ref{ssec:nonrel_state_OPE} \cite{Nishida:2007pj,Son:2008ye}.
This is equivalent to the statement that the total energy in a system with harmonic potential is larger than the zero-point energy of the motion in the center of mass.
The existence of two Schr\"{o}dinger-invariant QFTs with the same operator $\mathcal{O}$ and different scaling dimensions fits well with the discussion in section \ref{ssec:fermions_unitarity}.
In fact, the operator $\psi_{\downarrow} \psi_{\uparrow}$ has dimension $d$ in the free theory, and dimension 2 at the fixed point where fermions are at unitarity; these two values form a pair of solutions according to eq.~\eqref{eq:scaling_dimensions_nonrel_AdSCFT}.
It is worth noticing that despite various similarities with the relativistic case, the gravitational background dual to Schr\"{o}dinger-invariant theories proposed in this subsection does not have an analog of the Breitenholer-Freedman bound \cite{Moroz:2009kv}.

The above discussion focused on the study of geometric duals to Schr\"{o}dinger symmetry in Poincaré coordinates.
Even in the relativistic case, it is sometimes important to formulate AdS/CFT correspondence in global coordinates, \eg see \cite{Balasubramanian:1998sn,Balasubramanian:1998de,Marolf:2004fy}.
The formulation in global coordinates of geometries with non-Lorentzian symmetry at the boundary, including the case in  eq.~\eqref{eq:Son_AdS_metric}, was developed in \cite{Blau:2009gd}. 
Finally, let us mention that non-relativistic holography also appeared in the context of Horava-Lifshitz gravity, which is a renormalizable theory proposed as a UV completion of general relativity with anisotropic scaling between space and time \cite{Horava:2009uw}.
In this context, one can also find NC geometry at the boundary of Lifshitz holography \cite{Christensen:2013lma,Hartong:2014pma,Hartong:2015zia}.

\subsection{Non-relativistic fermions}
\label{ssec:nonrel_fermions}

We will provide an action formulation for a non-relativistic spin--$\frac{1}{2}$ fermion, which is linear in the derivatives and reproduces the EOM derived by Levy-Leblond \cite{LvyLeblond1967NonrelativisticPA} (reviewed in section \ref{ssec:nonrel_Dirac}).
We perform the null reduction of the Dirac action in 3+1 dimensions \cite{Auzzi:2017jry}.\footnote{For an approach based on representation theory, see \cite{de2006galilei,Geracie:2016bkg}. For a limiting procedure $c \rightarrow \infty$, see \cite{Fuini:2015yva}.  } 
In the massless case, the Weyl spinors composing the four-dimensional Dirac spinor $\Psi=(\Psi_{L}, \Psi_R)$ decouple and we can restrict our analysis to the left-handed part
\beq
S_{L}=\int d^4 x \sqrt{-g} \,  \mathcal{L}_L = \int d^4 x \sqrt{-g} \,  i \, \Psi_L^{\dagger} \bar{\sigma}^A \nabla_A \Psi_L  \, , \quad
\nabla_{M} \Psi_L =\left( \p_{M} + \frac{1}{4} \omega_{MAB} \sigma^{AB} \right) \Psi_L \, ,
\label{eq:Weyl_action}
\eeq
where $\sigma^A$ are the covariant Pauli matrices, $\omega_{MAB}$ is the torsionless spin connection and $\sigma^{AB}= \frac{1}{2} [\sigma^A, \sigma^B]$ are the Lorentz generators.
To perform the null reduction, we consider the background \eqref{eq:nullred_metric_sec4} with null isometry and we express the covariant Pauli matrices in light-cone coordinates $\sigma^{\pm} = \frac{1}{\sqrt{2}} (\sigma^3 \pm \sigma^0)$.
Finally, we decompose the Weyl fermion as
\beq
\Psi_L (x^M)=\begin{pmatrix}
\xi (x^{\mu}) \\ \chi (x^{\mu})
\end{pmatrix}  e^{im u}\, ,
\label{eq:ansatz_fermions_nullred}
\eeq
where $\xi, \chi$ are complex Grassmann fields that encode the spin representation $l=1/2$ of the Bargmann group in eq.~\eqref{eq:ansatz_fields_nullred_sec4}.
Plugging the previous ansatz inside the Weyl Lagrangian in eq.~\eqref{eq:Weyl_action}, after several manipulations one gets
\begin{subequations}
\beq
\mathcal{L}_L =
 \begin{pmatrix}  \xi^{\dagger} & \chi^{\dagger}  \end{pmatrix} 
  \begin{pmatrix} A  & B \\ C & D \end{pmatrix} 
 \begin{pmatrix}  \xi \\ \chi  \end{pmatrix} \, , 
 \label{eq:Lagrangian_ferm2}
\eeq
\beq
\begin{aligned}
A &= -\sqrt{2}\le m+\frac14 \tau_{\mu \nu} e^\mu_1 e^\nu_2\ri\, , & \\
B &= (e_1^\mu-i e^\mu_2) \le i \tilde{D}_\mu - \frac{i}{4} \tau_{\mu \nu} v^\nu \ri \, ,
 \quad
C =(e_1^\mu+i e^\mu_2) \le i \tilde{D}_\mu -i \frac{ 3}{4} \tau_{\mu \nu} v^\nu \ri \,  , & \\
D &= \sqrt{2} \left[  v^\mu \le i \tilde{D}_\mu 
+\frac{i}{4} h^{\rho \sigma} \p_\mu h_{\rho \sigma} \ri
-\frac{i}{2} \le v^\mu v^\nu \p_\mu \tau_\nu - \p_\mu v^\mu \ri
+ \frac14 F_{\mu \nu} \,  e^\mu_1 e^\nu_2  
\right] \, . &
\end{aligned}
\eeq
\end{subequations}
where we defined the covariant derivatives and the field strengths as
\begin{subequations}
\beq
\tilde{D}_\mu \xi \equiv 
\le \p_\mu + \frac{i}{2} \tilde{\omega}_{\mu 1 2} 
+ i m \, m_\mu \ri \xi  \, ,  \qquad
\tilde{D}_\mu \chi \equiv
\le \p_\mu -\frac{i}{2} \tilde{\omega}_{\mu 1 2} 
+ i m \, m_\mu \ri \chi  \, ,  
\eeq
\beq
\tilde{\omega}_{\mu  a b}  = \frac{1}{2}
\left[ e^{\nu}_{\,\,\,a} \left(  \p_{\mu} e^{b}_{\,\,\, \nu} -  \p_{\nu} e^{b}_{\,\,\, \mu}  \right)  -   e^{\nu}_{\,\,\, b} \left(  \p_{\mu} e^{a}_{\,\,\, \nu} -  \p_{\nu} e^{a}_{\,\,\, \mu}   \right)  
 -e^{\nu}_{\,\,\,a} e^{\rho}_{\,\,\, b}    e^c_{\,\,\, \mu} \left( \p_{\nu} e^{c}_{\,\,\, \rho}-
   \p_{\rho} e^{c}_{\,\,\, \nu}  \right) \right] \, ,
\eeq
\beq
F_{\mu\nu} \equiv \p_{\mu} m_{\nu} - \p_{\nu} m_{\nu} \, , \qquad
\tau_{\mu\nu} \equiv \p_{\mu} \tau_{\nu} - \p_{\nu} \tau_{\nu} \, .
\eeq
\end{subequations}
Notice that the previous objects are torsionful despite they arise from the null reduction of the torsionless connection on the parent manifold.
The reason is that they do not correspond to a specific component of the higher-dimensional connection, but are built by combining various terms arising during the reduction.

The Lagrangian~\eqref{eq:Lagrangian_ferm2} is linear in all the derivatives and provides the general action for a Schr\"{o}dinger-invariant spin--$\frac{1}{2}$ fermion coupled to type I NC geometry.
The field $\xi$ is auxiliary and can be integrated out using the Euler-Lagrange equations  
\beq
\xi=\frac{i(e^\mu_1-i e^\mu_2)\le \tilde{D}_\mu+\frac14 v^\nu \tau_{\mu \nu}\ri \chi}
{\sqrt{2} \le m+\frac{\tau_{\mu \nu} e_1^\mu e_2^\nu}{4}\ri} \, .
\label{eq:EOM_chi} 
\eeq
Replacing it into eq.~\eqref{eq:Lagrangian_ferm2},
we could obtain a cumbersome Lagrangian written only in terms of the dynamical field $\chi$.

One can get a better understanding of the theory by restricting to flat space \eqref{eq:flat_NC_space} with vanishing gauge potential $m_{\mu}=0$.
The left-handed Weyl action becomes
\beq
\begin{aligned}
\mathcal{L}_L  =  - \sqrt{2} m \xi^{\dagger} \xi 
- \sqrt{2}i \chi^{\dagger} \p_{t} \chi
 +  i \chi^{\dagger} (\p_1 + i \p_2) \xi
+  i \xi^{\dagger} (\p_1 - i \p_2) \chi  \, ,
\end{aligned} 
\eeq
which is the configuration space version of the Levy-Leblond equations  \eqref{eq:nonrel_Dirac_sec3}, which was written in momentum space.
After using the EOM for the auxiliary field $\xi$, one finds a Schr\"odinger action for the dynamical field $\chi$
\beq
S_L = \int d^3 x \, \le i \chi^{\dagger} \p_t \chi - \frac{1}{2m} |\p_i \chi|^2 \ri \, .
\eeq

\subsection{Galilean electrodynamics}
\label{ssec:GED}

In this section, we will derive an action for GED, which comprises and extends the electric and magnetic limits of electromagnetism discussed in section \ref{ssec:Galilean_em}.
Besides providing an off-shell formulation to the problem in flat space, we will also extend the formalism to the case of an arbitrary manifold \cite{Festuccia:2016caf}.
Finally, we will define a quantum version of the theory and describe its renormalization properties \cite{Chapman:2020vtn}.
While we will not focus on holography in this section, let us mention for the interested reader that GED also arises as the linearized action on D--branes in non-relativistic open string theory \cite{Gomis:2020fui,Ebert:2021mfu}.

\paragraph{GED from a limit.}
Let us recover the electric and magnetic limits of electromagnetism from a different perspective compared to section \ref{ssec:Galilean_em}, \ie by working on the gauge transformations and the EOM of a relativistic $\mathrm{U}(1)$ gauge connection $A_{\mu}$. 
After restoring the factors in the speed of light, the gauge transformations read
\beq
A'_t = A_t + \frac{1}{c} \p_t \Lambda \, , \qquad
A'_i = A_i + \p_i \Lambda \, ,
\eeq
while the EOM without sources $\p_{\mu} F^{\mu\nu} = 0$ are given by
\beq
\p_i \le  \p_i A_t - \frac{1}{c} \p_t A_i \ri = 0 \, , \qquad 
\frac{1}{c} \p_t \le \p_i A_t - \frac{1}{c} \p_t A_i \ri + \p_j F_{ji} = 0 \, ,
\eeq
where $F_{ij} = 2 \p_{[i} A_{j]}$ is the usual field strength.
We can perform the following limits:
\begin{itemize}
\item \textbf{Timelike (electric) limit.}
\beq
A_t = - a_t \, , \qquad
A_i = \frac{1}{c} a_i \, , \qquad
\Lambda = \frac{1}{c} \lambda \, , \qquad
c \rightarrow \infty \,\, \mathrm{with} \,\, 
a_0, a_i, \lambda \,\, \mathrm{fixed} \, .
\eeq
Defining $f_{ij} = 2 \p_{[i} a_{j]}$, the EOM reduce to
\beq
\p^i \p_i a_t = 0 \, , \qquad
-\p_t \p_i a_t + \p^j f_{ji} = 0 \, , 
\label{eq:electric_EOM_sec5}
\eeq
which match with eq.~\eqref{eq:electric_limit_EOM_Levy} in the sourceless case after adding the Bianchi identities.
The gauge transformations reduce to $\delta_{\lambda} a_t=0$ and $\delta_{\lambda} a_i= \p_i \lambda$.
\item \textbf{Spacelike (magnetic) limit.}
\beq
A_t = - a_t \, , \qquad
A_i = c a_i \, , \qquad
\Lambda = c \lambda \, , \qquad
c \rightarrow \infty \,\, \mathrm{with} \,\, 
a_0, a_i, \lambda \,\, \mathrm{fixed} \, .
\eeq
This leads to the EOM (here $E_i= \p_t a_i - \p_i a_t $ is the electric field)
\beq
\p_i E^i = 0 \, , \qquad
\p^j f_{ji} = 0 \, , 
\label{eq:magnetic_EOM_sec5}
\eeq
which match with eq.~\eqref{eq:magnetic_limit_EOM_Levy}, together with the limit on the Bianchi identity.
The gauge transformations become $\delta_{\lambda} a_t=-\p_t \lambda$ and $\delta_{\lambda} a_i= \p_i \lambda$, leaving the electric field invariant.
\end{itemize}
The previous two limits can be recovered from GED, which arises from a non-relativistic limit of the Maxwell action minimally coupled to a real scalar field $\phi$.
After taking
\beq
\phi = c \varphi \, , \qquad
A_t = - c \varphi - \frac{1}{c} a_t \, , \qquad
A_i =  a_i \, , \qquad
\Lambda = c \lambda \, , \qquad
c \rightarrow \infty \,\, \mathrm{with} \,\, 
\varphi, a_0, a_i \,\, \mathrm{fixed} \, ,
\eeq
one gets the action
\beq
S_{\rm GED} = \int d^{d+1} x \, \le -\frac{1}{4} f^{ij} f_{ij} - E^i \p_i \varphi + \frac{1}{2} (\p_i \varphi)^2 \ri \, .
\label{eq:GED_action_sec5}
\eeq
The non-relativistic electrodynamics has no propagating degrees of freedom and the mediation is instantaneous.\footnote{This can be seen, for instance, by the fact that there are gauge choices such that the propagator for the gauge field does not depend on the energy $\omega$ conjugate to the time coordinate.  }
The scalar $\varphi$ is inert under gauge transformations and is non-dynamical, but plays an essential role in building an off-shell formulation.
The EOM are given by eq.~\eqref{eq:electric_EOM_sec5}, complemented with
\beq
\p_t^2 \varphi - \p_i E_i = 0 \, , 
\eeq
that arise from the variation of $\varphi$.
Therefore, one can just recover the electric limit as a subcase of the action \eqref{eq:GED_action_sec5}.\footnote{However, the electric limit and GED are different theories, because the latter contains additional degrees of freedom and symmetries. }
The magnetic limit is obtained by setting $\varphi=0$, which is another consistent subcase leading to eq.~\eqref{eq:magnetic_EOM_sec5}.

The on-shell group of symmetries of the three limits of electromagnetism is large.
In fact, the electric and magnetic theories are invariant under the infinite-dimensional Galilean conformal algebra \cite{Bagchi:2009my,Martelli:2009uc}, since the EOM are invariant under time-dependent translations and spatial dilatations.
GED is also on-shell invariant under an infinite-dimensional group because time-dependent translations are still allowed.
The set of symmetries encoded by the off-shell action \eqref{eq:GED_action_sec5} is much smaller, in fact, it only consists of the following generators \cite{Festuccia:2016caf}
\beq
\begin{aligned}
& H = \p_t \, , \qquad  P_i = \p_i \, , \qquad
J_{ij} = x_i \p_j - x_j \p_i \, , & \\
& G_i = t \p_i \, , \qquad
D_1 = t \p_t \, , \qquad
D_2 = x^i \p_i \, .
\end{aligned}
\eeq
The remarkable feature is that there are two independent dilatations (commuting with each other), implying that GED is a Galilean field theory that has a Lifshitz invariance \eqref{eq:Lifhsitz_transformations} for any value of the dynamical exponent $z$.
This means that GED can be coupled to matter theories with any choice of $z$ without breaking the scaling symmetry.
In the following, we will be mainly interested in the coupling with a Schr\"{o}dinger scalar, see eq.~\eqref{eq:scalar_Schrodinger_term} below.

\paragraph{GED from null reduction.}
GED can be derived using null reduction.
This procedure does not only provide a simple way to check the consistency of the previous construction, but straightforwardly allows to generalize the action \eqref{eq:GED_action_sec5} to curved space.
One starts from the Maxwell's action on a $(d+2)$--dimensional Lorentzian manifold 
\beq
S_{\rm Maxw} = - \frac{1}{4} \int d^{d+2} x\, \sqrt{-g} \, F_{MN} F^{MN} \, ,
\eeq
with metric \eqref{eq:nullred_metric_sec4} and the following decomposition for the gauge field
\beq
A_M (x^N) = \le \varphi(x^{\nu}) , a_{\mu} (x^{\nu}) \ri \, .
\label{eq:ansatz_nullred_GED}
\eeq
Notice that this ansatz fits the general form \eqref{eq:ansatz_fields_nullred_sec4} for spin $l=1$ in the massless case ($m=0$), as one expects for the gauge field mediating the interaction.
After null reduction, the GED action in curved space reads
\beq
S_{\rm GED} = \int d^{d+1} x \, e \, \left[ - \frac{1}{4} h^{\mu \rho} h^{\nu \sigma} f_{\mu\nu} f_{\rho \sigma} - h^{\mu\nu} \hat{v}^{\rho} f_{\rho\nu} \p_{\mu} \varphi - \hat{\Phi} h^{\mu\nu} \p_{\mu} \varphi \p_{\nu} \varphi + \frac12 (\hat{v}^{\mu} \p_{\mu} \varphi)^2 \right] \, ,
\label{eq:curved_GED}
\eeq
where we used the boost-invariant NC data \eqref{eq:boost_invariant_NCdata}.
This expression reduces to \eqref{eq:GED_action_sec5} in flat space, \ie when the conditions \eqref{eq:flat_NC_space} and $m_{\mu}=0$ are imposed.

\paragraph{Renormalization of scalar GED.}
One can add propagating modes to GED (see comment below eq.~\eqref{eq:GED_action_sec5}) by coupling the system to matter fields, \emph{e.g.}, to a Schr\"{o}dinger scalar $\phi.$
The minimal coupling term reads
\beq
S_{\rm min} = \int d^3 x \, \left[ \frac{i}{2} \le \bar{\phi} D_t \phi -  \phi D_t \bar{\phi} \ri - \frac{1}{2  \mathcal{M}} D_i \bar{\phi} D^i \phi  \right] \, ,
\label{eq:scalar_Schrodinger_term}
\eeq
where the covariant derivatives acting on $\phi$ ($\bar{\phi}$) are defined by 
\beq
D_t \equiv \p_t \mp i e A_t \, , \qquad
D_i \equiv \p_i \mp i e A_i 
\label{eq:standard_cov_div}
\eeq
and $ \mathcal{M} \equiv m - e \varphi$ is a covariant mass.
This object can be constructed because the field $\varphi$ (associated with the null direction of the parent gauge connection) is a scalar under gauge transformations, and 
it is dimensionless.
The sum of the actions \eqref{eq:GED_action_sec5} and \eqref{eq:scalar_Schrodinger_term} gives the minimal realization of scalar Galilean Electrodynamics, $S_{\rm sGED} \equiv S_{\rm GED} + S_{\rm min}$. 

The covariant mass $\mathcal{M}$ is responsible for making the theory a $\sigma$--model, since it is necessary to series-expand the kinetic term to build standard Feynman rules and compute loop corrections.
It turns out that such a theory is non-renormalizable, since an infinite number of marginal deformations, built with the dimensionless scalar $\varphi$, are produced at quantum level.
A renormalizable scalar GED action $S_{\rm sGED} + \Delta S_{\rm sGED}$ is given by \cite{Chapman:2020vtn}
\beq
\label{eq:action_extra_shira}
\Delta S_{\rm sGED} = \int dt \, d^2 x \le \mathcal{J} [\mathcal{M}] \, \p^i \mathcal{M} \p_i \mathcal{M} \, \bar{\phi} \phi - \frac{\l}{4} \mathcal{V} [\mathcal{M}] \, (\bar{\phi} \phi)^2 - \mathcal{E}[\mathcal{M}] \, (\p^i \p_i \mathcal{M} - e^2 \bar{\phi} \phi) \, \bar{\phi} \phi \ri \, .
\eeq
While there is a non-renormalization theorem (based on selection rule \ref{sel_rule1}) which protects the electric charge $e$ from running, the functionals $\mathcal{J}[\mathcal{M}], \mathcal{V}[\mathcal{M}]$ and $\mathcal{E}[\mathcal{M}]$ renormalize non-trivially \cite{Chapman:2020vtn}.
The freedom governed by these functionals allows us to find conformal manifolds of fixed points where the Schr\"{o}dinger symmetry is preserved.
The appearance of conformal manifolds is surprising because they are ubiquitous in supersymmetric examples \cite{Leigh:1995ep,Green:2010da}, but very rare in other cases.
Here we see that giving up Lorentz symmetry, in favour of the causal structure of the Galilean group, is another way to generate a conformal manifold.
In section \ref{ssec:SGED} we will show that, quite surprisingly, supersymmetry does not significantly improve the renormalization properties observed in the GED case.

\subsection{Non-relativistic field theories with SU$(1,n)$ symmetry}
\label{ssec:Lambert_QFT}

Null reduction is a powerful tool to generate Schr\"{o}dinger-invariant field theories, but its applications are broader.
Following \cite{Lambert:2021nol}, in this section we will reduce a Lorentzian CFT along a null direction in a conformally compactified $2n$--dimensional Minkowski spacetime, resulting in a deformation of the lower-dimensional symmetries away from Schr\"{o}dinger group.
We will show that the Killing vectors span the algebra $\mathrm{SU}(1,n)$, then we will find a corresponding state/operator map and build QFTs invariant under these symmetries.
Other than providing a novel class of non-relativistic theories, these models received attention in the context of non-Lorentzian $M$--theory, where SUSY-invariant non-Abelian gauge theories with SU(1,3) invariance emerge \cite{Lambert:2019jwi}.

\paragraph{Conformal compactification.}
Let us start from $2n$--dimensional Minkowski spacetime in light-cone coordinates $\hat{x}^M = (\hat{x}^+, \hat{x}^-, \hat{x}^i)$ with $i=\lbrace 1, \dots , 2n-2 \rbrace$:
\begin{equation}
ds^2_M = \hat{\eta}_{MN} d\hat{x}^M \hat{x}^N  = - 2 d\hat{x}^+ d\hat{x}^- + d\hat{x}_i d\hat{x}^i \, .
\end{equation} 
We perform the following coordinate transformation\footnote{Notice that we swapped all the $\pm$ indices with respect to reference \cite{Lambert:2021nol}, to facilitate the comparison with the notation for null reduction used in section \ref{ssec:null_red}. }  
\beq
\begin{aligned}
& \hat{x}^- = 2 R \tan \left( \frac{x^-}{2 R} \right) \, ,  & \\
& \hat{x}^+ = x^+ + \frac{1}{4 R} x^i x^i \tan \left( \frac{x^-}{4 R} \right) \, ,  & \\
& \hat{x}^i = x^i - R \, \Omega_{ij} x^j \tan \left( \frac{x^-}{2 R} \right)   \, ,  &
\end{aligned}
\label{eq:coord_change_Lambert}
\eeq
where $R$ is a length scale which makes the arguments of the trigonometric functions dimensionless, and $\Omega_{ij}$ is a constant antisymmetric matrix satisfying 
\beq
\Omega_{ij} \Omega_{jk} = - \frac{1}{R^2} \delta_{ik} \, .
\label{eq:norm_Omega_Lambert}
\eeq
We then perform an additional Weyl transformation $ds^2_{\Omega} = \cos^2 \le x^-/2R \ri ds^2_M$ that leads to the line element
\beq
ds^2_{\Omega} = g_{MN} dx^M dx^N = - 2 dx^- \le dx^+ + \frac{1}{2} \Omega_{ij} x^j dx^i \ri + dx_i dx^i \, .
\label{eq:Lambert_metric}
\eeq
Notice that the transformations \eqref{eq:coord_change_Lambert} made the range of $x^- \in [\pi R, \pi R]$ finite. 
We can thus perform for a KK reduction along this direction of $2n$--dimensional Minkowski space.
While this procedure is similar to the null reduction discussed in section \ref{ssec:null_red}, here there is an additional step given by the conformal rescaling of the original spacetime, which is ultimately related to the deformation $\Omega_{ij}$.
We refer to this new mechanism as a \textit{deformed null reduction}. 
One recovers the standard null reduction in Minkowski space when $R \rightarrow \infty,$ which implies $\Omega_{ij} \rightarrow 0$ from eq.~\eqref{eq:norm_Omega_Lambert}.
One can show that the metric \eqref{eq:Lambert_metric} can be realized as the conformal boundary of AdS$_{2n+1}$ by selecting a particular slicing, thus providing a holographic interpretation for this construction.

\paragraph{Symmetries of the deformed null reduction.}
Next, we identify the symmetry group preserved by the deformed null reduction described above.
A representation of the generators of the conformal group $\mathrm{SO}(2n,2)$ in terms of differential operators is given by
\beq
\begin{aligned}
& \mathcal{P}_M = -i \hat{\p}_M \, , \qquad
\mathcal{J}_{MN} = -i \le \hat{x}_M \hat{\p}_N - \hat{x}_N \hat{\p}_M  \ri  \, , & \\
& \mathcal{D} = -i \hat{x}^M \hat{\p}_M \, , \qquad
\mathcal{K}_M = -i \le \hat{x}_N \hat{x}^N \hat{\p}_M - 2 \hat{x}_M  \hat{x}^N \hat{\p}_N  \ri \, . &
\end{aligned}
\eeq
These conformal Killing vectors are preserved under the coordinate transformation \eqref{eq:coord_change_Lambert} followed by the Weyl rescaling that brings to the metric \eqref{eq:Lambert_metric}.
Performing the KK reduction along the null direction $x^-$ restricts to a set of generators which span a subgroup $H \subset \mathrm{SO}(2n,2)$, corresponding to the centralizer of $P_-$.
A possible basis for these generators is
\beq
\begin{aligned}
& M = \mathcal{P}_- + \frac{1}{4} \Omega_{ij} \mathcal{J}_{ij} + \frac{1}{8 R^2} \mathcal{K}_+ \, , \qquad
B = \frac{R}{2} \Omega_{ij} \mathcal{J}_{ij} \, , \qquad
J^{\alpha} = \frac{1}{2} L^{\alpha}_{ij} \mathcal{J}_{ij} \, , \qquad H = \mathcal{P}_+ \, , 
 & \\
&  P_i = \mathcal{P}_i + \frac{1}{2} \Omega_{ij} \mathcal{J}_{j+} \, ,  \qquad
G_i = \mathcal{J}_{i-} - \frac{1}{4} \Omega_{ij} \mathcal{K}_j \, , \qquad
 D = \mathcal{D} - \mathcal{J}_{-+} \, , \qquad
C = \frac{1}{2} \mathcal{K}_- \, , &
\end{aligned}
\label{eq:generators_map_Lambert}
\eeq
where $\alpha \in \lbrace 1, \dots , n(n-2) \rbrace $ (valid when $n \geq 3$) and $L^{\alpha}_{ij}$ are constant matrices.
The generators $\lbrace M, H, P_i, G_i, D, C \rbrace$ comprise a deformed version of the Schr\"{o}dinger case (see section \ref{ssec:Schroed_group}), while $\lbrace B, J^{\alpha} \rbrace$ form the rotation subgroup $\mathrm{U}(1) \times \mathrm{SU}(n-1)$, which is only non-trivial when $n \geq 3$.
All together, the full set of generators \eqref{eq:generators_map_Lambert} form the $\mathrm{SU}(1,n)$ group, times a $\mathrm{U}(1)$ factor corresponding to the centralizer $P_-$.

Since the limit $R \rightarrow \infty$ makes the coordinate transformations \eqref{eq:coord_change_Lambert} trivial,  one expects to recover the usual null reduction in Minkowski space, whose set of conformal Killing vectors commuting with $P_-$ generate the Schr\"{o}dinger group.
Indeed, the Schr\"{o}dinger group is obtained by taking the limit $R \rightarrow \infty$ inside eq.~\eqref{eq:generators_map_Lambert}, and then adding back by hand the subset of rotations inside $\mathrm{SO}(2n-2)$ which are broken by the deformation $\Omega_{ij}$ at finite $R$.\footnote{One does not immediately obtain the Schr\"{o}dinger group when $R \rightarrow \infty$, because there are certain elements of the conformal group $\mathrm{SO}(2n,2)$ which only commute with the null momentum $P_-$ after taking the limit. }

\paragraph{State/operator map.}
We show that the state/operator correspondence described in section \ref{ssec:nonrel_state_OPE} is not spoilt by the $\Omega$--deformation.
One can still define local operators using the stabilizer of the origin $(x^+, x^i)=(0, \mathbf{0})$ inside the algebra $\mathfrak{su}(1,n)\oplus \mathfrak{u}(1)$.
Any other point in the spacetime can be reached by using eq.~\eqref{eq:local_operator_space}, where now the deformed Hamiltonian $H$ and momentum generators $P_i$ are taken from eq.~\eqref{eq:generators_map_Lambert}.
Given a local operator $\mathcal{O}(0),$ we associate the state $\ket{\mathcal{O}} \equiv \mathcal{O}(0) \ket{0}$.  
We declare that an operator has scaling dimension $\Delta_{\mathcal{O}}$ if 
\beq
[D, \mathcal{O}(0)] = i \Delta_{\mathcal{O}} \mathcal{O}(0)   \quad \Rightarrow \quad  D \ket{\mathcal{O}} = i \Delta \ket{\mathcal{O}} \, .
\eeq
Even in the presence of the deformation $\Omega$, the commutation rules of the algebra $\mathfrak{su}(1,n)$ identify $\lbrace H, P_i \rbrace $ as raising operators, and $\lbrace C, G_i \rbrace $ as lowering operators.
Therefore, the lowest weight (which we call primary) of a representation still satisfies the constraints \eqref{eq:primary_Schroed}.
At this point, we consider the following similarity transformation on the Hilbert space and on the space of operators:
\beq
\ket{\bar{\mathcal{O}}} = e^{- H} e^{\frac{1}{2} C} \ket{\mathcal{O}} \, , \qquad
\bar{\mathcal{O}}(0) = e^{-H} e^{\frac{1}{2} C} \mathcal{O}(0) e^{-\frac{1}{2} C} e^{H} \, .
\eeq
If $\mathcal{O}(0)$ is a primary, then we find
\beq
\ket{\bar{\mathcal{O}}} = e^{-H} e^{\frac{1}{2} C} \mathcal{O}(0) \ket{0} = e^{-H} \mathcal{O}(0) \ket{0} \, ,
\eeq
which coincides with the state in eq.~\eqref{eq:state_Nishida}.
Following the same arguments outlined in section \ref{ssec:nonrel_state_OPE}, one concludes that the non-relativistic state/operator correspondence maps the dilatation operator to an oscillator Hamiltonian, even when $\Omega_{ij} \ne 0$.
Of course, in the limit $R \rightarrow \infty$ one trivially recovers the map defined for the Schr\"{o}dinger case.

\paragraph{Field theories with SU$(1,n)$ symmetry.}
We define QFTs invariant under the $\mathrm{SU}(1,n)$ symmetry group of the deformed null reduction.
Generally speaking, on Minkowski space with coordinates $x^M$ and compact direction $x^- \in [-\pi R, \pi R]$ we can perform a Fourier expansion
\beq
\mathcal{O} (x^M) = \sum_{k} e^{i \frac{k}{R} x^-} \mathcal{O}_k (x^{\mu}) \, ,
\eeq
where $k$ can be either integer or half-integer depending on the boundary conditions.
In the standard null reduction, one only retains a fixed momentum $P_-=m$ as in eq.~\eqref{eq:ansatz_fields_nullred_sec4}, but here we will momentarily keep the full set of KK modes.
In the deformed null reduction, the coordinate transformation 
\eqref{eq:coord_change_Lambert} is followed by a Weyl rescaling, under which a scalar field with dimension $\Delta$ changes as $ \hat{\mathcal{O}} (x^M) =  \left[ \cos \le x^-/2R \ri \right]^{\Delta} \mathcal{O}(x^M) $.
Therefore, the natural ansatz for a scalar field undergoing the deformed null reduction reads
\beq
\hat{\Phi}(x^M) = \left[\cos \le \frac{x^-}{2R} \ri \right]^{\Delta} \, \sum_k e^{i \frac{k}{R} x^-} \varphi_k (x^{\mu}) \, .
\label{eq:ansatz_similLambert_nullredscalar}
\eeq
Let us apply the above technology to a relativistic free scalar field in $2n$ dimensions
\beq
S = -  \int d^{n-2}x dx^- dx^+  \, \sqrt{-g} \, 
g^{MN} \partial_M \hat{\Phi}^{\dagger} \partial_N \hat{\Phi}  \, .
\label{eq:free_scalar_sec5} 
\eeq
After plugging the metric \eqref{eq:Lambert_metric} and the ansatz \eqref{eq:ansatz_similLambert_nullredscalar}, one finds
\beq
\begin{aligned}
S   =  \pi R \sum_{k} \int d^{2n-2}x  dx^+  \, & \left[
 2 i \frac{k}{R} \varphi^{\dagger}_k (\p_{+} \varphi_k)  
 +  (\p_{i} \varphi^{\dagger}_k) (\p_{i} \varphi_k)  \right. \\
& \left. - 2 \Omega_{ik} x^k (\p_{i} \varphi^{\dagger}_k) (\p_{+} \varphi_k) 
+ \frac{|x|^2}{4 R^2}  (\p_{+} \varphi^{\dagger}_k) (\p_{+} \varphi_k) 
    \right]  \, .
    \label{eq:nullred_freescalar_ope2}
 \end{aligned}
\eeq
The action realizes $\mathrm{SU}(1,n)$ spacetime symmetry for each KK mode $k$ independently.

There are some possible generalizations of this model.
First, one can perform a similar deformed null reduction for fermions and gauge fields, too \cite{Lambert:2021nol,Smith:2023jjb}.
In this regard, one can find QFTs coming with SU$(1,n)$ symmetry coming from null reduction in five-dimensional supersymmetric cases \cite{Lambert:2019jwi,Lambert:2019fne,Lambert:2020jjm,Lambert:2020zdc}. 
Secondly, a novelty of this approach compared to the usual null reduction is that by keeping all the KK modes, one can go back and uplift the non-relativistic model to a Lorentzian theory in one higher dimension (as claimed in \cite{Lambert:2021nol}).
Finally, notice that the metric \eqref{eq:Lambert_metric} arises from a conformal compactification of Minkowski space, therefore it seems to provide a description of the operator picture.
It would be interesting to apply the previous technology to the state picture and find the mapping between the corresponding geometries, in the same spirit as the relativistic case \cite{progress}.

\subsection{Non-relativistic trace anomalies}
\label{ssec:nonrel_trace_anomalies}

The breaking of classical Weyl invariance at quantum level manifests through the trace anomaly, \ie a non-vanishing contribution to the trace of the energy-momentum tensor \cite{Duff:1993wm}.
In the relativistic case, the Weyl anomaly only appears in even spacetime dimensions.
It is instructive to recall the four-dimensional case
\beq
T^{\mu}\,_{\mu} = a E_4 - c W^2_{\mu\nu\rho\sigma} + a' \square R \, ,
\label{eq:rel_anomaly_4d}
\eeq
where $a,c$ are the central charges of the CFT at the fixed point, $R$ the Ricci scalar, $E_4$ the Euler density, and $W_{\mu\nu\rho\sigma}$ the Weyl tensor.\footnote{The anomaly also contains the parity-odd Pontryagen density. In this review, we only focus on parity-even terms. An analysis of parity-odd anomalies for non-relativistic theories (either Lifshitz or Galilean) was performed in certain cases in references \cite{Arav:2014goa,Arav:2016xjc}. }
The anomaly term proportional to $a'$ is scheme-dependent, \ie it can be eliminated by the addition of a local counterterm \cite{Bonora:1985cq}.
The Euler density and the Weyl tensor are called type A and type B anomalies, respectively \cite{Deser:1993yx}.
They are distinguished by their behaviour under Weyl transformations: the former has non-vanishing Weyl variation, while the latter is Weyl-invariant.
The central charges can be analytically continued along the entire RG flow.
The function generalizing the coefficient of type A anomalies satisfies monotonicity properties in two and four dimensions, a result which formalizes the irreversibility of RG flows \cite{Zamolodchikov:1986gt,CARDY1988749,Osborn:1989td, Jack:1990eb, Osborn:1991gm, Komargodski:2011vj}.\footnote{Other approaches to prove monotonicity theorems along an RG flow were developed \eg in \cite{Casini:2011kv,Jafferis:2011zi,Myers:2010xs,Casini:2004bw,Casini:2012ei,Casini:2023kyj}.}

In this section, we will review the classification of the terms entering the trace anomaly for a Schr\"odinger field in 2+1 dimensions coupled to a NC geometry \cite{Jensen:2014hqa,Arav:2016xjc,Auzzi:2015fgg,Auzzi:2016lrq}. Several results were previously derived for Lifshitz-invariant theories from holography \cite{Chemissany:2012du} and in field theory \cite{Arav:2014goa}.
We will then compute the trace anomaly in explicit Schr\"{o}dinger-invariant theories \cite{Auzzi:2016lxb,Auzzi:2017jry,Auzzi:2017wwc,Pal:2017ntk,Fernandes:2017nvx}.
Other computations of trace anomalies in non-relativistic systems were carried out in \cite{Daza:2018nvg,maki2019virial}.

\paragraph{Classification of the trace anomaly.}
To compute the trace anomaly, one needs to introduce a curved background which acts as a source for the currents.
Type I NC geometry provides the natural candidate to which a Schr\"{o}dinger field can couple.
The variation of the generating functional $\mathcal{W}$ defines the currents in the following way
\beq
\delta \mathcal{W} = \int d^{d+1} x \, e \, \left[ - \delta m_{\mu} \langle J^{\mu} \rangle + \delta \bar{v}^{\mu} \langle \mathcal{P}_{\mu} \rangle - \delta \tau_{\mu} \langle \mathcal{E}^{\mu} \rangle - \frac{\delta \bar{h}^{\mu\nu}}{2} \langle T_{\mu\nu} \rangle \right] \, ,
\label{eq:variationW_sec5}
\eeq
where $J^{\mu}$ is the current associated to particle number symmetry, $\mathcal{P}_{\mu}$ the momentum density, $\mathcal{E}^{\mu}$ the energy current and $T_{\mu\nu}$ the spatial stress tensor.
In the previous expression, we performed a variation of the metric data such that $\delta m_{\mu}$ and $\delta \tau_{\mu}$ are arbitrary, while $\delta v^{\mu}$ and $\delta h^{\mu\nu}$ are constrained to satisfy eq.~\eqref{eq:constraints_NC}. For this reason, we denoted with $\delta \bar{v}^{\mu}$ and $\delta \bar{h}^{\mu\nu}$ their unconstrained variations along the spatial directions.

Given a generic Weyl transformation parametrized by 
\beq
\tau_\mu \rightarrow e^{2 \sigma} \tau_\mu \, , \qquad
v^\mu \rightarrow e^{-2 \sigma} v^\mu \, , \qquad
h_{\mu \nu} \rightarrow e^{2 \sigma} h_{\mu \nu} \qquad
h^{\mu \nu} \rightarrow e^{-2 \sigma} h^{\mu \nu} \, ,
\label{eq:weyl_sec5}
\eeq
where $\sigma(x^{\mu})$ is a spacetime-dependent parameter, one finds from eq.~\eqref{eq:variationW_sec5} that the Ward identity associated to Weyl invariance is
\beq
2 \tau_{\mu} \langle \mathcal{E}^{\mu} \rangle - h^{\mu\nu} \langle T_{\mu\nu} \rangle = 0 \, . 
\label{eq:Ward_identity_T}
\eeq
Alternatively, one can combine the terms in the left-hand side of eq.~\eqref{eq:Ward_identity_T} to define a new stress tensor $\mathcal{T}^{\mu}_{\,\,\, \nu} \equiv 2 v^{\mu} \mathcal{P}_{\nu}  - h^{\mu\rho}   T_{\rho\nu} $, in terms of which the Ward identity simply becomes $\langle \mathcal{T}^{\mu}_{\,\,\,\mu} \rangle =0$. This redefinition was proposed in \cite{Hartong:2014pma,Hartong:2014oma}.

Eq.~\eqref{eq:Ward_identity_T} is the non-relativistic counterpart of the tracelessness of the classical relativistic energy-momentum tensor.
The determination of the Weyl anomaly is a cohomology problem \cite{Bonora:1983ff}.
The classification can be performed intrinsically, \ie by directly working with NC data \cite{Arav:2016xjc}.
However, in the Schr\"{o}dinger case one can exploit the null reduction technology to simplify the construction of scalar terms \cite{Jensen:2014hqa}.
A Weyl transformation on the type I NC background is equivalent to a Weyl transformation in the parent manifold which is independent of the $u$ coordinate $ n^M \p_M \sigma =0 , $ where the vector $n^M$ was defined in eq.~\eqref{eq:null_isometry_sec4}.
The transformations in the set (\ref{eq:weyl_sec5}) can be obtained from the null reduction of the relativistic variations
\beq
g_{MN} \rightarrow e^{2 \sigma} g_{MN} \, , 
\qquad
g^{MN} \rightarrow e^{-2 \sigma} g^{MN} \, ,
\qquad
n^M \rightarrow n^M \, , \qquad 
n_M \rightarrow  e^{2 \sigma} n_M \, .
\label{eq:weyl_nullred}
\eeq
The space of terms with uniform scaling dimension and invariant under the symmetries of the non-relativistic theory split into distinct sectors invariant under Weyl transformations \cite{Arav:2016xjc}.
These sectors are only characterized by the number of appearances of $n^M$, and the cohomological problem can be studied separately in each sector.

Let us now focus on the case of a 2+1 dimensional Schr\"{o}dinger-invariant theory.
The classification of the trace anomaly drastically changes if an integrability structure exists, \ie when the background is a TTNC geometry (see definition in section \ref{ssec:NC_geometry}).
If Frobenius condition is imposed, the possible scalars entering the anomaly collapse to a single sector, composing a finite set. 
Denoting the anomaly with $\mathcal{A}$, the only independent term is the type B anomaly coming from the null reduction of the Weyl tensor \cite{Arav:2016xjc,Auzzi:2015fgg}
\beq
\mathcal{A} = b W^2_{MNPQ} + \mathcal{A}_{\rm ct} \, ,
\eeq
where $\mathcal{A}_{\rm ct}$ are scheme-dependent contributions that can be eliminated with a counterterm.

If Frobenius condition is \textit{not} imposed, there are infinite sectors for the classification of the anomaly, which can be studied separately \cite{Arav:2016xjc,Auzzi:2016lrq}.\footnote{Notice that NC geometry only acts as a background to define the sources to define all the components of the energy-momentum tensor. To avoid an obstruction in accessing all the components, one should leave the NC background generic. Nonetheless, when considering physical theories, one can impose causality conditions afterwards.  }
The minimal sector without appearances of $n^M$ has anomaly $\mathcal{A}_0$ given by the null reduction of the 3+1 dimensional relativistic formula in eq.~\eqref{eq:rel_anomaly_4d}. 
The next sector with a single appearance of $n$ has vanishing trace anomaly $\mathcal{A}_{1} = 0$.
In sectors with more appearances of $n^M$, one can build by dimensional analysis several type B anomalies, but the cohomological problem in these sectors is not studied and therefore it is not known if any other type A anomaly survives.
The natural candidate to define a monotonic quantity along a non-relativistic RG flow is the coefficient of the null-reduced Euler density.
One can either apply perturbative arguments based on local RG equations \cite{Pal:2016rpz,Auzzi:2016lrq}, or adopt non-perturbative dilaton techniques \cite{Arav:2017plg}.
Despite the similarities, some of the ingredients that guarantee the irreversibility in the relativistic case are missing, such as the positivity of the Zamolodchikov metric on conformal manifolds or the Reeh-Schlieder theorem.
Therefore, a Schr\"{o}dinger version of the $a$--theorem is still missing.

\paragraph{Trace anomaly in explicit theories.}
Computing the trace anomaly in explicit examples does not only provide a consistency check of the cohomology, but also determines the central charges of the model under consideration.

One way to compute the Weyl anomaly from a specific action is to perform the heat kernel (HK) technique, \eg see \cite{Vassilevich:2003xt,Mukhanov:2007zz}.
The HK method can only be applied to hermitian and elliptic differential operators (\ie, they need to have a finite number of vanishing or negative eigenvalues).  
In the non-relativistic case, this requires an extension of the Schr\"{o}dinger differential operator
\beq 
\mathcal{D}_{\rm Schr} \equiv - 2 i m \p_t - \p_i^2 
\label{eq:Schroed_op}
\eeq
to Euclidean signature, which is subtle.
\textbf{(1)} The naive  $t \rightarrow -it$ continuation gives \beq 
\mathcal{D}_{\rm HK} \equiv 2m \p_t - \p_i^2 \, , 
\eeq
which is nothing but the heat kernel operator itself, as it is well-known in mathematical physics \cite{cannon_browder_1984}.
The application of the HK method to the operator $\mathcal{D}_{\rm HK} $ leads to a differential equation without diffusion of the physical time coordinate and with an ill-defined Seeley-De Witt series expansion, since the solution is a Dirac delta distribution at all times.
The differential operator $\mathcal{D}_{\rm HK}$ was used in \cite{Pal:2017ntk} to argue that the trace anomaly vanishes for theories with Schr\"{o}dinger fields.
\textbf{(2)} Another possibility is to perform a double analytic continuation $ t \rightarrow - i t, \,  m \rightarrow i m,$ and then interpret the corresponding Euclidean operator as
\beq
\mathcal{D}_{E} \equiv  2 m \sqrt{-\p_t^2} - \p_i^2 \, .
\label{eq:Euclidean_Schr1}
\eeq
This can be written as a sum of elliptic operators after Laplace-transforming $\sqrt{-\p_t^2}$ to the second-order differential operator $-\p_i^2$, as explained in \cite{Solodukhin:2009sk}.
The trace anomaly $\mathcal{A}_0$ computed  \eqref{eq:Euclidean_Schr1} in the sector without appearances of $n^M$ presents the same coefficients as the relativistic parent theory, except for an overall $1/m$ normalization depending on the $\mathrm{U}(1)$ mass \cite{Auzzi:2016lxb,Auzzi:2017jry}.
The operator \eqref{eq:Euclidean_Schr1} provides the unique prescription which transforms the parabolic operator \eqref{eq:Schroed_op} into an elliptic one, but the price to pay is the introduction of a non-local term $\sqrt{-\p_t^2}$, which makes the recipe non-analytical.
For this reason, it is not clear if the Euclidean computation presented in \cite{Auzzi:2016lxb,Auzzi:2017jry} can be used for  the physical case of the Schr\"{o}dinger operator with real time coordinate.\footnote{However, let us comment that the same trace anomaly was computed in \cite{Fernandes:2017nvx} using the Fujikawa approach, without any analytic continuation to Euclidean space. }

It would be desirable to better understand the analytic properties of Schr\"{o}dinger theories. 
For instance, another standard technique to compute anomalies consists of a perturbative computation involving Feynman diagrams.
An analysis of vacuum diagrams, based on arguments similar to the selection rule \ref{sel_rule1}, led the authors of \cite{Pal:2017ntk} to conclude that the anomaly vanishes.
However, the degree of divergence of the diagrams (essential to apply the selection rule \ref{sel_rule1}) and the role of contact terms are not discussed.
A computation of triangle diagrams based on the dimensional regularization discussed in \cite{Arav:2016akx} may provide a way to shed light on the trace anomaly in explicit cases.

\section{Non-relativistic supersymmetry}
\label{sec:SUSY}

While there is a unique gradation of the Poincaré algebra, more freedom is allowed in defining a supersymmetric extension of the Bargmann algebra.
We discuss this topic in section \ref{ssec:nonrel_SUSY_algebra}, focusing on the case obtained with null reduction.
We define in section \ref{ssec:nonrel_superspace} a non-relativistic superspace, which provides a covariant organization of the fields inside supermultiplets and superfields.
We then analyze the renormalization properties of two supersymmetric and Galilean-invariant theories in 2+1 dimensions: the non-relativistic analog of the WZ model (section \ref{ssec:Galilean_WZ}) and the supersymmetric extension of GED (section \ref{ssec:SGED}).

Let us stress that in this section we will mainly focus on the gradation of the Schr\"{o}dinger algebra that can be obtained by null reduction.
Several supersymmetric extensions of the Lifshitz group have been studied (\eg see \cite{Chapman:2015wha,Arav:2019tqm,Yan:2022dqk}), but we will not consider them here.

\subsection{Non-relativistic supersymmetry algebra}
\label{ssec:nonrel_SUSY_algebra}

Two gradations of the Bargmann algebra were first constructed in 3+1 dimensions \cite{Puzalowski}.
Using the terminology of \cite{Bergman:1995zr}, the gradation $\mathcal{S}_1 \mathcal{G}$ includes a single two-component complex supercharge, while $\mathcal{S}_2 \mathcal{G}$ contains two supercharges.
The two superalgebras can be obtained by contraction of $\mathcal{N}=1$ and $\mathcal{N}=2$ super-Poincar\'e algebras with a central extension, respectively. Alternatively, $\mathcal{S}_2 \mathcal{G}$ can be obtained by null reduction of the super-Poincar\`e algebra in 4+1 dimensions. It turns out that $ \mathcal{S}_1 \mathcal{G} \subset \mathcal{S}_2 \mathcal{G}$.
SUSY theories with either one of these two superalgebras have been studied in 3+1 dimensions \cite{Puzalowski,CLARK198491,deAzcarraga:1991fa,Meyer:2017zfg}.

In this review, we will focus on non-relativistic supersymmetric theories in 2+1 dimensions with $\mathcal{S}_2 \mathcal{G}$ invariance, which is the interesting case  where the anticommutators of two supercharges give both a time and a space translation.
To build this superalgebra, at least two copies of the supercharges are needed \cite{Gomis:2004pw}.
Concretely, the $\mathcal{S}_2 \mathcal{G}$ algebra in 2+1 dimensions is characterized by the following non-vanishing (anti)commutators
\begin{subequations}
\beq
\begin{aligned}
& [P_j, G_k] = i \delta_{jk} M \, , \qquad
[H, G_j] = i P_j \, ,  \\ 
& [P_j, J] = -i \epsilon_{jk} P_k \, , \qquad
[G_j, J] = -i \epsilon_{jk} G_k  \, ,  \qquad \qquad j,k=1,2  \label{eq:Bargmann} 
\end{aligned}
\eeq
\beq
\begin{aligned}
& [J,Q_1] =  \frac12 Q_1 \, , \quad
[J,Q_2] = - \frac12 Q_2 \, , \quad
[Q_1, G_1 - i G_2] = -i Q_2 \, , & \\
&  \lbrace Q_1, Q_1^{\dagger}  \rbrace = \sqrt{2} H \, , \quad
\lbrace Q_2, Q_2^{\dagger}  \rbrace = \sqrt{2} M \, , & \\
& \lbrace Q_1, Q_2^{\dagger} \rbrace = -  (P_1 -i P_2) \, , \quad
\lbrace Q_2, Q_1^{\dagger} \rbrace = -  (P_1 + i P_2)  \, ,
\end{aligned}
\label{eq:commu_superGalileo2} 
\eeq
\end{subequations}
where $P_j$ are the spatial components of the momentum, $H$ is the Hamiltonian, $ G_j $ are the generators of Galilean boosts, $ J $ is the planar angular momentum, $M$  is the $\mathrm{U}(1)$ central charge and $Q_{\a}$ with $\alpha \in \lbrace 1,2 \rbrace$ are two complex supercharges. 
This is the non-relativistic $ \mathcal{N}=2 $ super-Bargmann algebra in 2+1 dimensions; by removing $Q_1$ from \eqref{eq:commu_superGalileo2}, we obtain the $ \mathcal{S}_1 \mathcal{G}$ algebra.
The derivation of the above superalgebra in 2+1 dimensions by means of a contraction was performed in the context of three-dimensional NC supergravity \cite{Andringa:2013mma,Bergshoeff:2015uaa}.

We show that the previous superalgebra can be obtained by null reduction of the $\mathcal{N}=1$ super-Poincar\'e algebra in 3+1 dimensions \cite{PhysRevD.1.2901}, following the notation in \cite{Auzzi:2019kdd}.
The bosonic part of the parent Poincaré subalgebra reduces to the Bargmann subalgebra \eqref{eq:Bargmann} by identifying some components of the linear and angular momenta with the central charge and the boost operators, see eq.~\eqref{eq:map_generators_Son}.
To perform the reduction of the fermionic part, we rewrite the right-hand side of the four-dimensional anticommutator 
$\{\mathcal{Q}_\alpha , \bar{\mathcal{Q}}_{\dot{\beta}} \} = i \sigma^M_{\alpha \dot{\beta}} \partial_M$
 in terms of light-cone derivatives $\partial_{\pm} = \frac{1}{\sqrt{2}}( \partial_3 \pm \partial_0)$ to get 
\beq
\lbrace \mathcal{Q} , \bar{\mathcal{Q}} \rbrace = i \begin{pmatrix}
\sqrt{2} \p_+  &   \p_1 - i \p_2 \\
 \p_1 + i \p_2 & - \sqrt{2} \p_{-} 
\end{pmatrix} \, .
\label{SUSY algebra relativistica}
\eeq 
According to the general decomposition \eqref{eq:ansatz_fields_nullred_sec4} of a field under null reduction (here we denoted with $x^-$ the null direction along which the reduction is performed), we identify
\beq
\p_+ \rightarrow \p_t \, , \qquad
\p_- \rightarrow i m \, .
\label{eq:derivatives_reduction_sec6}
\eeq 
The matching with eq.~\eqref{eq:commu_superGalileo2} is achieved after reinterpreting the four-dimensional two-spinor components as three-dimensional complex Grassmann fields with $\mathcal{Q}_{\alpha} \to Q_\alpha$, 
 $\bar{\mathcal{Q}}_{\dot{\beta}} \to Q_\beta^\dagger$.

One can extend the analysis by including the scaling symmetry and perform in a similar way the null reduction of the four-dimensional superconformal algebra $\mathrm{SU}(2,2|1)$.
This enlarges the non-relativistic superalgebra with the dilatation operator $D$, the special conformal generator $C$, its complex fermionic superpartner $S$, and the bosonic generator $R$ of $\mathrm{U}(1)$ R-symmetry.
The (anti)commutation relations are given by the Schr\"{o}dinger subalgebra \eqref{eq:Schroedinger_algebra} together with its fermionic extension in eqs.~\eqref{commu_superGalileo2_sec36} and \eqref{eq:superconformal_algebra_sec36}.
All together, they form the $\mathcal{N}=2$ super-Schr\"{o}dinger algebra in 2+1 dimensions \cite{Duval:1984cj,Duval:1993hs,Julia:1994bs,Duval:1990hj}.
This is the full set of classical symmetries satisfied by the supersymmetric CS model in section \ref{ssec:SUSY_CS} and by the actions that we will consider in this section.

\subsection{Non-relativistic superspace}
\label{ssec:nonrel_superspace}

Superspace provides a convenient way to covariantly treat SUSY-invariant theories by extending ordinary spacetime with Grassmann coordinates $(\theta^{\alpha}, \bar{\theta}^{\dot{\alpha}})$ which are associated to the action of supercharges.
In the relativistic case, $d$--dimensional superspace is built as the coset
\beq
\text{superspace} = \frac{\text{super-Poincaré}}{\mathrm{SO}(d,1)} \, ,
\eeq
and SUSY transformations correspond to translations in superspace.
A similar intrinsic procedure can be applied to the non-relativistic case by quotienting the super-Bargmann algebra by the subgroup of spatial rotations and Galilean boosts.
Here we build $\mathcal{N}=2$ Bargmann superspace (which realizes the $\mathcal{S}_2 \mathcal{G}$ algebra discussed in section \ref{ssec:nonrel_SUSY_algebra}) in 2+1 dimensions by null reduction of the four-dimensional $\mathcal{N}=1$ parent superspace, since this procedure allows to directly inherit the quotient algebra already implemented in the relativistic case \cite{Auzzi:2019kdd}.\footnote{Non-relativistic superspace was first introduced in 3+1 dimensions \cite{CLARK198491,deAzcarraga:1991fa}. Alternative constructions based on an intrinsic approach in 2+1 dimensions can be found in \cite{Bergman:1995zr,Nakayama:2009ku}.  }

\paragraph{Construction of superspace.}
An explicit differential representation of the super-Poincar\'e algebra is given by the following supercharges and SUSY covariant derivatives
\begin{subequations}
\beq
{\cal Q}_{\alpha} = i \frac{\p}{\p \theta^{\alpha}} - \frac12 \bar{\theta}^{\dot{\beta}}  \p_{\alpha \dot{\beta}}  \, , \qquad
\bar{\cal Q}_{\dot{\alpha}} = - i \frac{\p}{\p \bar{\theta}^{\dot{\alpha}}} + \frac12 \theta^{\beta} \p_{\beta \dot{\alpha}} \, ,
\label{eq:4d_supercharges}
\eeq
\beq
{\cal D}_{\alpha} = \frac{\p}{\p \theta^{\alpha}} - \frac{i}{2} \bar{\theta}^{\dot{\beta}} \p_{\alpha \dot{\beta}}   \, , \qquad
\bar{\cal D}_{\dot{\alpha}} =  \frac{\p}{\p \bar{\theta}^{\dot{\alpha}}} - \frac{i}{2} \theta^{\beta}  \p_{\beta \dot{\alpha}}  \, ,
\label{eq:4d_derivatives}
\eeq
\end{subequations}
which act on local superfields $\Psi(x^M, \theta^\alpha, \bar{\theta}^{\dot{\alpha}})$.
To perform the null reduction, we introduce light-cone coordinates and rewrite the combinations $\p_{\alpha \dot{\beta}}  = \sigma^M_{\alpha \dot{\beta}} \p_M$ inside eqs.~\eqref{eq:4d_supercharges}--\eqref{eq:4d_derivatives} in terms of the derivatives $\p_{\pm}, \p_1, \p_2$.
Next, we apply the recipe \eqref{eq:ansatz_fields_nullred_sec4} to all the ordinary fields.
Since SUSY requires each field component of a multiplet to be an eigenfunction of the $\p_-$ operator with the same eigenvalue $m$, we collectively impose
\beq
\Psi(x^M, \theta^\alpha, \bar{\theta}^{\dot{\alpha}}) = e^{im x^{-}} \tilde{\Psi}(t \equiv x^+, x^{i}, \theta^{\alpha}, \bar{\theta}^{\alpha} \equiv (\theta^\alpha)^\dagger)  \, .
\label{eq:decomposition_superfield}
\eeq
Acting on these superfields with the supercharges and derivatives in eqs.~\eqref{eq:4d_supercharges} and \eqref{eq:4d_derivatives} rewritten with light-cone derivatives, and performing the identifications \eqref{eq:derivatives_reduction_sec6}, we obtain
\beq  
\label{eq:nonrelQD}
\begin{cases}
Q_1 = i \frac{\p}{\p \theta^1} - \frac12  \bar{\theta}^2 (\p_1 - i \p_2) - \frac{1}{\sqrt{2}} \bar{\theta}^1  \p_t \\
\bar{Q}_1 = -i \frac{\p}{\p \bar{\theta}^1} + \frac12 \theta^2 (\p_1 + i \p_2) + \frac{1}{\sqrt{2}} \theta^1 \p_t \\
Q_2 = i \frac{\p}{\p \theta^2} - \frac12  \bar{\theta}^1 (\p_1 + i \p_2) - \frac{i}{\sqrt{2}} \bar{\theta}^2 M \\
\bar{Q}_2 = -i  \frac{\p}{\p \bar{\theta}^2} + \frac12 \theta^1 (\p_1 - i \p_2) - \frac{i}{\sqrt{2}} \theta^2 M  \\
\end{cases} \,\,
\begin{cases}
D_1 =  \frac{\p}{\p \theta^1} - \frac{i}{2} \bar{\theta}^2 (\p_1 - i \p_2) - \frac{i}{\sqrt{2}} \bar{\theta}^1 \p_t \\
\bar{D}_1 =   \frac{\p}{\p \bar{\theta}^{1}} - \frac{i}{2} \theta^2 (\p_1 + i \p_2) - \frac{i}{\sqrt{2}} \theta^1 \p_t \\
D_2 =  \frac{\p}{\p \theta^2} - \frac{i}{2} \bar{\theta}^1 (\p_1 + i \p_2) - \frac{1}{\sqrt{2}} \bar{\theta}^2 M \\
\bar{D}_2 =   \frac{\p}{\p \bar{\theta}^2} - \frac{i}{2} \theta^1 (\p_1 - i \p_2) - \frac{1}{\sqrt{2}} \theta^2 M  \\
\end{cases}
\eeq
These operators provide a representation of the non-relativistic SUSY algebra \eqref{eq:commu_superGalileo2} in a three-dimensional $\mathcal{N}=2$ Bargmann superspace, with associated superfields $\tilde{\Psi}$.

One can realize irreducible representations of the superalgebra by imposing additional constraints.
(Anti)chiral superfields $\Phi (\bar{\Phi})$ can be obtained either by null reduction of the relativistic four-dimensional (anti)chiral superfields or directly in three-dimensional superspace by imposing 
\beq
\bar{D}_\a \Phi = 0 , \qquad \qquad D_\a \bar{\Phi} = 0  \, ,
\eeq
where the covariant derivatives are defined in \eqref{eq:nonrelQD}. 
The explicit solution for a non-relativistic chiral superfield in component fields reads\footnote{The normalization of the fermionic terms is chosen to get a canonically normalized kinetic term in actions built with this chiral superfield.}
\beq
\Phi (x, \theta, \bar{\theta}) = \phi(x_L) + \theta^1 \psi_1(x_L) + \theta^2 \, 2^{1/4} \sqrt{m} \, \psi_2(x_L) - \frac{1}{2} \theta^{\alpha} \theta_{\alpha} F(x_L)  \, , \quad  x_L^{\alpha\beta} =  x^{\alpha\beta} - \frac{i}{2} \theta^{\alpha}  \bar{\theta}^{\beta} \, , 
\label{eq:nonrel_chiral_superfield}
\eeq
where $\varphi$ is a dynamical complex scalar, $(\xi,\chi)$ combine into a Weyl fermion according to the ansatz \eqref{eq:ansatz_fermions_nullred}, and $F$ is a complex auxiliary scalar.
These component fields form a supermultiplet under the super-Bargmann symmetry.
The antichiral superfield is obtained from \eqref{eq:nonrel_chiral_superfield} by hermitian conjugation.
The previous decomposition resembles the relativistic case, but the main difference is that the Grassmann field $\xi$ is non-dynamical.
When imposing a reality constraint, one obtains a vector superfield $V$ (also called prepotential).
Since its explicit expression is more cumbersome, we refer the interested reader to \cite{Baiguera:2022cbp} for more details on its decomposition in component fields.

\paragraph{Supersymmetric actions.}
Manifestly supersymmetric actions can be constructed by using the Berezin integration on spinorial coordinates. In the relativistic superspace, for a generic superfield $\Psi$ we define
\beq 
\label{eq:relberezin}
\int d^4x d^4\theta \, \Psi = \int d^4x \, {\cal D}^2 \bar{\cal D}^2 \Psi \Big|_{\theta = \bar{\theta}=0} \, ,
\eeq
in terms of the covariant derivatives \eqref{eq:4d_derivatives}.
Performing the null reduction with the ansatz \eqref{eq:decomposition_superfield}, we obtain the prescription for the Berezin integrals in the non-relativistic superspace
\beq 
\label{eq:Berezin_integration_null_red}
\begin{aligned}
& \int d^4x d^4\theta \, \Psi = \int d^4x \, {\cal D}^2 \bar{\cal D}^2 \Psi \Big|_{\theta = \bar{\theta}=0} \;  \longrightarrow  \\
&  \int d^3x D^2 \bar{D}^2 \tilde{\Psi} \Big|_{\theta = \bar{\theta}=0} \; \times \frac{1}{2\pi} \int_0^{2\pi}  dx^- \,  e^{imx^-}   \equiv \int d^3x d^4\theta \, \tilde{\Psi}  \; \times \frac{1}{2\pi} \int_0^{2\pi}  dx^- \,  e^{imx^-} \, ,
\end{aligned}
\eeq
where in the right-hand side $d^3x \equiv dt \,dx^1 dx^2$, and the spinorial derivatives are given in eq.~\eqref{eq:nonrelQD}. 
We immediately notice that when $m \neq 0$ the reduction becomes trivial after performing the $x^-$ integral. 
Non-vanishing expressions only arise if the super-integrand $\Psi$ is uncharged with respect to the mass generator. In the construction of SUSY invariant actions, this is equivalent to requiring the action to be invariant under an extra global U(1) symmetry \cite{deAzcarraga:1991fa}.  

\paragraph{Quantum theory.}
At quantum level, the generating functional is defined by
\beq
\mathcal{Z} [J, \bar{J}, J_V] = \int [\mathcal{D}\Phi \mathcal{D}\bar{\Phi} \mathcal{D} V ] \, 
\exp \left[ i S + i \int d^3 x \, \le \int d^2 \theta J \Phi + \int d^2 \bar{\theta} \bar{J} \bar{\Phi} + \int d^4 \theta J_V V  \ri   \right] \, ,
\label{eq:SUSY_path_integral}
\eeq
$J,\bar{J}$ are (anti)chiral superfields and $J_V$ is a vector superfield, all of them acting as sources. Correlation functions can be obtained by repeated application of functional derivatives 
\beq
\label{eq:sources}
\frac{\delta J (z_i)}{\delta J(z_j)} =  \bar{D}^2 \, \delta^{(7)} (z_i - z_j) \, , \quad
\frac{\delta \bar{J} (z_i)}{\delta \bar{J}(z_j)} =  D^2  \, \delta^{(7)} (z_i - z_j) \, , \quad
\frac{\delta J_V (z_i)}{\delta J_V (z_j)} =  \delta^{(7)} (z_i - z_j)  \, ,
\eeq
where $ z \equiv (x^{\mu}, \theta^{\alpha}, \bar{\theta}^{\dot{\alpha}}) $ and $\delta^{(7)} (z_i - z_j) \equiv \delta^{(3)} (x_i - x_j) \delta^{(2)} (\theta_i - \theta_j)\delta^{(2)} (\bar{\theta}_i - \bar{\theta}_j)$. The covariant derivatives acting on the delta functions arise due to the constrained nature of the (anti)chiral supercurrents.

Loop corrections can be directly studied in terms of superfields.
Propagators and vertices are extracted in the same way described in section \ref{ssec:loop_corrections}, except for additional covariant derivatives coming from the application of eq.~\eqref{eq:sources}.
Supergraphs can be reduced to ordinary Feynman diagrams by performing D-algebra, which is the manipulation of covariant derivatives to perform explicitly the
integrations along the Grassmannian coordinates.\footnote{The D-algebra technology is pedagogically explained in \cite{Gates:1983nr}. The adaptation to the super-Bargmann case is presented in \cite{Auzzi:2019kdd,Baiguera:2022cbp}. }
For instance, D-algebra can be used to show that the selection rule \ref{sel_rule1} generalizes to Feynman diagrams built with superfields.

\subsection{Galilean Wess-Zumino model}
\label{ssec:Galilean_WZ}

The non-relativistic superspace developed in section \ref{ssec:nonrel_superspace} can be used to build actions manifestly invariant under the $\mathcal{N}=2$ super-Schr\"{o}dinger group.
This method was used in \cite{Auzzi:2019kdd} to find a Galilean-invariant Wess-Zumino (WZ) model, whose action reads
\beq
\label{non-rel WZ action in superfield formalism} 
S = \int   d^3 x d^4 \theta \le \bar{\Phi}_1 \Phi_1 +  \bar{\Phi}_2 \Phi_2 \ri + g \int   d^3 x  d^2 \theta \,  \Phi_1^2 \Phi_2 + \mathrm{h.c.}
\eeq
where $\Phi_1, \Phi_2$ are non-relativistic chiral superfields of the form \eqref{eq:nonrel_chiral_superfield} and the Berezin integration was defined in \eqref{eq:Berezin_integration_null_red}.
According to the anisotropic dimensional scaling of Galilean theories, the coupling constant $g$ is dimensionless and the theory is classically scale invariant.
In fact, the full symmetry is given by the super-Schr\"{o}dinger group discussed at the end of section \ref{ssec:nonrel_SUSY_algebra}.

The main feature of this action, compared to its relativistic counterpart, is that the integrand must be uncharged with respect to the global $\mathrm{U}(1)$ symmetry associated with the central extension of the Bargmann algebra.
For this reason, a non-vanishing superpotential only exists  when at least two species of (anti)chiral superfields are chosen.
Eq.~\eqref{non-rel WZ action in superfield formalism} represents the simplest choice, where the masses of the two matter superfields are $M_1=m$ and $M_2 = -2m.$

On general grounds, supersymmetry poses strong constraints on the dynamics of a theory and often allows one to obtain exact results.
In the relativistic WZ model, the existence of a non-renormalization theorem which states that the superpotential is quantum exact, forces all its loop corrections to vanish \cite{Grisaru:1979wc,Seiberg:1993vc}. Perturbative corrections are only allowed for the K\"{a}hler potential.
The non-renormalization theorem for the superpotential also applies to the non-relativistic version of the model. However, in this case, there are additional constraints that are due to the retarded nature of the (anti)chiral propagator, contrarily to the causal Feynman propagator of the relativistic model. 
In fact, the particular structure of its poles forces the K\"{a}hler potential to be one-loop exact \cite{Auzzi:2019kdd}. 

We then conclude that the Galilean WZ model is one-loop exact and the $\beta$-function of the theory is fully determined by ($\mu$ is an arbitrary energy scale)
\beq
\beta_g=\frac{d g } {d \log \mu} =\frac{g^3}{ 4 \pi m} \, .
\label{betafunction_WZ}
\eeq
The theory is IR-free at low energies, and the non-vanishing of the beta function signals the breaking of the original scale invariance, due to quantum effects.
It is interesting to observe that the Galilean WZ model provides an example of a theory where the non-relativistic limit (performed here as a null reduction) and quantization (\ie the limit $\hbar \rightarrow 0$) do not commute, since the relativistic WZ model is not one-loop exact.

\subsection{Supersymmetric Galilean electrodynamics}
\label{ssec:SGED}

The ${\cal N}=2$ supersymmetric generalization of GED (we will refer to it as SGED) can be obtained by performing the null reduction of a four-dimensional ${\cal N}=1$ supersymmetric gauge theory, directly in superspace. 
In this section, we will only consider the Abelian case, \ie the gauge group is taken to be $\mathrm{U}(1)$. 
Using this method, the action reads \cite{Baiguera:2022cbp}
\beq
\label{eq:null_SGED_action}
S_{\rm nSGED} =  \int d^3x d^2\theta \;  W^2 + \int d^3x d^4\theta \; \bar{\Phi} e^{g V} \! \Phi \, ,
\eeq
where $W^2 \equiv \tfrac12 W^\a W_\a$ is written in terms of the field strength $W_\a = i \bar{D}^2 D_\alpha V$, $V$ is a real scalar prepotential and $g$ a dimensionless coupling.
This theory is invariant by construction under the super-Schr\"{o}dinger group composed by the generators in eqs.~\eqref{eq:Schroed_action_sec2}, \eqref{commu_superGalileo2_sec36} and \eqref{eq:superconformal_algebra_sec36}.
In addition, the model enjoys local invariance under $\mathrm{U}(1)$ supergauge transformations.

While the action \eqref{eq:null_SGED_action} is formally the same as supersymmetric electrodynamics in relativistic superspace, subtleties are hidden in the Berezin integration \eqref{eq:Berezin_integration_null_red}, which is defined in terms of the non-relativistic covariant derivatives \eqref{eq:nonrelQD}. 
Recall from section \ref{ssec:nonrel_superspace} that any superspace integrand has to be invariant under the $\mathrm{U}(1)$ particle number symmetry.
This implies that the real prepotential $\mathcal{V}$ defined in the parent manifold has vanishing mass eigenvalue, \ie
\beq
\mathcal{V} (x^M, \theta^{\alpha}, \bar{\theta}^{\dot{\alpha}}) = V (x^{\mu}, \theta^{\alpha}, \bar{\theta}^{\beta}) \, .
\label{eq:ansatz_prepotential}
\eeq
In other words, $V$ comes from a four-dimensional vector superfield that is independent of the null direction $x^{-}$. 
Since the $\theta \bar{\theta}$ component is the gauge field $A_{\alpha\beta}$, the previous ansatz is the supersymmetric generalization of eq.~\eqref{eq:ansatz_nullred_GED}. 
From a physical point of view, this is equivalent to saying that the prepotential acts as an instantaneous mediator of interactions between the matter (super)fields.
We will refer to eq.~\eqref{eq:null_SGED_action} as the  \textit{null SGED} action, since it is the result of applying the null reduction method.

Due to the redundancy encoded by supergauge transformations, the quantum theory is only well-defined after applying the Faddeev-Popov procedure \cite{FADDEEV196729}.
Feynman rules can be read directly in terms of superfields from the gauge-fixed action.
The analysis of the loop corrections reveals that the coupling constant $g$ enjoys a non-renormalization theorem, due to the extension of the selection rule \ref{sel_rule1} to this case \cite{Baiguera:2022cbp}.
However, one can show that there is an infinite set of divergent diagrams which make the model non-renormalizable.

\paragraph{Renormalizable SGED.}
Surprisingly, SUSY is not enough to protect the theory from generating infinitely many marginal deformations.
This forces us to modify the original null SGED into a non-linear sigma model
\beq
\label{eq:action_newSGED}
S_{\rm SGED} =  \int d^3 x  d^2 \theta \, W^2 + \int d^3 x  d^4 \theta \; \bar{\Phi} e^{g V} \Phi \, {\cal F}( \bar{D}_2 D_2 V) \, ,
\eeq
where ${\cal F}$ is a generic smooth function of its argument.
It should be noted that the prepotential $V$ and the components $D_2, \bar{D}_2$ of the supersymmetric covariant derivative are dimensionless in the anisotropic Schr\"{o}dinger counting, and that the combination $\bar{D}_2 D_2 V$ is also supersymmetric and supergauge invariant.
Therefore, the new term provides a marginal deformation of the original theory \eqref{eq:null_SGED_action} consistent with its classical symmetries.
This represents the supersymmetric version of the arbitrary functions $\mathcal{J}, \mathcal{V}, \mathcal{E}$ that enter the GED action at quantum level (see  eq.~\eqref{eq:action_extra_shira}). 
Upon Taylor expansion of ${\cal F}$, this action exhibits an infinite number of new couplings weighted by the derivatives
\beq \label{eq:Fderivatives}
{\cal F}^{(n)} \equiv \frac{d^n {\cal F}(x)}{d x^n}\Big|_{x=0} \, .
\eeq
One can study the quantum corrections to the action in eq.~\eqref{eq:action_newSGED} by means of the supergauge covariant approach \cite{Gates:1983nr}.
The result is that the new SGED theory is renormalizable at one-loop, and the beta functions of the coupling constants read
\beq
\beta_{\mathcal{F}^{(n)}} = \frac{d \mathcal{F}^{(n)}}{d \log \mu} = - g^{n+2} \frac{n! \; n}{16 \pi m (\sqrt{2} m)^{n} \, \mathcal{F}^{(n)}} \, ,
\label{eq:beta_Fn}
\eeq
where $\mu$ is a renormalization scale and $m$ the mass of the (anti)chiral superfields.
In particular, all the beta functions vanish when $g=0$, where no wavefunction renormalization occurs.
At this value of the coupling, the gauge transformations of matter fields become trivial and the minimal coupling $e^{gV}$ disappears, while a $V$ dependence survives in the analytic function $\mathcal{F}$. At $g=0$ the theory exhibits a non-trivial IR fixed point where it enjoys the full super-Schr\"{o}dinger invariance and corresponds to an interacting field theory with action
\beq
S_{\rm fixed} = \int d^3 x  d^2 \theta \, W^2 + \int d^3 x  d^4 \theta \; \bar{\Phi} \Phi \, {\cal F}( \bar{D}_2 D_2 V) \, .
\eeq
There is an infinite-dimensional non-relativistic superconformal manifold parametrized by arbitrary values of the couplings $\mathcal{F}^{(n)}.$
The RG flow of the $\mathcal{F}^{(n)}$ couplings and the topology of the conformal manifold that we have found are peculiar to the one-loop approximation. At higher orders, one expects different couplings to mix at different orders, possibly leading to a richer spectrum of fixed points. In that case, the existence of a conformal manifold would constrain the functional form of the ${\cal F}$ function non-trivially. 

A natural extension of the previous analysis is the study of non-Abelian Galilean electrodynamics, which was constructed in \cite{Bagchi:2015qcw,Bagchi:2022twx,Banerjee:2022uqj}.
Its supersymmetric generalization would provide a Schr\"{o}dinger version of super Yang-Mills (SYM) theories, which play a fundamental role in high-energy physics.
It would be interesting to analyze the loop corrections to  these models to test whether an infinite number of marginal deformations is still produced at quantum level, and if a conformal manifold of fixed points still arises.
In particular, $\mathcal{N}=4$ SYM is superconformal invariant at the full quantum level and allows to perform precise matchings of the AdS/CFT correspondence.
In recent years, it has been shown that certain decoupling limits of $\mathcal{N}=4$ SYM lead to quantum mechanical models with non-relativistic symmetries and a well-defined holographic dual \cite{Harmark:2007px}.
This represents a parallel route to the null reduction of SYM theories that we will pursue in section \ref{sec:SMT}.


\section{Spin Matrix Theories and non-relativistic holography}
\label{sec:SMT}

Most of the quantitative understanding of AdS/CFT duality with gauge group SU$(N)$ is in the planar regime $N \rightarrow \infty$, where the number of perturbative diagrams on the CFT side reduces, and loop corrections of the string theory are suppressed.
However, this approach is blind to non-perturbative objects such as black holes and D-branes, which are essential to access the full quantum description of string theory.
To capture finite--$N$ effects, most of the efforts have focused on quantities protected by SUSY. 
Some examples are giant gravitons \cite{Biswas:2006tj}, localization methods \cite{Pestun:2007rz} and index techniques \cite{Hosseini:2017mds,Cabo-Bizet:2018ehj,Choi:2018hmj,Benini:2018ywd}. 

Here we present a different approach, based on the SMT limit of AdS/CFT duality \cite{Harmark:2014mpa}.  
This regime corresponds to a decoupling limit obtained either by approaching zero-temperature critical points in the grand-canonical ensemble, or by zooming in towards unitarity bounds in the microcanonical ensemble, as we review in section \ref{ssec:decoupling_limits_SMT}.
We show in section \ref{ssec:nonrel_nature_SMT} that the above-mentioned decoupling limits have a non-relativistic nature.
SMTs are quantum-mechanical models which generalize the nearest-neighbour integrable spin chain arising in the planar limit and can be defined without reference to the decoupling limits. 
We follow this perspective in section \ref{ssec:SMT_spin_chains}.
Then we systematically compute in section \ref{ssec:effective_SMT} the interacting Hamiltonian for all the effective quantum-mechanical theories arising from the decoupling limits, and focus on the SU(2) subsector in section \ref{ssec:examples_SMT}.
A large class of these theories, which has $\mathrm{SU}(1,1)$ symmetry subgroup, admits a QFT description in terms of a local field, including a superfield formulation when supersymmetry is preserved (see section \ref{ssec:QFT_SMT}). 
We finally define the SMT limit on the gravity side and make contact with holography in section \ref{ssec:nonrel_holography}.

\subsection{Spin Matrix Theory from decoupling limits}
\label{ssec:decoupling_limits_SMT}

We consider $\mathcal{N}=4$ SYM defined on $\mathbb{R} \times S^3$, which is the conformal boundary of global AdS.
Non-Abelian gauge theories showcase interesting thermodynamics on compact manifolds, characterized by a deconfinement phase transition at the Hagedorn temperature, \ie when the string partition function becomes divergent \cite{ATICK1988291,Aharony:2003sx,Sundborg:1999ue,Kristensson:2020nly}. 
Inspired by this fact, we will define a prescription to identify certain zero-temperature critical points of $\mathcal{N}=4$ SYM.

\paragraph{Grand-canonical formulation.}
In the grand-canonical ensemble, we introduce chemical potentials associated with the Cartan charges of the symmetry group $\mathrm{PSU}(2,2|4)$ of $\mathcal{N}=4$ SYM.
We denote with $D$ the dilatation operator, $\mathbf{S}_1, \mathbf{S}_2$ the Cartan generators for rotations, and with $\mathbf{Q}_1,\mathbf{Q}_2,\mathbf{Q}_3$ the Cartan generators for the $\mathrm{SU}(4)$ R-symmetry.
The role of these generators is summarized for convenience in table~\ref{tab:Cartan_charges}.

\begin{table}[ht]   
\begin{center}   
\begin{tabular}  {|c|c|c|} \hline  
\textbf{Generators} & \textbf{CFT side} & \textbf{Gravity side}  \\ \hline
\rule{0pt}{4.9ex} $D$ & Dilatation operator & Rescaling along radial direction of $\mathrm{AdS}_5$  \\
\rule{0pt}{4.9ex}  $\mathbf{S}_1, \mathbf{S}_2$ & $\mathrm{SO}(4)$ rotations & Angular momenta on $S^3 \subset \mathrm{AdS}_5$  \\ 
\rule{0pt}{4.9ex}  $\mathbf{Q}_1,\mathbf{Q}_2,\mathbf{Q}_3$  & $\mathrm{SU}(4)$ R-symmetry &  Angular momenta on $S^5$
 \\[0.2cm]
\hline
\end{tabular}   
\caption{Cartan charges of the $\mathrm{PSU}(2,2|4)$ group. In the middle column, we report the (bosonic) subgroup to which they belong, while in the right column, we give the dual interpretation on the gravity side (see section \ref{ssec:nonrel_holography}). } 
\label{tab:Cartan_charges}
\end{center}
\end{table}

The partition function in the grand-canonical ensemble is defined by
\beq
\mathcal{Z} (\beta, \Omega) = \Tr \le e^{- \beta (D- \vec{\Omega} \cdot \vec{J})} \ri \, ,
\label{eq:grand_canonical_partition_funct}
\eeq
where $\beta=1/T$ is the inverse temperature, and $\vec{\Omega} = (\omega_1, \omega_2, \Omega_1, \Omega_2, \Omega_3)$ is the set of chemical potentials associated with the generators $\vec{J}=(\mathbf{S}_1, \mathbf{S}_2, \mathbf{Q}_1,\mathbf{Q}_2,\mathbf{Q}_3)$.
The previous trace is performed over the singlet states on $\mathbb{R} \times S^3$, or equivalently over the gauge-invariant operators on $\mathbb{R}^4$.\footnote{The singlet condition on a $\mathbb{R} \times S^3$ comes from requiring that flux lines need to close on a compact space. }
The main idea, developed in \cite{Harmark:2006di,Harmark:2006ta,Harmark:2006ie,Harmark:2008gm}, is that deconfinement phase transitions persist even for non-vanishing chemical potential, \ie at certain critical points where $(T, \vec{\Omega})= (0, \vec{\Omega}^{(c)})$.
In this section, we mainly follow the notation used in \cite{Harmark:2007px,Harmark:2014mpa,Menculini:2020lww}.

Let us parametrize the chemical potentials as $\vec{\Omega} = \Omega (a_1, a_2, b_1, b_2, b_3)$ with $\Omega \rightarrow 1$ corresponding to the critical points. We then have 
\beq
J \equiv \vec{\Omega}^{(c)} \cdot \vec{J} = a_1 \mathbf{S}_1+ a_2 \mathbf{S}_2 + b_1 \mathbf{Q}_1+ b_2 \mathbf{Q}_2 + b_3 \mathbf{Q}_3 \, .
\label{eq:definition_J}
\eeq
To guarantee that the partition function gives sensible results in the limit $ \beta \rightarrow \infty$, it is convenient to explicitly write the loop expansion of the dilatation operator of $\mathcal{N}=4$ SYM
\beq
D = D_0 + \delta D = D_0 + \lambda D_2 + \lambda^{\frac{3}{2}} D_3 + \dots 
\label{eq:expansion_dilatation_operator}
\eeq
where $\lambda$ is the 't Hooft coupling, $D_0$ the tree-level dilatation operator, $\delta D$ its full quantum correction, and $D_2$ the one-loop contribution.
Next, we re-express the exponent in eq.~\eqref{eq:grand_canonical_partition_funct}
\beq
\beta \le D - \vec{\Omega} \cdot \vec{J} \ri = 
\beta \delta D + \beta \le  D_0 - J \ri + \beta \le 1- \Omega  \ri J \geq 0 \, ,
\label{eq:exponent_partition_function}
\eeq
where the above inequality is needed to achieve a well-defined zero-temperature limit.

When $\lambda=0,$ then $\delta D=0$ and we get a non-trivial partition function (\ie neither vanishing nor divergent) if
\beq
D_0 \geq J \, , \qquad
\tilde{\beta} \equiv \beta \le 1- \Omega \ri \,\, \mathrm{fixed} \, .
\label{eq:decoupling_limit1}
\eeq
Since all the Cartan charges of $\mathrm{PSU}(2,2|4)$ have integer or half-integer eigenvalues, it is clear that $D_0 - J =0$ or $D_0 - J \geq \frac{1}{2}$, and only the former modes contribute to the partition function after the limit is taken.

If we keep instead $\lambda$ fixed and non-zero, then the partition function vanishes in the limit $\beta \rightarrow \infty$, unless we require $\delta D=0$.
This condition is only valid for supersymmetric states.
In order to get an interacting theory with other surviving modes, we are forced to send $\lambda \rightarrow 0,$ but keeping at the same time $\beta \lambda$ fixed.
This can be achieved by redefining
\beq
\tilde{\lambda} \equiv \frac{\lambda}{1- \Omega} \, ,
\label{eq:decoupling_limit2}
\eeq
so that $\beta  \delta D \rightarrow \tilde{\beta}  \tilde{\lambda} D_2 $, where we sent $\lambda \rightarrow 0$ and applied eqs.~\eqref{eq:expansion_dilatation_operator} and \eqref{eq:decoupling_limit1}.
In summary, the combination of all the previous conditions reads
\beq
\boxed{
\beta \rightarrow \infty \, , \qquad \Omega \rightarrow 1 \, , \qquad
\tilde{\beta} = \beta (1 - \Omega) \,\, \mathrm{fixed} \, , \qquad
\tilde{\lambda} = \frac{\lambda}{1- \Omega} \,\, \mathrm{fixed} \, , \qquad
N \,\, \mathrm{fixed} \, , }
\label{eq:grand_canonical_limit}
\eeq
and the partition function at the zero-temperature critical points becomes
\beq
\mathcal{Z} (\tilde{\beta}, \tilde{\lambda}) = \mathrm{Tr}_{D_0 = J} \le e^{- \tilde{\beta} (D_0 + \tilde{\lambda} D_2)} \ri \, ,
\label{eq:result_partition_function}
\eeq
where $\tilde{\beta}$ and $\tilde{\lambda}$ play the role of effective temperatures and coupling constants, respectively.
The partition function is completely identified by the tree-level and one-loop contributions of the dilatation operator.
One can show that the action of $D_2$ closes in each sector, thus defining a set of consistent theories.

We refer to \eqref{eq:grand_canonical_limit} as a \textit{decoupling limit}, since all the states with $D_0 > J$ are suppressed and decouple from the partition function.
The fields surviving the limit transform into representations $R_s$ of a spin group, which is always a subgroup of $\mathrm{PSU}(2,2|4)$.
The field content in each sector can be re-expressed in terms of the letters of $\mathcal{N}=4$ SYM, living in the singleton representation of $\mathrm{PSU(2,2|4)}$.
As shown in \cite{Harmark:2007px}, there exist 12 non-trivial decoupling limits, giving rise to the sectors listed in table \ref{tab:BPS}.
Furthermore, the fields also transform under the adjoint representation of $\mathrm{SU}(N)$.
For this reason, we refer to the effective theories defined in the limit \eqref{eq:grand_canonical_limit} as \textit{Spin Matrix Theories}.
A thermodynamic analysis of the degrees of freedom surviving the near-BPS limits reveals that the theories corresponding to the spin groups collected by rows in table \ref{tab:BPS} effectively behave as 0+1, 1+1, or 2+1 dimensional models.

\begin{table}[ht]   
\begin{center}   
\begin{tabular}  {|c|c|} \hline   
\textbf{Spin group $G_s$}	& \textbf{Combination of Cartan charges $J$}  \\
\hline
$\SU(2)$ &  $\mathbf{Q}_1+\mathbf{Q}_2$ \\ 
$\SU(1|1)$ &  $\frac{2}{3} \mathbf{S}_1+\mathbf{Q}_1 + \frac{2}{3} \le \mathbf{Q}_2 + \mathbf{Q}_3 \ri$ \\ 
$\SU(1|2)$ &  $\frac{1}{2} \mathbf{S}_1+\mathbf{Q}_1 +\mathbf{Q}_2 + \frac{1}{2} \mathbf{Q}_3 $ \\ 
$\SU(2|3)$ &  $\mathbf{Q}_1+\mathbf{Q}_2+\mathbf{Q}_3$ \\ \hline
 $\SU(1,1)$ bosonic  & $\mathbf{S}_1 + \mathbf{Q}_1$   \\
  $\SU(1,1)$ fermionic  & $\mathbf{S}_1 + \frac{2}{3} ( \mathbf{Q}_1 +  \mathbf{Q}_2 + \mathbf{Q}_3 ) $   \\
	   $\SU(1,1|1)$ &  $\mathbf{S}_1 + \mathbf{Q}_1+ \frac{1}{2}(\mathbf{Q}_2+\mathbf{Q}_3)$ \\
	  $\mathrm{PSU}(1,1|2)$  & $\mathbf{S}_1 + \mathbf{Q}_1+\mathbf{Q}_2$   \\ \hline
	  $\SU(1,2)$ &  $\mathbf{S}_1 + \mathbf{S}_2$  \\
	   $\SU(1,2|1)$  & $\mathbf{S}_1 + \mathbf{S}_2 + \frac{1}{2} \mathbf{Q}_1 + \frac{1}{2} \mathbf{Q}_2$   \\
	   $\SU(1,2|2)$  & $\mathbf{S}_1 + \mathbf{S}_2 +  \mathbf{Q}_1$   \\
	  $\mathrm{PSU}(1,2|3)$  &  $\mathbf{S}_1 + \mathbf{S}_2 +  \mathbf{Q}_1 +\mathbf{Q}_2 + \mathbf{Q}_3$   \\ \hline
\end{tabular}
\caption{List of the 12 non-trivial decoupling limits \eqref{eq:grand_canonical_limit}. The horizontal blocks distinguish theories that are effectively 0+1, 1+1, and 2+1 dimensional from a counting of the degrees of freedom. }
\label{tab:BPS}
\end{center}
\end{table}

\paragraph{Microcanonical formulation.}
One can equivalently define the decoupling limits in the microcanonical ensemble, where the conditions are expressed in terms of the energy (scaling dimension) of the states (operators).
In the following, we will often identify $D=E$ due to the relativistic state/operator correspondence, assuming that the three-sphere has a unit radius.
The microcanonical formulation of the limit reads
\beq
\boxed{
D-J \rightarrow 0  \, , \qquad \lambda \rightarrow 0 \, , \qquad
\frac{1}{\lambda} \le D-J \ri \,\, \mathrm{fixed} \, , \qquad
N \,\, \mathrm{fixed} \, , }
\label{eq:micro_canonical_limit}
\eeq
which leads to
\beq
H = J + \tilde{g}^2 \lim_{\lambda \rightarrow 0} \frac{D-J}{\lambda} = 
J + \tilde{g}^2 H_{\rm int} \, , 
\label{eq:effective_SMT}
\eeq
where $\tilde{g}$ is an effective coupling constant, $J$ was defined in eq.~\eqref{eq:definition_J} and we identify $H_{\rm int} = D_2$.\footnote{The coupling constant $\tilde{g}$ is equivalent to $\tilde{\lambda}$ defined in eq.~\eqref{eq:decoupling_limit2}. We use a different name to stress the difference between micro- and grand-canonical formulations.  }
This equation defines the effective Hamiltonian for the SMTs arising in the decoupling limit \eqref{eq:micro_canonical_limit}.
Computing \eqref{eq:effective_SMT} for all the sectors listed in table \ref{tab:BPS} will be the task of section \ref{ssec:effective_SMT}.

The prescription \eqref{eq:micro_canonical_limit} can be understood as a low-energy limit in which we only focus on the modes close to the energy scale $E=J$, while the rest of the spectrum decouples.
In particular, the first condition in eq.~\eqref{eq:decoupling_limit1} can be interpreted as $E \geq J,$ which is a unitarity bound that guarantees that the states in the CFT have positive definite norm.
Equivalently, this is a BPS bound between the energy and a linear combination of the Cartan charges of $\mathcal{N}=4$ SYM.
For this reason, we will also refer to eq.~\eqref{eq:micro_canonical_limit} as near-BPS limits.

A crucial observation is that the decoupling limits \eqref{eq:grand_canonical_limit} and \eqref{eq:micro_canonical_limit} keep $N$ fixed, but not necessarily $N = \infty$.
This allows us to go beyond the planar limit and include non-perturbative effects, albeit inside a subsector of the original $\mathcal{N}=4$ SYM theory.

\subsection{Non-relativistic nature of Spin Matrix Theory}
\label{ssec:nonrel_nature_SMT}

There are several indications that the decoupling limits \eqref{eq:micro_canonical_limit} select a non-relativistic corner of $\mathcal{N}=4$ SYM.
In the following list, we anticipate some facts from sections \ref{ssec:effective_SMT} and \ref{ssec:nonrel_holography}:
\begin{itemize}
\item Any SMT Hamiltonian enjoys an emergent global $\mathrm{U}(1)$ symmetry, which is naturally interpreted as the particle number conservation typical of Schr\"{o}dinger theories (see section \ref{ssec:Schroed_group}).
In this regard, antiparticles decouple in all the SMT limits.
\item SMTs are models that can be obtained by a KK decomposition along the three-sphere of $\mathcal{N}=4$ SYM defined on $\mathbb{R} \times S^3$.
This procedure reduces the original QFT to a matrix QM, where only the tree level and one-loop parts of the dilatation operator contribute.
\item The gravity duals of SMT are well-defined limits of type IIB string theory on $\mathrm{AdS}_5 \times S^5$, where the speed of light is sent to infinity.
The target space geometry and the algebra of generators on the worldsheet are both non-relativistic.
\end{itemize}

We add another argument.
In the large $N$ limit of $\mathcal{N}=4$ SYM, one has access to an integrable spin chain description.
The ground state is the configuration where all the spins are aligned in the same direction.
Low-energy excitations are called \textit{magnons} and correspond to the flipping of a single spin.\footnote{Recent studies of SMT magnons from the gravity side were performed in \cite{Roychowdhury:2020yun}. }
They can be thought of as pseudo-particles propagating along the spin chain with momentum $p$ and satisfying the dispersion relation \cite{Beisert:2005tm}
\beq
\delta D = E- J = \sqrt{1- \frac{\lambda}{\pi^2} \sin^2 \le \frac{p}{2}\ri } -1 \, .
\eeq
Taking the limit $\lambda \rightarrow 0$ in eq.~\eqref{eq:micro_canonical_limit}, we get
\beq
E - J = \frac{\lambda}{2 \pi^2} \sin^2 \le \frac{p}{2} \ri \approx \frac{\lambda p^2}{8 \pi^2}   \, ,
\eeq
which is a non-relativistic dispersion relation for small momenta.
This is another signal that the truncation of $\mathcal{N}=4$ SYM into sectors with only one-loop contributions of the dilatation operator shows non-relativistic traits.

\subsection{Spin Matrix Theory from spin chains}
\label{ssec:SMT_spin_chains}

While SMTs were defined via the Hamiltonian \eqref{eq:effective_SMT} as near-BPS limits of $\mathcal{N}=4$ SYM, a priori one can take a different approach and build them independently \cite{Harmark:2014mpa}.
In general, SMTs are quantum-mechanical models whose defining ingredients are the representation $R_s$ of a semi-simple Lie (super)-group $G_s$ and the adjoint representation of $\SU(N).$
The Hilbert space is constructed starting by a vacuum state $\ket{0}$ by acting with ladder operators which are matrices carrying an index $s \in R_s$ for the spin group.  

In the bosonic sector, the ladder operators form the algebra of harmonic oscillators
\beq
(a_s)^i_{\,\, j} |0 \rangle = 0 \, , \qquad
\left[ (a^r)^i_{\,\, j} , (a^{\dagger}_s)^k_{\,\, l}  \right] = 
\delta^r_s \delta^k_j \delta^i_l \, , \qquad
\forall s,i,j \, ,
\eeq
and they define a Hilbert space spanned by the states
\beq
\tr \left( a^{\dagger}_{s_1} \dots a^{\dagger}_{s_l}  \right)
\tr \left( a^{\dagger}_{s_{l+1}} \dots \right) \dots
\tr \left( a^{\dagger}_{s_{k+1}} \dots a^{\dagger}_{s_L} \right) |0 \rangle  \, ,
\label{eq:combo_multitrace}
\eeq
where $L$ is referred to as the \textit{length} of the state, and the traces are performed over the colour indices of the adjoint representation of $\mathrm{SU}(N)$.
Notice that the multi-trace structure in eq.~\eqref{eq:combo_multitrace} enforces the $\mathrm{SU}(N)$ singlet condition. 

The fermionic sector works similarly, in that one defines anticommutation relations for the ladder operators
\beq
(b_s)^i_{\,\, j} |0 \rangle \, , \qquad
\left\lbrace (b^r)^i_{\,\, j} , (b^{\dagger}_s)^k_{\,\, l}  \right\rbrace = 
\delta^r_s \delta^k_j \delta^i_l \, , \qquad
\forall s,i,j \, .
\eeq
All the other (anti)commutators between ladder operators, either bosonic or fermionic in any combination, vanish.

Next, we define an interacting Hamiltonian for the system.
We demand that the interaction is quartic in the fields, (consisting of two creation and two annihilation operators), and that it commutes with all the generators of the spin group.
For simplicity, we apply these constraints in a bosonic case with a single copy of the ladder operators; this scenario can be straightforwardly generalized to multiple copies of bosons and fermions.
The interacting Hamiltonian reads
\begin{subequations}
\beq
H_{\rm int}^{\rm bos} = \frac{1}{N} \, U^{s' r'}_{s r}  \sum_{\sigma \in S(4)} 
T_{\sigma} (a^{\dagger}_{s'})^{i_{\sigma(1)}}_{\,\,\, i_3}
(a^{\dagger}_{r'})^{i_{\sigma(2)}}_{\,\,\, i_4}
(a^{s})^{i_{\sigma(3)}}_{\,\,\, i_1}
(a^{r})^{i_{\sigma(4)}}_{\,\,\, i_2} \, , 
\eeq
\beq
\sum_{\sigma \in S(4)} T_{\sigma} \sigma = (14) + (23) - (12) - (34) \, .
\eeq
\end{subequations}
The coefficients $U^{s' r'}_{s r}$ encode the information on the spin structure of the theory.
The reality of the Hamiltonian implies $(U^{s' r'}_{s r})^* = U^{s r}_{s' r'}$, while its symmetry under permutations forces $U^{s' r'}_{s r} = U^{r' s'}_{r s}$.
More compactly, the previous Hamiltonian can be written in terms of normal-ordered operators by using the properties of commutators and the cyclicity of the trace:
\beq
H_{\rm int}^{\rm bos} =  - \frac{1}{N} \, U^{s' r'}_{s r} : \tr \left( [a^{\dagger}_{s'}, a^s ] [a^{\dagger}_{r'}, a^{r}] \right) :  
\label{eq:dictionary_bos_term_SMT}
\eeq 
In the fermionic case, commutators are replaced by anti-commutators, \ie
\beq
H_{\rm int}^{\rm ferm} = - \frac{1}{N} U^{s' r'}_{s r} : \tr \left( \lbrace b^{\dagger}_{s'}, b^s \rbrace \lbrace b^{\dagger}_{r'}, b^{r}\rbrace \right) :  
\label{eq:dictionary_ferm_term_SMT}
\eeq
The interactions between mixed bosonic--fermionic operators are obtained similarly.

Notice that no assumption has been made on the value of $N$.
In fact, SMTs provide a generalization of the nearest-neighbours spin chain to finite $N$.
The integrable spin chain interpretation is recovered by taking $N= \infty$ in eqs.~\eqref{eq:dictionary_bos_term_SMT} and \eqref{eq:dictionary_ferm_term_SMT}.
However, when moving away from this case by including $1/N$ corrections, one is allowed to join or split the spin chains. 

Let us now make contact with the near-BPS limit \eqref{eq:micro_canonical_limit} of $\mathcal{N}=4$ SYM, which gives a precise identification of $U^{s'r'}_{sr}$ in any of the sectors listed in table \ref{tab:BPS}.
In the large $N$ limit, the Hilbert space collapses to states generated by single trace operators
\beq
| s_1 \dots s_L \rangle = \tr \left( a^{\dagger}_{s_1} \dots a^{\dagger}_{s_L}  \right) | 0 \rangle \, ,
\eeq
and the cyclicity property of the trace allows for an interpretation in terms of a spin chain with translation invariance.
When $N \rightarrow \infty,$ the action of the interacting Hamiltonian on the singlet states reads
\beq
H_{\rm int}  | s_1 s_2 \dots s_L \rangle =
2 \sum_{k=1}^L U^{m n}_{s_k \, s_{k+1}} | s_1 \dots s_{k-1} \, m  n \, s_{k+2} \dots s_L \rangle \, .
\eeq
This describes a nearest-neighbour spin chain Hamiltonian, which is characterized by the data $U^{s'r'}_{sr}.$
More explicitly, the coefficients can be computed by looking at the action on two-particle states\footnote{We denote the states as
$
| k l \rangle \equiv | \dots k l \dots \rangle  \, ,
$
where the dots refer to other sites of the spin chain which are left unchanged by the Hamiltonian.}
\beq
H_{\rm int} | k  l \rangle =
H_{\rm spin} | k  l \rangle = 
 2 U^{m n}_{k l} | m  n \rangle  \, .
 \label{eq:action_spin_chain_two_particle_states}
\eeq
Since the extension from large to finite $N$ is unique, the SMT in a given decoupling limit is defined unambiguously by the action of the spin chain Hamiltonian on the surviving degrees of freedom of the sector.

The coefficients $ U^{s' r'}_{s r}$ can be extracted from the expression of $D_2 = H_{\rm spin}$ inside the corresponding subsector of $\mathrm{PSU}(2,2|4)$, which was worked out in several papers \cite{Minahan:2002ve,Beisert:2002ff,Beisert:2003tq,Beisert:2003ys,Beisert:2004ry,Bellucci:2005vq,Bellucci:2006bv,Beisert:2007sk, Zwiebel:2007cpa,Beisert:2008qy,Beisert:2010jr}.
In general, the comparison is not immediate because the spin representation $R_s$ of the SMT Hamiltonian is different from the singleton representation used to compute $D_2$.
However, eq.~\eqref{eq:action_spin_chain_two_particle_states} provides a direct way to read the coefficients from the action of $D_2$ on a spin chain.

\subsection{Spin Matrix Theory Hamiltonians}
\label{ssec:effective_SMT}

In this section, we compute the SMT Hamiltonian \eqref{eq:effective_SMT} describing the effective degrees of freedom surviving the decoupling limits \eqref{eq:micro_canonical_limit} for all the possible spin groups listed in table \ref{tab:BPS}.
This is possible by focusing on the $\mathrm{PSU}(1,2|3)$ sector, which is the largest and contains all the other spin groups as subsectors.
The techniques to compute the Hamiltonian in any near-BPS limit were developed in \cite{Harmark:2019zkn,Baiguera:2020jgy,Baiguera:2020mgk,Baiguera:2021hky,Baiguera:2022pll} to which we refer for more details.
Here we mainly take notation and results from \cite{Baiguera:2022pll}.
The techniques that one can adopt are the following:
\begin{enumerate}
\item Compute the one-loop corrections $D_2$ to the dilatation operator of $\mathcal{N} = 4$ SYM, and then zoom in towards the unitarity bound \eqref{eq:micro_canonical_limit} of interest \cite{Minahan:2002ve,Beisert:2002ff,Beisert:2003tq,Beisert:2003ys,Beisert:2004ry,Bellucci:2005vq,Bellucci:2006bv,Beisert:2007sk, Zwiebel:2007cpa,Beisert:2008qy,Beisert:2010jr}.
The Hamiltonian in SMT language \eqref{eq:dictionary_bos_term_SMT}--\eqref{eq:dictionary_ferm_term_SMT} is obtained by translating the results between different representations, as explained in eq.~\eqref{eq:action_spin_chain_two_particle_states} and text below.
\item Expand $\mathcal{N}=4$ SYM defined on $\mathbb{R} \times S^3$ in KK modes along the three-sphere.
Promote the classical Hamiltonian obtained in this way at quantum level by requiring that no change of orderings is needed.
This approach was originally proposed in \cite{Harmark:2019zkn} and later developed in \cite{Baiguera:2020jgy,Baiguera:2020mgk,Baiguera:2021hky,Baiguera:2022pll}.
We refer to this procedure as \textit{spherical expansion}.
\item When a near-BPS limit \eqref{eq:micro_canonical_limit} preserves part of the original SUSY of PSU$(2,2|4)$, one can define a cubic supercharge $\mathcal{Q}$ whose anticommutator closes into the interacting Hamiltonian $ \lbrace \mathcal{Q} , \mathcal{Q}^{\dagger} \rbrace = H_{\rm int}.$
This method is based on extending a technique valid in the PSU(1,1$|$2) subsector (see \cite{Beisert:2007sk, Zwiebel:2007cpa}) to the SMT language  \cite{Baiguera:2021hky,Baiguera:2022pll}.
\item Build all the possible blocks (quadratic in the fields) that comprise an irreducible representation of the spin group characterizing the near-BPS limit \eqref{eq:micro_canonical_limit}.
The most general Hamiltonian quartic in the fields is then built by combining the previous blocks \cite{Baiguera:2020mgk}.
\end{enumerate}

\begin{figure}[ht]
\centering
\includegraphics[scale=0.5]{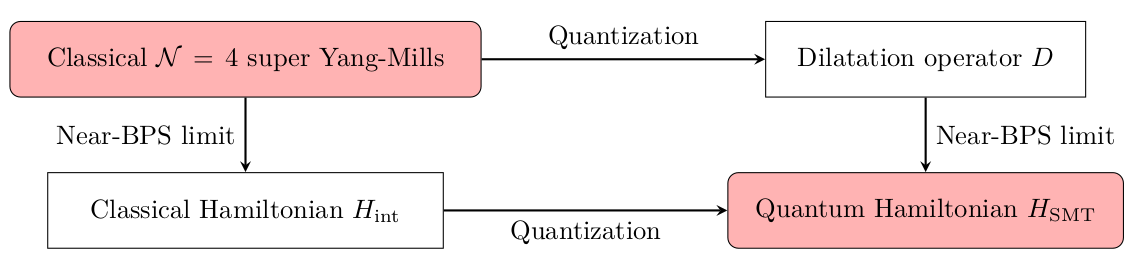}
\caption{\small Commutative diagram representing methods 1 and 2. The starting point is classical $\mathcal{N}=4$ SYM (top left), while the endpoint is a quantum SMT Hamiltonian (bottom down).
Following method 1, we move right-down by computing one-loop corrections $D_2$ to the dilatation operator and zooming in towards a near-BPS limit \eqref{eq:micro_canonical_limit}.
Following method 2, we move down-right by performing a spherical expansion and then promoting the theory at quantum level.}
\label{fig:commutative_diagram}
\end{figure}

It is important to notice that all the previous methods are equivalent, \ie they lead to the same quantum Hamiltonian.
This is particularly non-trivial by observing that the techniques 1 and 2 reverse the order of two procedures: the decoupling limit \eqref{eq:micro_canonical_limit} (which we will also interpret as a $c \rightarrow \infty$ limit in section \ref{ssec:nonrel_holography}) and the quantization.\footnote{For instance, exchanging the order between the non-relativistic limit and quantization leads to difference results in the Galilean WZ model reviewed in section \ref{ssec:Galilean_WZ}. }
Indeed, it was explicitly shown for several spin groups in table \ref{tab:BPS} that the diagram depicted in fig.~\ref{fig:commutative_diagram} is commutative \cite{Harmark:2019zkn,Baiguera:2020jgy,Baiguera:2020mgk,Baiguera:2021hky}.
For technical comments on the convenience of the various methods, we refer the reader to the introduction in reference \cite{Baiguera:2022pll}.

\paragraph{Interacting PSU(1,2$|$3) Hamiltonian.}
As mentioned above, the PSU$(1,2|3)$ sector comprises all the other spin groups listed in table \ref{tab:BPS}, thus giving a complete description of all the near-BPS limits \eqref{eq:micro_canonical_limit}.
The letter content of the PSU$(1,2|3)$ SMT is composed by 8 fields collectively denoted as 
\beq
V_I = \{\chi,\Phi_a,\zeta_a,A \} \, , 
\label{eq:letters_SU123}
\eeq
where $I=0,1,2,3$ and $a=1,2,3$.
Under the residual $\mathrm{SU}(3)$ R-symmetry, the field $\chi$ behaves as a fermion singlet, $\Phi_a$ are complex scalars transforming as a triplet, $\zeta_a$ compose a triplet of complex Grassmann fermions and $A$ is a singlet gauge component.\footnote{The actual letter content of the PSU$(1,2|3)$ sector is given by 3 scalars, 1 field strength and 5 fermions. The apparent mismatch with the fermionic components inside $V_I$ comes from the fact that $\chi$ encodes two chiral fermions, due to the existence of a Dirac equation relating them, see eq.~(2.2)--(2.3) in \cite{Baiguera:2022pll}.}
The letters also form representations of the $\mathrm{SU}(1,2)$ subgroup, which in general are labelled by two quantum numbers $(p,q)$ \cite{Bars:1989bb}.
The representation of interest for the PSU$(1,2|3)$ SMT is given by $(p,q)=(0,I-3)$, with the same index $I$ adopted in eq.~\eqref{eq:letters_SU123}.
Indeed, one can work out the generators in this representation, check that they satisfy the correct algebra, and study the action on the letters.
Finally, each letter carries two integer indices $(n,k)$ representing the number of covariant derivatives acting on the letters. Equivalently, they enumerate the KK modes arising from a spherical expansion of $\mathcal{N}=4$ SYM over the three-sphere inside $\mathbb{R} \times S^3$.

The total interacting Hamiltonian can expressed in terms of quadratic blocks \cite{Baiguera:2022pll}:
\begin{align}\label{eq:Hint_tot_cubic}
	\begin{split}
		H_{\rm int} &= H_D + H_F \, ,
		\\
H_D &= \frac{1}{N}	\sum_{n,k=0}^{\infty}
		\tr \left[ (\mathcal{B}_0^{\dagger})_{n,k} (\mathcal{B}_0)_{n,k}  + \sum_{a=1}^3 \sum_{I=1,2} (\mathcal{B}_I^{a\dagger})_{n,k} (\mathcal{B}_I^a)_{n,k}  + (\mathcal{B}_3^{\dagger})_{n,k} (\mathcal{B}_3)_{n,k} \right] \, ,  \\
H_F	&= \frac{1}{N} \sum_{n,k=0}^{\infty}
		\tr \left[ (\mathcal{F}_0^{\dagger})_{n,k} (\mathcal{F}_0)_{n,k}  + \sum_{a=1}^3 \sum_{I=1,2} (\mathcal{F}_I^{a\dagger})_{n,k} (\mathcal{F}_I^a)_{n,k}  + (\mathcal{F}_3^{\dagger})_{n,k} (\mathcal{F}_3)_{n,k} \right]   \, .
	\end{split}
\end{align}
The set of following $\mathcal{B}$ blocks composes the D-term part of the Hamiltonian: 
\begin{align}
	(\mathcal{B}_0)_{n,k}  & = \sum_{n',k'=0}^{\infty} \le  P^{(0,0)}_{n,k;n',k'} \lbrace \chi^{\dagger}_{n',k'} , \chi_{n+n',k+k'} \rbrace +
	\sum_{a=1}^3  P^{(0,1)}_{n,k;n',k'} [(\Phi^{\dagger}_a)_{n',k'} , (\Phi_a)_{n+n',k+k'}] \right. \nonumber \\ 
	& \left.  + \sum_{a=1}^3 P^{(0,2)}_{n,k;n',k'} \lbrace (\zeta^{\dagger}_a)_{n',k'} , (\zeta_a)_{n+n',k+k'} \rbrace  +  P^{(0,3)}_{n,k;n',k'} [A^{\dagger}_{n',k'} , A_{n+n',k+k'}]   \ri \, ,
\label{eq:block_B0} \\  \nonumber
	(\mathcal{B}_{1}^a)_{n,k} & \equiv
	\sum_{n',k'=0}^{\infty} \le  P^{(1,1)}_{n,k:n',k'}   \epsilon^{abc} 
	[ (\zeta_b)_{n+n',k+k'} , (\Phi^{\dagger}_c)_{n', k'} ]  \right. \\
	& \left.  
	-  P^{(1,2)}_{n,k;n',k'} [ (\zeta^{\dagger}_a)_{n',k'}  , A_{n+n',k+k'} ] + P^{(1,0)}_{n,k;n',k'}  [ (\Phi_a)_{n+n',k+k'} , \chi^{\dagger}_{n',k'}  ]  \ri \, ,
\label{eq:block_B1} \\
	(\mathcal{B}_{2}^a)_{n,k} & \equiv 
	\sum_{n',k'=0}^{\infty}  \le  P^{(2,1)}_{n,k;n',k'}  [(\Phi^{\dagger}_a)_{n',k'}  , A_{n+n',k+k'}]
	  +  P^{(2,0)}_{n,k;n',k'}  \lbrace (\zeta_a)_{n+n',k+k'} , \chi^{\dagger}_{n',k'}    \rbrace  \ri \, ,
\label{eq:block_B2} \\
(\mathcal{B}_3)_{n,k} & \equiv \sum_{n',k'=0}^{\infty} P^{(3,0)}_{n,k;n',k'}  [ A_{n+n',k+k'} , \chi^{\dagger}_{n',k'}  ] \, .
\label{eq:block_B3}
\end{align}
The following $\mathcal{F}_{n,k}$ blocks compose instead the F-term contributions:
\begin{align}
		(\mathcal{F}_0)_{n,k} & \equiv \frac{1}{2} \sum_{n'=0}^{n}\sum_{k'=0}^k P^{(0,0)}_{n',k';n-n',k-k'} \{\chi_{n-n',k-k'},\chi_{n',k'} \} \, , 
		\label{eq:block_F0}  \\
		(\mathcal{F}_1^a)_{n,k} & \equiv \sum_{n'=0}^n \sum_{k'=0}^k P^{(0,1)}_{n',k',n-n',k-k'} [(\Phi_a)_{n-n',k-k'},\chi_{n',k'}] \, ,
		\label{eq:block_F1}  \\  \nonumber
		(\mathcal{F}_2^a)_{n,k} & \equiv \frac{1}{2}  \sum_{n'=0}^{n}\sum_{k'=0}^k P^{(1,1)}_{n',k',n-n',k-k'} \epsilon^{abc} 	[(\Phi_c)_{n',k'},(\Phi_b)_{n-n',k-k'}]\\
 &	 + P^{(0,2)}_{n',k',n-n',k-k'} \{(\zeta_a)_{n-n',k-k'},\chi_{n',k'} \} \, ,
		\label{eq:block_F2} \\ 
		(\mathcal{F}_3)_{n,k} & \equiv \sum_{n'=0}^{n}\sum_{k'=0}^k P^{(1,2)}_{n',k',n-n',k-k'} [(\zeta_a)_{n-n',k-k'},(\Phi_a)_{n',k'}]  - P^{(0,3)}_{n',k';n-n'k-k'} [A_{n-n',k-k'},\chi_{n',k'}] \, .
	\label{eq:block_F3}
\end{align}
The previous blocks depend on the coefficient
\begin{equation}
P^{(i,j)}_{n,k,n',k'} = \sqrt{\frac{(k+n+i-1)!(k'+n' + j-1)!(n+n')!(k+k')!}{(k+k'+n+n' + i+j-1)!n!k!n'!k'!}} \, ,
\label{eq:Pij_coefficients}
\end{equation}
which carries information about the symmetry properties of the fields under the $\mathrm{SU}(1,2)$ subgroup.
From the point of view of the spherical expansion procedure, they can be related to Clebsch-Gordan coefficients defined on the three-sphere.

\paragraph{Properties of the Hamiltonian.}
Let us describe the main features of the Hamiltonian \eqref{eq:Hint_tot_cubic}.
\textbf{(1)} First of all, it is manifestly positive-definite because it is written as a quadratic combination of blocks.
Furthermore, the Hamiltonian can be interpreted as a norm in a linear space (identified by the irreducible representation of the blocks) which measures the distance from the point where the BPS bound is saturated \cite{Baiguera:2020mgk}.
\textbf{(2)} Secondly, the Hamiltonian has an emergent $\mathrm{U}(1)$ global symmetry associated to particle number conservation.
This is one of the hints for the non-relativistic nature of SMT, see section \ref{ssec:nonrel_nature_SMT}. In this regard, another non-Lorentzian feature corresponds to the fact that the summation over modes is restricted to positive integers, truncating negative values (that would appear in a Fourier expansion).
\textbf{(3)} Next, we study the symmetry structure. The blocks $W_I = \lbrace \mathcal{B}_I, \mathcal{F}_I \rbrace$ belong to the same $\mathrm{SU}(1,2)$ representation $(p,q)=(0,I-3)$ as the letters $V_I$ in eq.~\eqref{eq:letters_SU123}.
$\mathcal{B}_I$ and $\mathcal{F}_I$ can be organized into two $\mathcal{N}=3$ vector multiplets, each of them composed by three $\mathcal{N}=1$ chiral multiplets and one $\mathcal{N}=1$ vector multiplet.
This implies that $H_D, H_F$ in eq.~\eqref{eq:Hint_tot_cubic} are separately invariant under the full set of PSU(1,2$|$3) symmetries.
Therefore, any Hamiltonian of the form  $H_D+\Lambda H_F$ is invariant under the PSU$(1,2|3)$ spin group, for any choice of $\Lambda$.
The Hamiltonian $H_D+ H_F$ obtained from the decoupling limit \eqref{eq:micro_canonical_limit} might indicate the existence of a further enhanced symmetry that fixes $\Lambda =1$.\footnote{The enhancement of symmetries may be related to the observation that CPT invariance promotes $\mathcal{N}=3$ SUSY to $\mathcal{N}=4$ in the relativistic case \cite{DHoker:1999yni}. Since SMTs do not contain antiparticles, it is not obvious that a similar statement would also hold in the non-relativistic case.  }
\textbf{(4)} Finally, let us recall that the interacting Hamiltonian \eqref{eq:Hint_tot_cubic} is classical. The SMT Hamiltonian $H_{\rm SMT}$ in fig.~\ref{fig:commutative_diagram} is simply obtained by promoting the full expression at quantum level, without changes of ordering. 

A powerful tool to investigate supersymmetric field theories is superspace, whose construction for super-Schr\"{o}dinger invariant theories was reviewed in section \ref{ssec:nonrel_superspace}.
According to step 3 mentioned in this subsection, one can build a cubic supercharge $\mathcal{Q}$ that generates the interacting Hamiltonian via $\lbrace \mathcal{Q}, \mathcal{Q}^{\dagger} \rbrace = H_{\rm int}$.
This construction can be related to a similar cubic supercharge and superfield defined in flat superspace $\mathbb{C}^{2|3}$ \cite{Chang:2013fba,Chang:2022mjp} (see also section 6 of \cite{Baiguera:2022pll}).
In particular, the space of $\frac{1}{16}$--BPS operators of $\mathcal{N}=4$ SYM is isomorphic to the cohomology of such supercharge \cite{Grant:2008sk}.

Next, we show the power of the SMT approach in section \ref{ssec:examples_SMT} by dealing with the concrete case of $\mathrm{SU}(2)$ subsector.
We then investigate in \ref{ssec:QFT_SMT} a local formulation for the quantum-mechanical Hamiltonian \eqref{eq:Hint_tot_cubic}, focusing on the SU(1,1$|$1) subsector.

\subsection{Example: SU(2) Spin Matrix Theory}
\label{ssec:examples_SMT}

The surviving fields in the SU(2) decoupling limit \eqref{eq:micro_canonical_limit} are two complex scalars belonging to the fundamental spin--$\frac{1}{2}$ representation. Denoting the spin label as $s= \uparrow, \downarrow$, the interacting Hamiltonian reads
\beq
H_{\rm int} = - \frac{1}{2 N} \tr \le [\Phi^{\dagger}_{\uparrow}, \Phi^{\dagger}_{\downarrow}] [\Phi_{\uparrow}, \Phi_{\downarrow}]  \ri \, .
\label{eq:SU2_SMT}
\eeq
It can be obtained from eq.~\eqref{eq:Hint_tot_cubic} by setting $\Phi_{\uparrow} = (\Phi_1)_{0,0}, \Phi_{\downarrow}=(\Phi_2)_{0,0}$ and all the other fields to zero.

\begin{figure}
\centering
\includegraphics[scale=0.92]{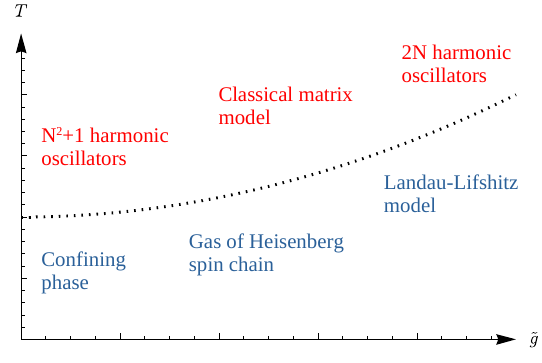}
\caption{Phase diagram of $\mathrm{SU}(2)$ SMT as a function of the temperature $T$ and the coupling constant $\tilde{g}$. The dotted line represents a deconfinement transition. Plot inspired on fig.~1 of \cite{Harmark:2014mpa}.  }
\label{fig:SU2_phases}
\end{figure}

The phase diagram of this theory was studied in \cite{Harmark:2014mpa}.
In the planar limit $N=\infty$, the system reduces to the ferromagnetic XXX$_{1/2}$ Heisenberg spin chain \cite{Minahan:2002ve}.
The Hagedorn temperature $T_H$ can be computed explicitly and matches the corresponding limit taken on the string theory living in AdS$_5 \times S^5$, including the case where a magnetic field is present \cite{Harmark:2006ie,Harmark:2006ta}.
At low temperature $T \ll T_H,$ the planar limit is a good approximation even when $N$ is allowed to be large but finite, \ie $N \gg 1$.  
The system is described by a confining phase when $\tilde{g} \ll 1$, a gas of Heisenberg spin chains at finite $\tilde{g}$, and an effective Landau-Lifshitz sigma-model with target space $S^2$ at strong coupling $\tilde{g} \gg 1$.
The latter regime will be recovered from a non-relativistic limit of string theory in section \ref{ssec:nonrel_holography}. 
At high temperature $T \gg T_H,$ the number of colours $N$ is allowed to be finite and one can distinguish again different phases depending on the coupling constant $\tilde{g}$.
When $\tilde{g} \ll 1$, a coherent state approach reveals a phase described by $N^2+1$ harmonic oscillators, transitioning to a matrix model for intermediate coupling, and finally to a system of $2N$ uncoupled harmonic oscillators when $\tilde{g} \gg 1.$ 
The phase diagram is depicted in fig.~\ref{fig:SU2_phases}.

\subsection{Local formulation}
\label{ssec:QFT_SMT}

As anticipated in table \ref{tab:BPS} and below eq.~\eqref{eq:result_partition_function}, increasing the symmetry of the spin group allows to find sectors that effectively behave as field theories with one or two spatial directions.
In this section, we show that this analogy is not only heuristic, but one can find a 1+1 dimensional local formulation of the SMTs with SU(1,1) subgroup symmetry in terms of a QFT \cite{Baiguera:2020jgy}.

We will focus on the SU(1,1$|$1) subsector, which preserves part of the original SUSY of $\mathcal{N}=4$ SYM.
The field content of the theory is given by a complex scalar $\Phi$ and a complex fermion $\psi$, obtained from the letters \eqref{eq:letters_SU123} by setting
\begin{equation}
 \Phi_{n} = (\Phi_1)_{n,0} \, ,  \qquad
  \psi_n=  \chi_{n+1,0}  \, ,
\label{eq:su111_dictionary}
\end{equation}
and all the other modes to zero.
Plugging this mapping in eq.~\eqref{eq:Hint_tot_cubic} leads to 
\begin{subequations}
\begin{equation}
\label{eq:su111_interaction}
H_{\rm int} = \frac{1}{2N} \left[ \sum_{l=1}^\infty \tr \le (\mathcal{B}_0^{\dagger})_l (\mathcal{B}_0)_l \ri + \sum_{l=0}^\infty \tr \le (\mathcal{B}_1^{\dagger})_l (\mathcal{B}_1)_l \ri \right] \,,
\end{equation}
\begin{equation}
\label{eq:blocks_su111}
 (\mathcal{B}_0)_l = \frac{1}{\sqrt{l}} \sum_{n=0}^\infty \left( [ \Phi_{n}^\dagger , \Phi_{n+l} ] + \frac{\sqrt{n+1}}{\sqrt{n+l+1}} \{ \psi^\dagger_{n},\psi_{n+l}\} \right) \,,
\quad
(\mathcal{B}_1)_l = \sum_{n=0}^\infty \frac{1}{\sqrt{n+l+1}}  [ \psi^{\dagger}_{n+l} , \Phi_{n} ] \,,
\end{equation}
\end{subequations}
where the above blocks are obtained from eqs.~\eqref{eq:block_B0}--\eqref{eq:block_B1} after using the dictionary \eqref{eq:su111_dictionary}.
The blocks $(\mathcal{B}_0)_l, (\mathcal{B}_1)_l$ comprise an irreducible representation of the SU(1,1$|$1) group, obtained by acting with raising operators on the lowest weight $(\mathcal{B}_1)_{l=0}$ \cite{Baiguera:2020mgk}. 

\paragraph{Semi-local formulation of the SU(1,1$|$1) sector.}
An analysis of the representations of SU(1,1$|$1) spin group reveals that we can build a superspace composed by the coordinates $(t,x,\theta,\theta^{\dagger})$, where $t \in \mathbb{R}$ is the original time direction of $\mathcal{N}=4$ SYM, $x \sim x+2 \pi $ a periodically identified spatial direction, and $\theta$ a complex Grassmannian coordinate.
In this superspace, we define a differential representation for the supercharge and the associated SUSY covariant derivative
\beq
\mathcal{Q} = \frac{\partial}{\partial \theta} - \frac{i}{2} \theta^{\dagger} \partial_x  \, , \quad
\mathcal{Q}^{\dagger} = \frac{\partial}{\partial \theta^{\dagger}} - \frac{i}{2} \theta \partial_x \, , \quad
 D = i \frac{\partial}{\partial \theta} - \frac{1}{2} \theta^{\dagger} \partial_x  \, , \quad
D^{\dagger} = - i \frac{\partial}{\partial \theta^{\dagger}} + \frac{1}{2} \theta \partial_x \, .
\eeq
(Anti)chiral superfields are annihilated by $D^{\dagger}(D)$.
In this case, we define a chiral fermionic superfield $\Psi$ and a chiral bosonic superfield $\mathcal{A}$ with auxiliary fields\footnote{The terminology \textit{bosonic} or \textit{fermionic} refers to the statistics of the bottom component of the superfield. }
\bea
& \Psi (t,x,\theta, \theta^{\dagger}) = \psi (t,x) + \theta \Phi(t,x) + \frac{i}{2} \theta \theta^{\dagger} \partial_x \psi (t,x) \, , & \\
& \mathcal{A} (t,x, \theta, \theta^{\dagger}) = A (t,x) + \theta \lambda (t,x) + \frac{i}{2} \theta \theta^{\dagger} \partial_x A (t,x) \, , 
\eea
where $\Phi, \psi$ are the dynamical scalar and fermion fields, $A$ is a gauge field and $\lambda$ a gaugino.
All of them transform in the adjoint representation of $\mathrm{SU}(N)$, \ie they carry matrix indices.
The fields $\lbrace A, \lambda \rbrace $ play the role of mediators of other interactions, and are remnants of the gauge invariance of $\mathcal{N}=4$ SYM action.
The antichiral superfields are obtained by hermitian conjugation of the chiral ones.
The action describing the SU(1,1$|$1) sector reads \cite{Baiguera:2020jgy}
\beq
S = \int dt dx \int d \theta^{\dagger} d \theta \, \tr \left[  -i \Psi^{\dagger} \le \mathcal{D}_0 + \mathcal{D}_x - \frac{1}{2} \ri \Psi + \mathcal{A}^{\dagger} \mathcal{A}  \right] \, ,
\label{eq:full_QFT_action_su111_superspace}
\eeq
where the covariant derivatives (with respect to the gauge superfield) are defined by
\beq
\mathcal{D}_0 \Psi \equiv \p_0 \Psi \, , \qquad
\mathcal{D}_x \Psi \equiv \partial_x \Psi - i \tilde{g} [\mathcal{A} , \Psi] - i \tilde{g} [\mathcal{A}^{\dagger}, \Psi] \, .
\eeq
This proposal is simple: the matter part generalizes the Dirac action to the case of a fermionic superfield, while the real part of the gauge superfield enters the covariant derivative as a minimal coupling.
There are some differences with four-dimensional $\mathcal{N}=1$ SUSY. 
First of all, in this superspace the term $\mathcal{A}^{\dagger} \mathcal{A}$ is non-dynamical.
Secondly, any holomorphic combination of the fermionic superfields vanishes due to their Grassmannian nature, \ie $\Psi^2 = 0 , (\Psi^{\dagger})^2 = 0.$

Next, we expand the action in components. Since the gaugino $\lambda$ is auxiliary, it can be integrated out to get
\begin{subequations}
\beq
S=  \int dt dx \, \tr \left\lbrace i \Phi^{\dagger} (\partial_0 + \partial_x) \Phi - \partial_x \psi^{\dagger} (\partial_0 + \partial_x) \psi + i A^{\dagger} \partial_x A  + \tilde{g} A j + \tilde{g} A^{\dagger} j^{\dagger} + \tilde{g}^2 [\Phi, \psi^{\dagger}] [\psi , \Phi^{\dagger}]  \right\rbrace \, ,
\label{eq:full_component_field_QFT_action_su111}
\eeq
\beq
j (t,x) = i \lbrace \partial_x \psi^{\dagger} , \psi \rbrace + [\Phi^{\dagger}, \Phi] \, .
\eeq
\end{subequations}
The current $j(t,x)$ is the version in configuration space of  the block $\mathcal{B}_0$ in eq.~\eqref{eq:blocks_su111}.
The expansion of the fields in momentum space can be obtained from the representation theory of the SU(1,1$|$1) group, leading to
\beq
\Phi(t,x) = \sum_{n=0}^\infty \Phi_n(t) e^{i (n+\frac{1}{2})x} \,, \quad
\psi (t,x) = \sum_{n=0}^\infty \frac{1}{\sqrt{n+1}} \psi_n(t) e^{i(n+1)x} \,, \quad
A(t,x) = \sum_{n=0}^{\infty} A_n (t) e^{inx} \, .
\label{eq:momentum_space_su111fields}
\eeq
Plugging the decompositions \eqref{eq:momentum_space_su111fields} inside the action \eqref{eq:full_component_field_QFT_action_su111}, we can integrate out $A_n(t)$ to obtain the Hamiltonian \eqref{eq:su111_interaction} after a Legendre transform.

The QFT in eq.~\eqref{eq:full_component_field_QFT_action_su111} is linear in both the time and space derivatives, with a structure similar to a complex chiral boson (except that here the field is Grassmann-valued) \cite{Sonnenschein:1988ug}.
Notice that the SMT limit enforces a summation of momenta \eqref{eq:momentum_space_su111fields} restricted to positive integers, which is however necessary to ensure the unitarity of the theory. 
This truncation of the modes implies that the equal-time (anti)commutators of the dynamical fields are non-trivial:
\begin{subequations}
\begin{equation}
[\Phi(t,x),\Phi(t,x')^\dagger ] = i S_{\frac{1}{2}}(x-x') \,, \qquad
\{\psi(t,x), i \partial_{x'} \psi(t,x')^\dagger \} = i S_{1}(x-x') \,,
\label{eq:ET_commutators_scalar}
\end{equation} 
\begin{equation}
S_j(x) = \sum_{n=0}^\infty e^{i (n+j) x} \, .
\end{equation}
\end{subequations}
For this reason, we say that the action \eqref{eq:full_QFT_action_su111_superspace} provides a \textit{semi-local} QFT description of the SU(1,1$|$1) sector.

\paragraph{Generalizations.}
It is natural to expect that the structure \eqref{eq:Hint_tot_cubic} of the Hamiltonian for all the sectors with SU(1,2) subgroup might be derived from a (semi-)local formulation, too.
In the general case, the information about the expansion over the three-sphere is encoded by two quantum numbers $(n,k)$, which suggests the existence of a 2+1 dimensional manifold where the effective QFT should be defined.
This is also confirmed by a counting of the degrees of freedom arising from thermodynamic considerations.
In this regard, it is worth noticing that the null reduction of $\mathcal{N}=4$ SYM, which provides a non-Abelian generalization of SGED (see section \ref{ssec:SGED}), also lives in 2+1 dimensions.
It would be interesting to find a connection between this framework and the SMT limits.

SMT limits are closely connected to the two-dimensional chiral algebra of $\mathcal{N}=4$ SYM.
In this regard, it is relevant to notice that the letters of the two-dimensional chiral algebra are ghost-like, similar to the behaviour of the fields observed in the (semi)local formulation in eq.~\eqref{eq:full_component_field_QFT_action_su111}.
Furthermore, the decoupling condition leading to the PSU(1,1$|$2) sector is equivalent to the Schur condition, and the spin group identically matches with the chiral algebra itself \cite{Beem:2013sza,Bonetti:2016nma}.
In the general case of the PSU(1,2$|$3) sector, the letters $V_I$ in eq.~\eqref{eq:letters_SU123} and the blocks $W_I$ mentioned defined in eqs.~\eqref{eq:block_B0}--\eqref{eq:block_F3} transform in the $(p,q)=(0,I-3)$ of the $\mathfrak{su}$(1,2) algebra.
However, the quantum statistical nature of the blocks is inverted compared to the letters (\ie for any fixed $I=0,1,2,3$, when the letter is bosonic, the block is fermionic, and viceversa). 
Therefore, the study of SMT in its largest sector may provide novel insights to understand the nature of the chiral algebra.

\subsection{Non-relativistic holography}
\label{ssec:nonrel_holography}

In this section, we make contact with holography and show that the SMT limit can be applied to the string theory living on AdS$_5 \times S^5$.
Here we only present a summary of the main results, but a detailed review of the gravity side can be found in \cite{Oling:2022fft} and references therein.

Non-relativistic string theory is a decoupled sector of string theory that can be obtained either \textbf{(1)} by a $c \rightarrow \infty$ limiting procedure or \textbf{(2)} by null reduction of relativistic string theory \cite{Gomis:2000bd,Andringa:2012uz,Harmark:2017rpg,Harmark:2018cdl}.
\textbf{(1)} In the former case, one selects two distinguished longitudinal coordinates in the target space geometry and then picks a critical Kalb-Ramond field to make the $c \rightarrow \infty$ limit regular on the worldsheet \cite{Gomis:2000bd,Andringa:2012uz}.
This procedure generalizes the construction of NC geometry as a limit of a Lorentzian geometry, where a single time direction plays a distinguished role in eq.~\eqref{eq:metric_PNR}, and a background gauge field \eqref{eq:ansatz_Cmu_Jensen} is needed to define a regular action.
The geometry obtained in this new case is called torsional string Newton-Cartan (TSNC) \cite{Bergshoeff:2019pij,Bidussi:2021ujm}.
\textbf{(2)} The second approach performs a null reduction of a Lorentzian target space with null isometry, resulting in a TSNC geometry with longitudinal spatial isometry.
The two methods are equivalent \cite{Harmark:2019upf}.
The SMT sector of non-relativistic string theory corresponds to a further $c \rightarrow \infty$ limit (which can also be seen as a tensionless limit $T \rightarrow 0$ of the string) that makes the worldsheet non-relativistic, \ie whose residual gauge transformations form a Galilean conformal algebra instead of Virasoro \cite{Harmark:2018cdl}.

Remarkably, the above-mentioned tensionless limit can be connected to the decoupling limit \eqref{eq:micro_canonical_limit}.
Using the AdS/CFT dictionary for strings in AdS$_5 \times S^5$, we get
\beq
\boxed{
E - J \rightarrow 0 \, , \qquad
T \rightarrow 0 \, , \qquad
\frac{E - J}{T^2} \,\, \mathrm{fixed} \, , \qquad
N \,\, \mathrm{fixed} \, . }
\label{eq:SMT_limit_strings}
\eeq
From a bulk perspective, $E$ is the energy corresponding to the global time coordinate of AdS space, the generators $\mathbf{S}_1, \mathbf{S}_2$ are the commuting angular momenta of $S^3 \subset \mathrm{AdS}_5$, and $\mathbf{Q}_1, \mathbf{Q}_2, \mathbf{Q}_3$ the commuting angular momenta of the $S^5$ factor. This is summarized in table \ref{tab:Cartan_charges}. The metric of the spacetime reads ($\ell$ is the AdS radius)
\beq
ds^2 = \ell^2 \le - \cosh^2 \rho dt^2 + d\rho^2 + \sinh^2 \rho d\Omega_3^2 + d\Omega_5^2 \ri \, .
\label{eq:metric_AdS5S5}
\eeq 
The tensionless limit \eqref{eq:SMT_limit_strings} on the background \eqref{eq:metric_AdS5S5} coincides with the above-mentioned $c \rightarrow \infty$ limit for the SMT regime of non-relativistic string theory if we identify \cite{Harmark:2017rpg}
\beq
T = \frac{1}{2 \pi c} \, , \qquad
c = \frac{1}{\sqrt{4 \pi g_s N}} \, ,
\eeq
where $g_s$ is the string coupling.
In particular, one can send $g_s \rightarrow 0$ while keeping $N$ fixed.

We can now implement the limit \eqref{eq:SMT_limit_strings} at the level of the geometry \eqref{eq:metric_AdS5S5}.
Let us denote $\mathbf{S} \equiv \mathbf{S}_1 + \mathbf{S}_2$ and $\mathbf{Q} \equiv  \mathbf{Q}_1 + \mathbf{Q}_2 + \mathbf{Q}_3$.
In general, one can always pick coordinates $(u,x^0,x^i)$ on AdS$_5 \times S^5$ such that \cite{Harmark:2020vll}
\beq
i \p_u = \frac{1}{2} \le  E - \mathbf{S} + \mathbf{Q} \ri  \, , \qquad
i \p_{x^0} = E - \mathbf{S} - \mathbf{Q} \, .
\eeq
These combinations are chosen to make the $u$ coordinate null on a NC submanifold, where the SMT limit forces the string to live.
After taking the limit \eqref{eq:SMT_limit_strings}, the momentum in the null direction is fixed to $P_u = \mathbf{Q}$, which is interpreted as the length of a spin chain on the QFT side of AdS/CFT duality.
The action of the string on the worldsheet simplifies by imposing a gauge-fixing condition and solving for the dynamical modes. One finally gets
\beq
S = - \frac{\mathbf{Q}}{2 \pi} \int d^2 \sigma \, \le m_{\mu} \p_0 X^{\mu} + \frac{1}{2} h_{\mu\nu} \p_1 X^{\mu} \p_1 X^{\nu}  \ri \, ,
\label{eq:sigma_model_SMT}
\eeq
where $\lbrace \sigma^0, \sigma^1 \rbrace $ are coordinates on the worldsheet, $X^{\mu}$ are the embedding fields, and we recognize the appearance of the data $m_{\mu}, h_{\mu\nu}$ of NC geometry, see section \ref{ssec:NC_geometry}.
This is the prototype for any sigma-model that arises from spin chains.
In particular, the simplest choice $\mathbf{Q} = \mathbf{Q}_1 + \mathbf{Q}_2$ leads to the SU(2) SMT sector, with sigma model \eqref{eq:sigma_model_SMT} given by
\beq
S_{\mathrm{LL}} = \frac{\mathbf{Q}}{2 \pi} \int d^2 \sigma \, 
\left[ \cos \theta \, \dot{\phi} - \frac{1}{4} \le (\theta')^2 + \sin^2 \theta (\phi')^2 \ri  \right] \, ,
\label{eq:LL_action}
\eeq
where $\dot{f} \equiv df/d\sigma^0,$ $f' \equiv df/d\sigma^1$, and $(\theta, \phi)$ parametrize an $S^2$ inside the factor $S^5$ of the geometry \eqref{eq:metric_AdS5S5}.
This is the Landau-Lifshitz model, which describes the XXX$_{1/2}$ Heisenberg spin chain in the large $\mathbf{Q}$ limit.
The direction $\sigma^1$ is a direction along which the string has winding, and corresponds on the field theory side to the position along the spin chain.
These results precisely match the SMT limit performed on the field theory side in section \ref{ssec:examples_SMT}, providing a solid test of AdS/CFT duality in this non-relativistic corner.

Let us stress that the SMT limit in eq.~\eqref{eq:micro_canonical_limit} sends $\lambda \rightarrow 0$ with $J = \mathbf{S}+ \mathbf{Q}$ fixed.
On the other hand, it is also possible to obtain the same SU(2) Landau-Lifshitz sigma model by a Penrose limit of strings on AdS$_5 \times S^5$ where $\mathbf{Q} \rightarrow \infty$, while $\lambda/\mathbf{Q}^2$ is fixed \cite{Kruczenski:2003gt}.
Penrose limits for any SMT sector of string theory were studied in \cite{Harmark:2020vll}; SMT limits of relativistic $pp$--wave backgrounds were investigated in \cite{Grignani:2009ny}, where a matching with the field theory side was found. 
Spinning solutions of the SMT string sigma models were studied in \cite{Roychowdhury:2020dke,Roychowdhury:2021wte}.

\section{Discussion}
\label{sec:discussion}

\subsection{Summary}

In this work we reviewed several aspects of non-relativistic QFTs, mainly focusing on models with Schr\"{o}dinger invariance.
After introducing some preliminary material in section \ref{sec:preliminaries}, we started in section \ref{sec:historical_review} with a historical overview of the early achievements in this reseach field. 
Those included a non-relativistic formulation of the
Dirac action and of electromagnetism, the study of anyons coupled to non-relativistic matter and an explicit computation of the trace anomaly for a Schr\"{o}dinger scalar with quartic interactions.
The tools of non-relativistic QFT allowed us to find exact solitonic solutions to the wave equations in the Jackiw-Pi model, to compute the scattering amplitude in the Aharonov-Bohm problem, and to show the existence of an enhanced $\mathcal{N}=2$ supersymmetry at a critical point in these Chern-Simons models.

We then moved to present modern methods and applications.
The main technologies (collected in section \ref{sec:modern_NR_QFT}) consisted of a $c \rightarrow \infty$ limit, which was usually compensated by the inclusion of additional critical background fields to make the limit regular; and of the null reduction of a relativistic theory defined on a Lorentzian manifold with null isometry.
These methods were used in section \ref{sec:modern_applications} to study condensed matter systems (such as fermions at unitarity and the quantum Hall effect), to find off-shell action formulations for various fields, and to compute trace anomalies.
We also discussed a deformed version of null reduction which led to the construction of QFTs invariant under SU$(1,n)$, which play an important role in the non-relativistic limit of $M$--theory.

The role of supersymmetry for Schr\"{o}dinger-invariant QFTs was discussed in section \ref{sec:SUSY}. 
We constructed a non-relativistic superspace to covariantly deal with the quantum corrections of two models in 2+1 dimensions obtained from the null reduction of a relativistic parent theory: a Galilean version of the Wess-Zumino model, and supersymmetric Galilean electrodynamics.
The causal properties of the propagator implied the existence of non-renormalization theorems.
As a consequence, the non-relativistic version of the Wess-Zumino turned out to be one-loop exact.
Surprisingly, supersymmetric Galilean electrodynamics (SGED) required the addition of an infinite class of marginal deformations to become renormalizable.
After this modification, SGED admitted a conformal manifold of fixed points where super-Schr\"{o}dinger invariance is achieved.

In section \ref{sec:SMT} we presented a different path towards non-relativistic physics, based on decoupling limits of the AdS/CFT correspondence that can be either interpreted as zero-temperature critical points (in the grand-canonical ensemble) or near-BPS limits (in the microcanonical ensemble).
The remarkable advantage of this approach is that in principle a precise matching of the duality can be attained without any assumption on the number of colours $N$, allowing to investigate non-perturbative phenomena.

\subsection{Future developments}

The results reviewed in this work open the way for several developments that we propose below.
While most of the theoretical foundations behind non-relativistic theories are solid and well-studied (see section \ref{sec:preliminaries}), the main disadvantage of Schr\"{o}dinger invariance is that it is less powerful than conformal symmetry.
As a result, more freedom is allowed and the structure of correlation functions with three or more external fields is not entirely fixed by Schr\"{o}dinger invariance.
This also makes the conformal bootstrap approach less constraining in the non-relativistic case.
While some attempts were performed in references \cite{Goldberger:2014hca,Chen:2020vvn,Chen:2022jhx}, it would be interesting to push this approach to sharpen the bounds that restrict the existence of non-relativistic CFTs. 

Another foundational problem is the relation between non-relativistic scale and conformal invariance, discussed in section \ref{ssec:Schroed_group}.
While in the relativistic case there are strong indications that dilatation invariance plus unitarity imply conformal invariance \cite{Polchinski:1987dy,Dymarsky:2013pqa}, including a proof in two and four dimensions, instead the situation is more obscure in the non-relativistic case \cite{Nakayama:2013is}.
The main technical obstructions are the absence of reflection positivity (also known as Reeh-Schlieder theorem) and the different causal structure of non-relativistic theories.
For these reasons, it is not obvious that the energy-momentum tensor of a Galilean-invariant theory with scale symmetry can be improved to become traceless.
Some perturbative attempts were developed in reference \cite{Nakayama:2009ww}, but no conclusive proof was found.
An alternative approach to this problem in the relativistic case is based on the application of monotonicity theorems along an RG flow \cite{Dymarsky:2013pqa}.
However, while there are candidates for a non-relativistic version of the $a$--theorem as discussed in section \ref{ssec:nonrel_trace_anomalies}, there are technical issues that make the proof of monotonicity challenging.
For instance, there is no positive-definite Zamolodchikov metric defined on conformal manifolds. 
Therefore, it would be important to find evidence either in favour or contrary to the enhancement of Galilean plus scale invariance into the Schr\"{o}dinger symmetry as a guiding principle to search for a proof or a counterexample to the problem.
In this regard, it is interesting to notice that a non-Lorentzian setting where scale invariance plus unitarity imply conformal invariance is provided by two-dimensional warped CFTs, as it was proven in \cite{Hofman:2011zj}.\footnote{Warped CFTs are a deformation of relativistic CFTs invariant under chiral dilatations. In two dimensions, their symmetry group is given by Virasoro times a Kac-Moody algebra.}

Euclidean space plays a fundamental role in QFT, since it is the ambient space where path integrals, correlation functions and other relevant quantities are properly defined.
In the relativistic case, it is possible to convert quantities from Lorentzian to Euclidean signature using a Wick rotation from real to imaginary time $t \rightarrow i t_E$.
In perturbative computations, this is usually a harmless procedure, since Cauchy theorem allows to change the contour of integration as long as we have an analytic function defined over a closed region.
This procedure is not trivial in the non-relativistic case because the propagator for a Schr\"{o}dinger field is not defined according to the Feynman prescription, but is purely retarded.
In particular, there is a different structure of the poles in the complex plane defined by the energy $\omega$ conjugate to the time direction.
Another manifestation of this difference arises when the Schr\"{o}dinger operator has to be continued to Euclidean space, as we discussed in section \ref{ssec:nonrel_trace_anomalies} to perform the heat kernel technique and to explicitly compute the trace anomaly in specific examples.
It would be desirable to get a deeper understanding of the causal structure and of the analytic continuation to imaginary time for Schr\"{o}dinger-invariant theories.

Let us now propose future developments related to some of the applications discussed in this review.
Chern-Simons theories are interesting because their topological nature allows us to couple them to non-relativistic matter.
These models enjoy an enhancement to $\mathcal{N}=2$ supersymmetry in 2+1 dimensions (see section \ref{ssec:SUSY_CS}) which makes possible to find exact results, such as the static solutions to the equations of motion of the Jackiw-Pi model (see section \ref{ssec:Jackiw_Pi}).
In this context, it would be nice to test whether the methodologies developed to study decoupling limits of $\mathcal{N}=4$ SYM in section \ref{ssec:decoupling_limits_SMT} can also be applied to investigate the low-energy limit of field theories in other dimensions.
The most relevant example is $\mathcal{N}=6$ superconformal Chern-Simons theory in 2+1 dimensions (known as ABJM) \cite{Aharony:2008ug}, 
which admits a subsector with SU(2)$\times$SU(2) symmetry described by two decoupled Heisenberg spin chains \cite{Grignani:2008is}.
One may also hope to derive these models from null reduction and study their properties using a superfield formulation.

A deformed version of null reduction was used in section \ref{ssec:Lambert_QFT} to derive a class of QFTs with SU$(1,n)$ symmetry.
First of all, these models represent novel non-relativistic field theories, and the investigation of their classical and quantum properties would be an interesting problem by itself.
Furthermore, a deformed version of the non-relativistic state/operator correspondence was proposed in \cite{Lambert:2021nol}, by identifying the energy eigenvalues of a Hamiltonian in a harmonic potential in the state picture with the scaling dimension of the fields in the operator picture.
Finding the geometric map that realizes the state/operator correspondence in this non-relativistic corner can shed light on the conformal properties of these models \cite{progress}.
Moreover, several SMT decoupling limits of $\mathcal{N}=4$ SYM collected in table \ref{tab:BPS} admit SU$(1,2)$ symmetry as a subgroup, thus fitting inside the class of novel QFTs mentioned above.
It is natural to expect that the structure \eqref{eq:Hint_tot_cubic} of the Hamiltonian for all the sectors with SU(1,2) subgroup might be derived from a local formulation, along the lines of the methods developed in section \ref{ssec:QFT_SMT}.
The existence of a map between state and operator picture for field theories with SU$(1,n)$ invariance may also provide a natural background to define a local formulation for these SMTs.

One may want to extend the analysis of non-relativistic supersymmetry to curved space, including the case of NC supergravity \cite{Andringa:2013mma,Bergshoeff:2015uaa,Bergshoeff:2015ija}.
The general structure of terms that must be included in a relativistic action, and the backgrounds to which a QFT should be coupled to preserve supersymmetry, were analyzed in \cite{Festuccia:2011ws,Dumitrescu:2012ha}.
It would be intriguing to study the analogous problem in the non-relativistic setting, as initiated in \cite{Bergshoeff:2020baa}. 
This would provide a first step towards a non-relativistic version of localization \cite{Pestun:2007rz,Kapustin:2009kz}.

An important goal of the SMT approach discussed in section \ref{sec:SMT} is to investigate a non-relativistic sector of AdS/CFT duality. 
A great advantage of the positive-definite structure of the SMT Hamiltonian \eqref{eq:Hint_tot_cubic} is that the strong-coupling regime $\tilde{g} \gg 1$ implies at leading order that only configurations with $H_{\rm int}=0$ survive, resulting in the 
constraints
\beq
(\mathcal{B}_I)_{n,k} = (\mathcal{F}_I)_{n,k} = 0 \, , \quad
\mathrm{where} \,\, I=0,1,2,3  \, .
\label{eq:SMT_constraints}
\eeq
A subset of these conditions, together with the computation of certain $\frac{1}{8}$--BPS solutions, were found in \cite{Mandal:2006tk,Grant:2008sk}.
A remarkable feature of the PSU(1,2$|$3) sector is the presence of black hole solutions, including $\frac{1}{16}$--BPS configurations \cite{Gutowski:2004yv,Chong:2005hr,Kunduri:2006ek,Chong:2005da}.
In this context, it has been a long-term puzzle to find contributions to the entropy of order $N^2$ from a field theory computation \cite{Kinney:2005ej,Chang:2013fba}, and recent advances were mainly based on the computation of superconformal indices \cite{Benini:2018ywd,Murthy:2020rbd,Goldstein:2020yvj}.
It was observed in \cite{Choi:2018hmj} that the superconformal index predicts the existence of a $\frac{1}{8}$--BPS black hole in the PSU(1,1$|$2) subsector, but this solution has never been found in the literature.
The main novelty of the approach presented in section \ref{ssec:effective_SMT} is the identification of the blocks as irreducible representations of the $\mathfrak{psu}$(1,2$|$3) algebra.
This feature makes the symmetries of the constraints manifest and should help in searching for the space of solutions of eq.~\eqref{eq:SMT_constraints}, including an alternative understanding of the microstates from a SMT perspective.
For instance, a common feature shared only by PSU(1,1$|$2) and PSU(1,2$|$3) sectors is the simultaneous presence of D-- and F--terms in the Hamiltonian.
While this may be a coincidence, it is worth exploring the relation between this observation and the existence of black hole solutions.
Since the SMT Hamiltonian \eqref{eq:Hint_tot_cubic} describes an effective theory where the exact BPS conditions are broken, this framework can also reveal the interesting physics of near-BPS black holes \cite{Larsen:2019oll}.

One can further push the AdS/CFT correspondence in this non-relativistic limit to find other implications on the gravity side for various regimes in the number $N$ of colours of the gauge group.
In the strict planar limit $N=\infty$, one has access from the field theory side to an integrable spin chain description that directly compares with sigma models in non-relativistic string theory, as reviewed in section \ref{ssec:nonrel_holography}.
In this regime, one can compute the spectrum of any possible SMT by employing integrability techniques, extending the analysis performed in the SU(2) case \cite{Harmark:2008gm}.
Moving to perturbative corrections in $1/N$, the connection between the sigma models on the string theory side and the continuum limit of the SMT spin chain can be extended to include the interactions between strings \cite{Harmark:2019upf}, which correspond to joining and splitting of spin chains from the field theory perspective.
The most interesting regime is the case at finite $N$, where non-perturbative effects play a major role.
A great success of the SMT framework was the discovery that the strongly-coupled ($\tilde{g} \gg 1$) regime of the SU(2) sector 
admits an effective description in terms of giant gravitons described by a Dirac-Born-Infeld action \cite{Harmark:2016cjq}.
It is natural to expect that one could extend this analysis to the other decoupling limits listed in table \ref{tab:BPS}.

Therefore, many open problems still need to be solved, and we expect many other interesting features of non-relativistic theories to be discovered in the upcoming years!

\section*{Acknowledgements}
I am grateful to Roberto Auzzi, Oren Bergman, Troels Harmark, Jelle Hartong, Yang Lei, Giuseppe Nardelli, Silvia Penati, Ziqi Yan and especially Shira Chapman and Gerben Oling for a careful reading and valuable comments on a preliminary version of this work.
I also thank Shira Chapman for sharing the lecture notes \cite{Shiralectures} from the \textit{1st school on Non-relativistic Quantum Field Theory, Gravity and Geometry}, which inspired section \ref{sec:preliminaries} of this review.
I am supported by the Israel Science Foundation (grant No. 1417/21), by the German Research Foundation through a German-Israeli Project Cooperation (DIP) grant “Holography and the Swampland”, by Carole and Marcus Weinstein through the BGU Presidential Faculty Recruitment Fund and by the ISF Center of Excellence for theoretical high energy physics. 
I acknowledge support by an Azrieli fellowship funded by the Azrieli foundation.
This version of the article has been accepted for publication, after peer review, but is not the Version of Record. The Version of Record is available online at: \href{https://link.springer.com/article/10.1140/epjc/s10052-024-12630-y}{https://link.springer.com/article/10.1140/epjc/s10052-024-12630-y}.

\appendix

\section{Conventions}
\label{app:conventions}

Except where explicitly stated, we set $\hbar=1$ through the entire manuscript.
The speed of light $c$ is kept explicit whenever we are performing any limit or expansion, but is otherwise also set to 1.

\subsection{Acronyms}

\begin{longtable}{rcl}
UV &     & ultraviolet \\[1.2mm]
IR &     & infrared \\[1.2mm]
QM        &             & quantum mechanics  \\[1.2mm]
QFT         &             & quantum field theory  \\[1.2mm]
CFT         &             & conformal field theory  \\[1.2mm]
SMT &     & Spin Matrix Theory \\[1.2mm]
NRCFT         &             & non-relativistic conformal field theory  \\[1.2mm]
OPE      &             & operator product expansion  \\[1.2mm]
1PI    &             & one-particle irreducible  \\[1.2mm]
CS &     & Chern-Simons \\[1.2mm]
EOM &     & equations of motion \\[1.2mm]
GED &     & Galilean electrodynamics \\[1.2mm]
RG &     & renormalization group \\[1.2mm]
AB &     & Aharonov-Bohm \\[1.2mm]
SUSY &     & supersymmetry \\[1.2mm]
SGED &     & supersymmetric Galilean electrodynamics \\[1.2mm]
NC &     & Newton-Cartan \\[1.2mm]
TNC &     & torsional Newton-Cartan \\[1.2mm]
TTNC &     & twistless torsional Newton-Cartan \\[1.2mm]
PNR &     & pre non-relativistic \\[1.2mm]
DLCQ &     & discrete light-cone quantization \\[1.2mm]
KK &     & Kaluza-Klein \\[1.2mm]
FQH &     & fractional quantum Hall \\[1.2mm]
IQH &     & integer quantum Hall \\[1.2mm]
WZ &     & Wess-Zumino \\[1.2mm]
HK &     & heat kernel \\[1.2mm]
SGED &     & supersymmetric Galilean electrodynamics \\[1.2mm]
STNC &     & string torsional Newton-Cartan \\[1.2mm]
\end{longtable}

\bibliographystyle{JHEP}

\bibliography{bibliography}

\end{document}